\documentclass[twocolumn, aps,pre,superscriptaddress,showkeys]{revtex4-1}
\usepackage{graphicx,color,colortbl}
\usepackage{epstopdf}
\usepackage{bm}
\usepackage{eucal}
\usepackage{mathrsfs}

\usepackage[dvipsnames]{xcolor}
\usepackage{hyperref}
\definecolor{hypcolor}{named}{BlueViolet}
\hypersetup{colorlinks=true, citecolor=hypcolor, urlcolor=hypcolor, linkcolor=hypcolor}

\usepackage{amsmath}
\usepackage{amsthm}
\usepackage{epsfig}
 
\newcommand{\be}{\begin{equation}} 
\newcommand{\ee}{\end{equation}}
\newcommand{\bea}{\begin{eqnarray}} 
\newcommand{\eea}{\end{eqnarray}}
\newcommand{\bes}{\begin{subequations}} 
\newcommand{\ees}{\end{subequations}}

\begin{document}
\title{Cluster States and $\pi$-Transition \\ in the Kuramoto Model with Higher Order Interactions}
\author{Alejandro Carballosa}
\affiliation{Group of Nonlinear Physics, Fac. Physics, University of Santiago de Compostela, 15782 Santiago de Compostela, Spain}
\affiliation{Galician Center for Mathematical Research and Technology (CITMAga), 15782 Santiago de Compostela, Spain}
\author{Alberto P. Mu\~nuzuri}
\affiliation{Group of Nonlinear Physics, Fac. Physics, University of Santiago de Compostela, 15782 Santiago de Compostela, Spain}
\affiliation{Galician Center for Mathematical Research and Technology (CITMAga), 15782 Santiago de Compostela, Spain}
\author{Stefano Boccaletti}
\affiliation{CNR, Consiglio Nazionale delle Ricerche, Istituto dei Sistemi Complessi, 50019, Sesto Fiorentino, Italy}

	\author{Alessandro Torcini}
	\affiliation{Laboratoire de Physique Théorique et Modélisation, UMR 8089, CY Cergy Paris Université, CNRS, 95302 Cergy-Pontoise, France}
	\affiliation{CNR, Consiglio Nazionale delle Ricerche, Istituto dei Sistemi Complessi, 50019, Sesto Fiorentino, Italy}		
	\affiliation{INFN, Sezione di Firenze, Via Sansone 1, I-50019 Sesto Fiorentino (FI), Italy}
    \author{Simona Olmi}
    \email[corresponding author: ]{simona.olmi@fi.isc.cnr.it}
	\affiliation{CNR, Consiglio Nazionale delle Ricerche, Istituto dei Sistemi Complessi, 50019, Sesto Fiorentino, Italy}
	\affiliation{INFN, Sezione di Firenze, Via Sansone 1, I-50019 Sesto Fiorentino (FI), Italy}
	
	\date{\today}

\begin{abstract}
We have examined the synchronization and de-synchronization transitions observable in 
the Kuramoto model with a standard pair-wise first harmonic interaction plus
a higher order (triadic) symmetric interaction for unimodal and bimodal Gaussian distributions of the natural frequencies
$\{ \omega_i \}$. These transitions have been accurately characterized thanks to a self-consistent mean-field approach joined with extensive numerical simulations.
The higher-order interactions favour the formation of two cluster states, which emerge from the incoherent regime via continuous (discontinuos) transitions for unimodal (bimodal) distributions. Fully synchronized initial states
give rise to two symmetric equally populated bimodal clusters, each characterized by either positive or negative natural frequencies. These bimodal clusters are formed at an angular distance $\gamma$, which increases for decreasing pair-wise couplings until it reaches $\gamma=\pi$ (corresponding to an anti-phase configuration),
where the cluster state destabilizes via an abrupt transition: the $\pi$-transition. The uniform clusters that reform immediately after (with a smaller angle $\gamma$) are composed by oscillators with positive and negative $\{ \omega_i \}$. For bimodal distributions we have obtained detailed phase diagrams involving all the possible
dynamical states in terms of standard and novel order parameters. In particular,
the clustering order parameter, here introduced, appears quite suitable to characterize the 
two cluster regime. As a general aspect,  hysteretic (non hysteretic) synchronization transitions, mostly mediated by the emergence of standing waves, are observable for attractive (repulsive) higher-order interactions.
\end{abstract}

\maketitle

\section{Introduction}

The Kuramoto model \cite{Kuramoto2012} represents the paradigmatic framework in which to investigate the synchronization 
phenomenon occurring when a large heterogeneous population of oscillators lock their phases at unison. 
The Kuramoto model has been successfully employed to recover the main synchronization features observable in
many contexts, ranging from biological scenarios \cite{Winfree1967,Buck1988,White1998,Waters2005} to human behaviours \cite{Neda2000}.
Due to its success, the model has been thoroughly studied in many different variations and set-ups \cite{Acebron2005,Arenas2008}. A definitive breakthrough in the studies dedicated to Kuramoto model is represented by the exact reduction methodology developed by  Ott and Antonsen in 2008 \cite{ott2008}. The Ott-Antonsen (OA) approach allows us to rewrite, for Lorentzian distributed frequencies, the mean-field evolution of the network in terms of a complex macroscopic field, the so-called  Kuramoto order parameter. This methodology has generated a renewal of interest in the field, allowing for low-dimensional investigation of phase oscillator networks.

The Kuramoto model has been firstly analyzed in a globally coupled configuration with an unimodal distribution of the natural frequencies. In this context, by varying the coupling strength, one observes a transition from an asynchronous (AS) to  a partially synchronized (PS) regime, where a finite fraction of oscillators are locked and rotate coherently with a common frequency.
For a sufficiently strong coupling the system can eventually become fully synchronized (FS) with all oscillators phase locked \cite{Acebron2005}.
Plenty of attention has been also devoted to the case where the frequency distribution is bimodal, i.e characterized by two symmetric peaks \cite{Kuramoto2012,Crawford1994}. In particular, for bimodal Lorentzian distributions, the phase diagram can be obtained analytically in the thermodynamic limit by employing the OA approach \cite{Martens2009}. In particular in \cite{Martens2009} it has been shown that, besides the previous mentioned regimes, there is also room for multistability and a limit-cycle solution of the order parameter, termed standing wave (SW) and characterized by two counter-rotating clusters with the same angular velocity. Travelling waves (TWs) are also expected in this set up whenever the bimodal distribution becomes asymmetric \cite{Martens2009}.The presence of hysteretic transitions \cite{Pazo2009} and other complex, time dependent states \cite{Bi2016} has been also shown to be a consequence of introducing a bimodal distribution of frequencies.  

Another generalization of the original Kuramoto model consisted in the modification of the coupling function by including either higher order harmonics \cite{winfree1980,daido1996} or, more recently, by extending the pair-wise interaction to many body terms \cite{Battiston2020, boccaletti2023}. For each extra harmonics appearing in the coupling term, the interaction of the single phase oscillator with all the others is mediated by an extra macroscopic field, termed Kuramoto-Daido order parameter \cite{daido1995,daido1996}. If the interaction is pair-wise, the Kuramoto-Daido order parameters appear linearly in the phase evolution equation, while many body interactions give rise to the emergence of non-linear combinations of these order parameters \cite{Tanaka2011,komarov2014, komarov2015,Battiston2020, boccaletti2023}.

Higher harmonics, as well as many body terms, emerge naturally by considering the dynamics of $N$ coupled dynamical systems in the proximity of a super-critical Hopf bifurcation \cite{ashwin2016} (or of a mean field Complex Ginzburg-Landau equation \cite{leon2019}), whenever such dynamics can be reduced to that of coupled phase oscillators, i.e. for sufficiently small coupling terms. 
In particular, at the lowest orders besides the usual Kuramoto-Sakaguchi model, a bi-harmonic coupling function as well as three body interactions terms always emerge. The latter ones can be either symmetric or asymmetric with respect to the  reference oscillator, thus implying that its phase appears either as a higher order or a simple harmonic interaction term, respectively \cite{Battiston2020, boccaletti2023}. The many body asymmetric interactions have been largely examined, because, in this case, the OA approach can be fruitfully employed to reduce the mean-field dynamics to the evolution of few macroscopic indicators, thus allowing for the derivation of analytic results \cite{Skardal2020}.

The inclusion of higher harmonics in the coupling function gives rise to the emergence of a large number of coexisting stable or multi-stable 
phase-locked states at different locations on the unitary circle, termed cluster states \cite{winfree1980,daido1995,Tanaka2011,komarov2014}.
A similar phenomenology has been also reported when considering only three body symmetric interactions; in such a case, the mean-field interaction
is quadratic in the modulus of the Kuramoto order parameter $R_1$ and bi-harmonic in the single oscillator phase \cite{Tanaka2011, komarov2015, Skardal2019}.
Therefore, bi-clusters in phase opposition are expected in this case. It should be noted that, if the bi-clusters are equally populated, the level of synchronization in the network, as measured by $R_1$, will 
be exactly zero. As a matter of fact the bi-clusters always coexist with the asynchronous regime, which remains stable in the thermodynamic limit \cite{Tanaka2011}. 
Finite  size fluctuations may be responsible for the destabilization of the asynchronous regime and the emergence  of asymmetrically populated clusters with an associated 
finite level of synchronization \cite{komarov2015}.
By considering as initial states the asymmetrically populated bi-clusters, one observes that they loose stability via an abrupt de-synchronization towards
the asynchronous regime, where the system remains forever \cite{Skardal2019}. The origin of this de-synchronization transition has been recently
related to a collective mode instability \cite{Xu2020}. It is worth mentioning that the bi-clusters (CSs), here analyzed, are different from the previously mentioned 
TW solutions, since they are characterized by coexisting phase-locked clusters, as for the TWs, but where all oscillators are frequency locked. 

Furthermore, in \cite{Skardal2020, Dai2021}, it has been shown that, by considering both pairwise and higher order interactions in the system,  
the latter ones can stabilize synchronized states even for repulsive pair-wise coupling, a situation prohibited in the classical Kuramoto model. 
Furthermore, it was recently shown that, even with both pairwise and triadic coupling being repulsive, the two-cluster state can exist when the triadic strength is sufficiently large \cite{Kovalenko2021}. In social terms, this means that some level of agreement can exist in the network despite all the individuals being contrarians to the mainstream.

In this work we aim at studying the combined effect of introducing a bimodal frequency distribution in the Kuramoto model with the presence of higher order interactions, that can be either attractive or repulsive. In particular, by adapting the approach developed in \cite{komarov2014}
for a bi-harmonic coupling function to our set-up, we have performed a mean-field analysis based on the identification of self-consistent stationary solutions for the Kuramoto order parameter. The mean-field results, joined with direct numerical simulations, allow us to identify all the
possible dynamical regimes and to characterize the transitions separating them, thus being able to obtain a quite detailed two dimensional phase diagram in terms of the pair-wise and higher order coupling terms.

This paper is structured as follows. In Section II we describe the model and the employed coherent indicators, as well as the simulation protocols. Section III is devoted to report the mean-field formulation and to the investigation and comprehension of the possible synchronization and de-synchronization transitions occurring for unimodal and bimodal Gaussian distributions of the natural frequencies, for attractive and repulsive higher-order interactions. The bimodal Kuramoto model with higher order interactions is further investigated in Section IV via numerical simulations. In particular, in this Section we report all the emergent dynamical regimes,
as well as the phase diagrams obtained for fully synchronized and random initial states.
The sub-section IV C is particularly devoted to the detailed characterization of the $\pi$-transition associated to the clustered regimes.  A summary of the obtained results and a brief discussion are reported
in Section V. A simplified version of the studied model depending on only one coupling
parameter is studied and deeply analyzed in the Appendix.

\section{Methods}

\subsection{Models}
 
In the original Kuramoto  model \cite{Kuramoto2012}, the ensemble of $N$ oscillators is coupled all-to-all
and the evolution equation for a generic phase oscillator is given by
\begin{equation}
    \frac{d\theta_i(t)}{dt} = \omega_i+\frac{\lambda}{N}\sum^N_j\sin{(\theta_j-\theta_i)}
\quad i=1,\dots,n      
    \label{eq:1}
\end{equation}
with $\omega_i$ being the natural frequency of the $i$-th oscillator and $\lambda$ the coupling strength.

For globally coupled phase oscillators the evolution equation
can be generally written in terms of complex mean-field variables,
termed Kuramoto-Daido order parameters, which take into account the 
interaction with the rest of the network, specifically :
\begin{equation}
  Z_m =  R_m {\rm e}^{j \Psi_m} = \frac{1}{N} \sum_{i=1}^N {\rm e}^{j m \theta_i}  \quad .
    \label{R}
\end{equation}
For $m=1$ one obtains the so-called Kuramoto order parameter,
whose modulus $R_1$  measures the level of synchronization among the phase oscillators:
$R_1 \propto 1/\sqrt{N}$ for an asynchronous state, while $R_1$ is finite
(one) in a partially (fully) synchronized situation.

Since the coupling interaction in Eq. \eqref{eq:1} is limited to the first harmonic 
and pair-wise, this can be rewritten simply in terms of $Z_1$ appearing 
linearly in the equation, as follows  
\begin{equation}
    \frac{d\theta_i(t)}{dt} = \omega_i+
    R_1 \sin(\Psi_1 - \theta_i) 
\quad i=1,\dots,n      \quad.
    \label{eq:1mf}
\end{equation}
 
In the absence of interactions, it is expected that the phase of each oscillator evolves according to its natural frequency $\omega_i$. Consequently, it is natural to study how the system behaves as a function of the chosen frequency distribution $g(\omega)$. The most widely studied is a symmetric, unimodal distribution, peaked at some value $\omega_0$ \cite{Acebron2005}.  In this case, for increasing values of $\lambda$, the system passes from being asynchronous (AS), to partially synchronized (PS), and finally, for large enough values of $\lambda$, to a fully synchronized (FS) regime.

In this paper we consider an extension of the classical Kuramoto model by adding a term for many body interactions; specifically 
we will consider three body (or triadic)  symmetric interactions. Therefore, the triadic interaction term will depend simultaneously on
the phase differences between the reference oscillator $i$ and all the other possible couples $j$ and $k$, i.e. on
$(\theta_i - \theta_j)$ and $(\theta_i - \theta_k)$.
For both pair-wise and triadic interactions we will consider an all-to-all configuration, where all possible pairs and triangles are formed.
Moreover we consider different weights for each interaction, being $\lambda_1$ the pair-wise interaction strength and $\lambda_2$ the triadic interaction strength,
namely:
\begin{eqnarray}
  \frac{d\theta_i(t)}{dt} &=& \omega_i+\frac{\lambda_1}{N}\sum^N_j\sin{(\theta_j-\theta_i)} \nonumber \\
  &+&\frac{\lambda_2}{2N^2}\sum^N_{j=1}\sum^N_{k=1}\sin{(\theta_j+\theta_k-2\theta_i)}  \quad .      
    \label{eq:2}
\end{eqnarray}

This model can be rewritten in terms of the mean-field variable $Z_1$ as follows
\begin{eqnarray}
\frac{d\theta_i(t)}{dt} &=& \omega_i+
     \lambda_1 R_1 \sin(\Psi_1 - \theta_i) \nonumber \\
    &+& \frac{\lambda_2}{2} R_1^2 \sin(2(\Psi_1-\theta_i))    \quad i=1,\dots,N \quad ;  
    \label{eq:mf2}
\end{eqnarray} 
where, due to the triadic symmetric interaction, a quadratic term $R_1^2$ is now present, as well as a bi-harmonic interaction \cite{Tanaka2011,komarov2015}.
It should be noted that for asymmetric triadic interactions, as the one studied in \cite{Skardal2020},  one would observe a single harmonic term for the reference oscillator phase joined 
to a nonlinear mean-field term of the type $R_1 R_2$, therefore a quite different dynamics should be expected in such a case.

The natural frequencies $\{\omega_i\}$ are usually assumed to be distributed as a bimodal
Gaussian, which can be written as the superposition of two Gaussians with centers $\pm\omega_0$ and standard deviation $\Delta$ as follows:
\begin{equation}
    g_B(\omega) = \frac{1}{2 \sqrt{2\pi} \Delta}\left(e^{\frac{-(\omega+\omega_0)^2}{2\Delta^2}}+e^{\frac{-(\omega-\omega_0)^2}{2\Delta^2}}\right)  \quad .
\end{equation}
In the following, the simulations will be performed by randomly generating $\{\omega_i\}$ from the above distribution with the additional constraint
that 50\% of the oscillators are associated to the negatively (positively) peaked distribution, thus ensuring equally distributed oscillators. 
Moreover, we will also report some results for unimodal Gaussian distributed natural frequencies, 
$$g_U(\omega) =\frac{1}{\sqrt{2\pi} \Delta}e^{\frac{-(\omega)^2}{2\Delta^2}}$$ 
where the distribution is centered in zero and has standard deviation $\Delta$.

\subsection{Coherence indicators}
\label{coh}

\begin{figure}[h!]
\centering
    \includegraphics[width=1.0\linewidth]{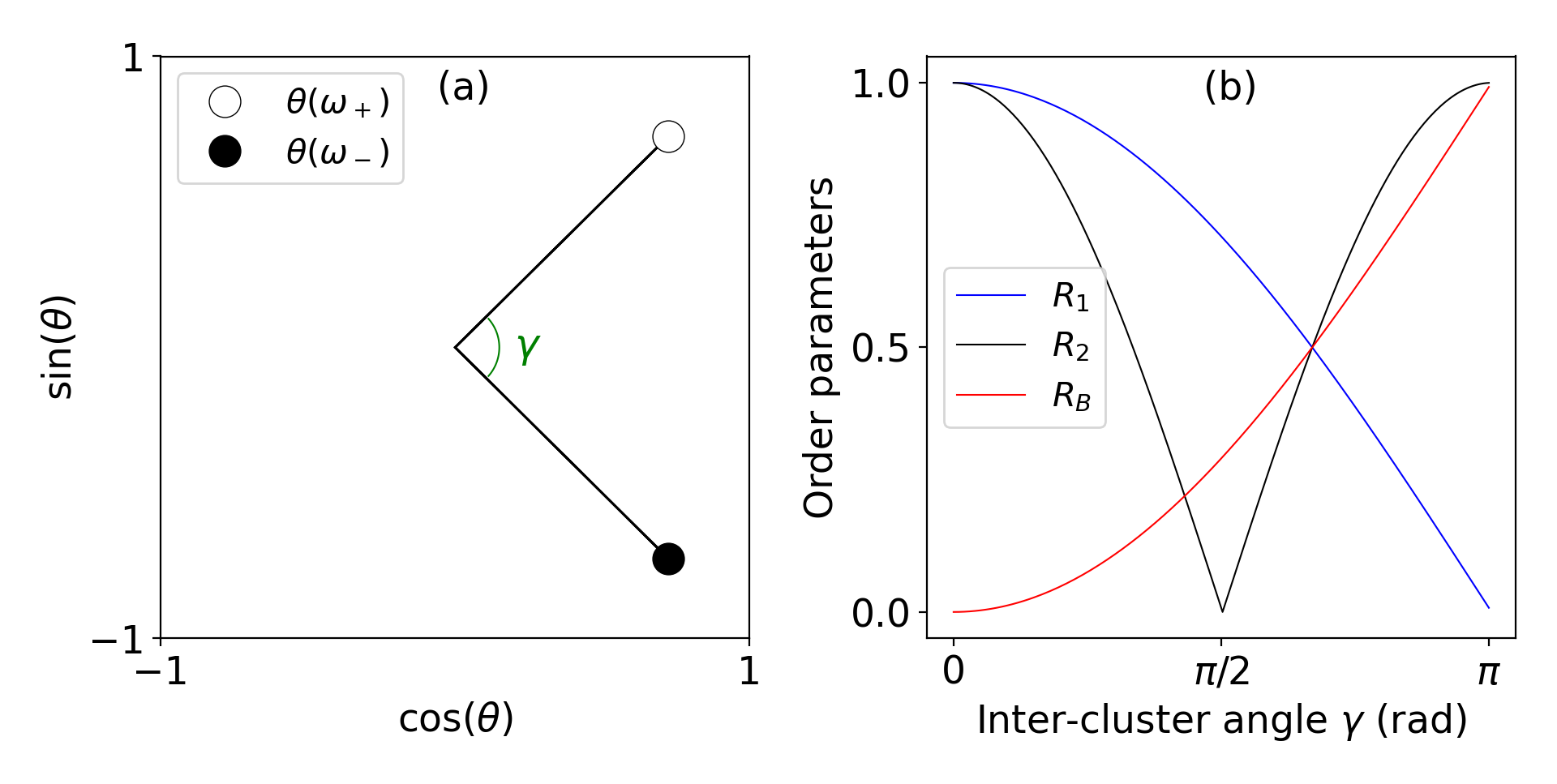}
    \caption{Order parameters (b) against the inter-cluster angle $\gamma$ (a). $R_1$ is the Kuramoto order parameter, $R_2$ is the order two
    Kuramoto-Daido parameter and $R_B$ is the clustering order parameter defined in eq. \ref{eq:Rs}.}
    \label{fig:angles}
\end{figure}

In order to measure the level of synchronization of the oscillators, we will employ the modulus of the Kuramoto order parameter $R_1$,
as well as the modulus of the  Kuramoto-Daido parameter $R_2$ that is particularly suited to reveal the presence of two phase clusters, as expected for the present model.
However, while the $R_2$ parameter is indeed able to clearly discriminate two phase clusters in phase opposition, whenever the two clusters 
are located on the circle at an intermediate angle $\gamma$ between 0 and $\pi$, this indicator shows a non monotonic dependence on the angle,
as shown in Fig. \ref{fig:angles}. It is therefore desiderable to introduce a new order parameter monotonically varying with the
angular separation of the two clusters, particularly suited to characterize bi-clusters :
\begin{align}
    R_B =\frac{R_1^{(+)}+R^{(-)}_1}{2}-R_1 ;
    \label{eq:Rs}
\end{align}
with $R_1^{(+)}$ ($R_1^{(-)}$) being the modulus of the Kuramoto order parameter limited to the oscillators with positive (negative)
natural frequencies. This indicator will be exactly zero whenever the oscillators are fully synchronized forming a single cluster,
then it will vary smoothly with the angle $\gamma$ between the two clusters, reaching the value one when $\gamma =\pi$, analogously to the indicator $R_2$,
as illustrated in Fig. \ref{fig:angles}

\subsection{Simulation protocols}

As the system shows many complex, multistable behaviors, we have examined the system evolution by considering 
many different initial conditions. These initial conditions are characterized by the different percentage $\eta$ of initially phase-locked oscillators. 
Referring to $\eta$, we define the following two simulation protocols: protocol (I) refers to simulations started with $\eta = 0$, 
where all the oscillators have random initial phases; protocol (II) refers to an initial state where all the oscillators phases are set
to the same value, therefore $\eta=1$. Furthermore, protocols (I A) and (II A) will denote 
simulations started from  $\eta = 0.0$ and $\eta = 1.0$ respectively, whenever one of the coupling parameters is
varied quasi-adiabatically. In particular, once initialized the phases, the simulations are performed at a constant
value of the control parameter, let us say $\lambda_1$, for a certain duration $T_s$, after discarding a transient $T_t$.
Then by considering the last configuration of the previous simulation as the new initial
condition, the control parameter is increased (decreased) by a small amount $\Delta \lambda_1$ 
and the simulation is repeated for the same transient and duration. This operation is repeated until the desired maximal (minimal)
value of $\lambda_1$ is reached. 
In the next sections, the effect of $0 < \eta < 1$ on the dynamical states will be also examined.

The network evolution equations \eqref{eq:2} have been integrated by employing a 4th order Runge-Kutta scheme either with fixed
time step $dt = 0.01 - 0.001$ or via an adaptive Runge-Kutta-Fehlberg method. Regarding the terminology, in the following sections we will refer to the synchronizing (de-synchronizing) transition as the forward (backward) transition, where we mean that the transition occurs by increasing (decreasing) the coupling. We will also refer to the instantaneous angular velocity of each oscillator $\dot{\theta_i}$ and to its time average by $\langle \dot{\theta_i}\rangle$.

\section{Mean-field Analysis}
\label{mean}

\begin{figure*}
\centering
    \includegraphics[width=0.49\linewidth]{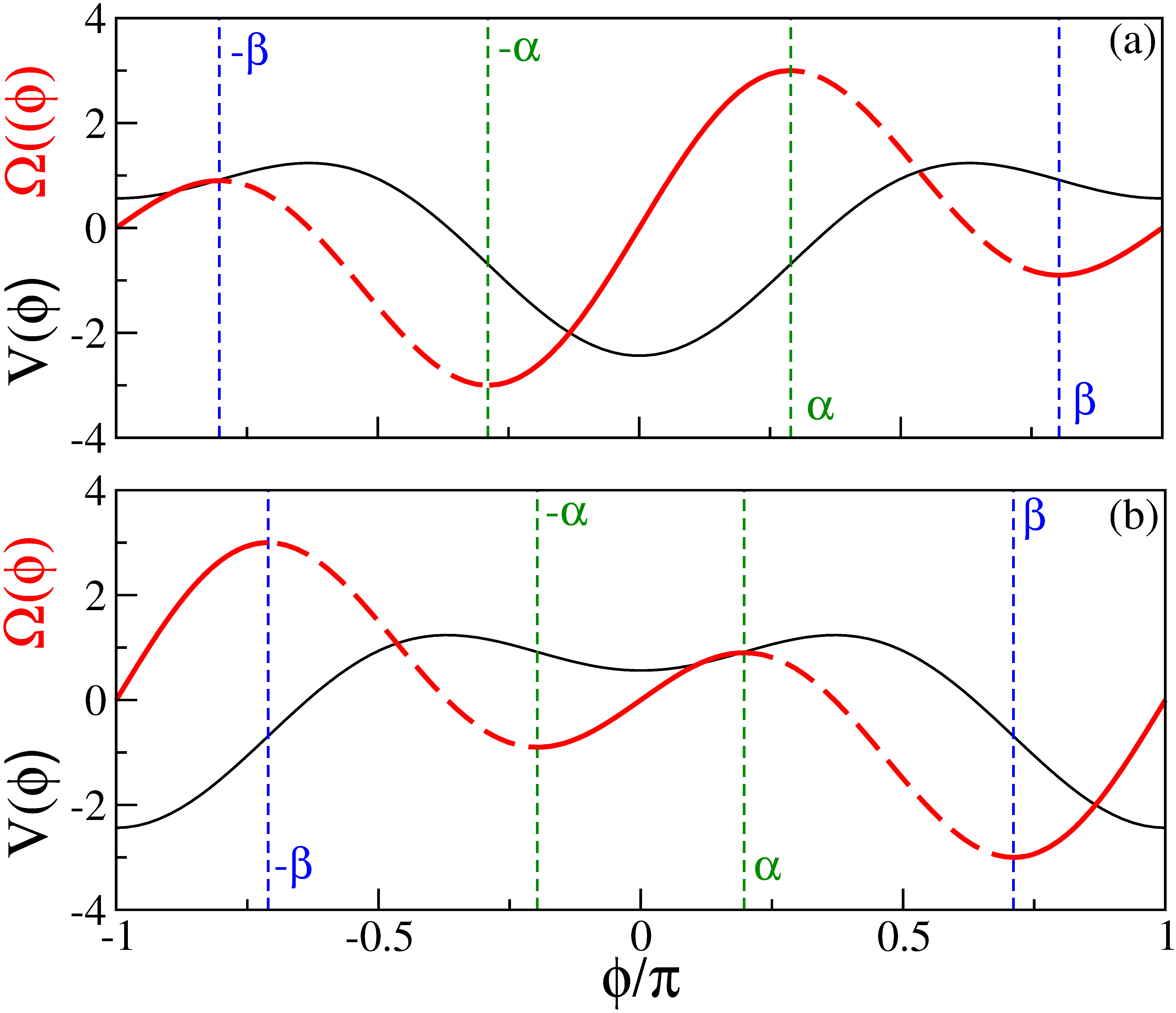}
    \includegraphics[width=0.49\linewidth]{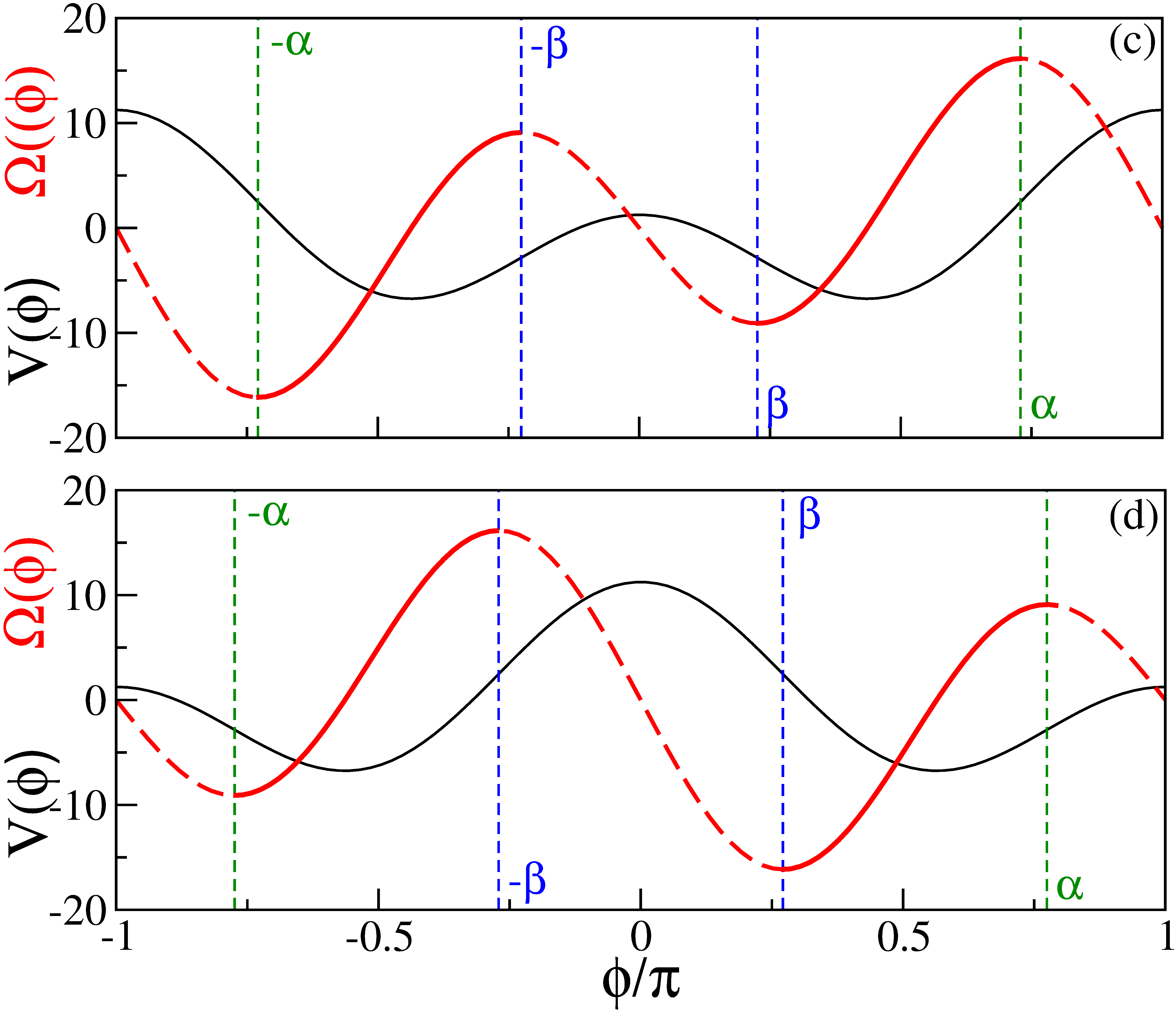}
    \caption{Potential $V(\phi)$ \eqref{eq:pot} and frequencies $\Omega(\phi)$ \eqref{locked} 
    versus the phase $\phi$ for positive values of $\lambda_2=15$ (a-b) and negative 
    one $\lambda_2=-100$ (c-d). Black (red) lines refer to $V(\phi)$ ($\Omega(\phi)$),
    while dashed green (blue) vertical lines denote the angles $\pm \alpha$ ($\pm \beta$).
    The red solid (dashed) curves identify the stable (unstable) branches associated to phases locked oscillators.
    Parameters are set in both panels to $R_1=0.5$ and $\Omega=0$ for the potential, while panel (a) (panel (b))
    refers to $\lambda_1 =3$ ($\lambda_1=-3$) and panel (c) (panel (d)) to $\lambda_1=10$ ($\lambda_1=-10$).}
    \label{fig:potential}
\end{figure*}

In this Section we will analyze the model in terms of a self-consistent
mean-field approach specifically inspired by the analysis performed in \cite{komarov2014}
for a bi-harmonic coupling function,
but essentially analogous to that employed by Kuramoto in his
seminal work \cite{Kuramoto2012}. In particular, 
by assuming that the system is partially synchronized, i.e. that $R_1$ is finite
and the clustered oscillators are uniformly rotating with a common angular velocity $\omega_c$,
the mean-field equation \eqref{eq:mf2} can be rewritten as follows in terms of the rescaled 
phase $\phi = (\theta - \Psi_1) \in [-\pi:\pi]$,
\begin{equation}
\frac{d\phi}{dt} = \Omega -
     \lambda_1 R_1 \sin(\phi) - \frac{\lambda_2}{2} R_1^2 \sin(2 \phi )    ;  
    \label{eq:mfbon}
\end{equation}
where $\Psi_1 = \omega_c t$ is the global phase and $\omega = \Omega + \omega_c$ is the natural frequency of the phase oscillator. The above equation \eqref{eq:mfbon} can be interpreted
as the dynamical equation for an overdamped oscillator moving in a potential landscape given by
\begin{equation}
V(\phi) = - \Omega \phi -  \lambda_1 R_1 \cos(\phi) - \frac{\lambda_2 R_1^2}{4} \cos(2 \phi ) \enskip .
\label{eq:pot}
\end{equation}
For even frequency distributions and frequency locked clusters we can assume that
$\Omega=0$; in such a case the potential is symmetrical
to a change of the phase sign, i.e. $V(\phi)=V(-\phi)$.
By following \cite{komarov2014}, we can affirm that the potential always reveals two coexisting minima whenever
\begin{equation}
 |u| = \left| \frac{\lambda_1}{\lambda_2 R_1} \right| < 1 \quad ;
\label{eq:pot}
\end{equation}
otherwise it displays a single minimum, analogously to the usual Kuramoto model.

We will mostly focus on the region where the two minima coexist,
by separately analysing the cases for positive and negative $\lambda_2$-values.
For $|u|<1$ the potential presents three extrema located at 
\begin{equation}
\phi =0 \enskip;\enskip  \phi=\pm \pi \quad and\enskip  \phi=  \pm {\bar \phi}= \pm \arccos(-u) \enskip,
\label{eq:pot}
\end{equation}
since the phases $\pm \pi$ coincide. 

As shown in Fig. \ref{fig:potential} (a-b), for $\lambda_2 >0$,
the potential displays two minima of different heights in $\phi=0$ and $\phi=\pm \pi$, and two maxima in $\pm {\bar \phi}$.
Therefore, depending on the initial conditions and on the protocol used for the simulations, one or two clusters may emerge in the system. In particular we notice that, for increasingly larger positive (negative) values of $\lambda_1$, the minimum corresponding to the cluster in $\phi=0$ ($\phi=\pm\pi$) becomes deeper 
and therefore more stable, as evident when looking at $V(\phi)$ (black solid lines) reported in Fig. \ref{fig:potential} (a) and (b). Furthermore, for $\lambda_1 \le - \lambda_2 R_1$ ($\lambda_1 \ge  \lambda_2 R_1$) the minimum in $\phi=0$ ($\phi = \pm \pi$) disappears and the potential shows an unique minimum.
For a negative $\lambda_2$-value the potential presents two maxima at $\phi=0$ and $\pm \pi$ and
two symmetric minima of same height at $\pm {\bar \phi}$, as shown in Fig. \ref{fig:potential} (c) and (d) (black solid lines).

The locked phases correspond to stationary solutions of \eqref{eq:mfbon} given by
\begin{equation}
 \Omega =      \lambda_1 R_1 \sin(\phi) + \frac{\lambda_2}{2} R_1^2 \sin(2 \phi )  =  Y(\phi, R_1) \quad  .
    \label{locked}
\end{equation}
These solutions will be stable whenever $\partial Y/\partial \phi = V^{"} (\phi) > 0$,
i.e. when the potential is concave in proximity of some minimum.
The curves $\Omega =  Y(\phi, R_1)$  are shown in Fig. \ref{fig:potential}:
solid (dashed) red lines denote the stable (unstable) regions.
The function $\Omega = Y(\phi, R_1)$ presents extrema in correspondence of the angles 
\begin{equation}
\pm \alpha = \pm \arccos {\frac{-u + \sqrt{u^2 + 8}}{4}  }
\label{alpha}
\end{equation}
and
\begin{equation}
 \pm \beta = \pm \arccos{\frac{-u - \sqrt{u^2 + 8}}{4}  } \quad .
\label{beta}
\end{equation}
We have denoted the absolute value of the natural frequencies at the extrema as
$\Omega_1 = |\Omega(\pm \alpha)|$ and $\Omega_2 = |\Omega(\pm \beta)|$ and
the minimal frequency among the two as $\Omega_{min}$.

Looking at the curves $\Omega =  Y(\phi, R_1)$ shown in Fig. \ref{fig:potential}
for $|u| <1$, it is evident that the system will always display two stable branches,
which can be identified as follows:
\begin{itemize}

\item{For $\lambda_2 >0$ :} the first branch corresponds to phases (frequencies) in the interval
${\Phi_1} \in [-\alpha : \alpha]$  ($[-\Omega_1:\Omega_1]$) and the second one to phases (frequencies)
in the interval ${\Phi_2} \in [-\pi : -\beta] \cup  [\beta:\pi]$ ($[-\Omega_2:\Omega_2]$);

\item{For  $\lambda_2 <0$ :}  the first  branch corresponds to phases (frequencies) in the interval
${\Phi_1} \in [-\alpha : -\beta]$  ($-\Omega_1:\Omega_2]$) and the second one to phases (frequencies) in the
interval ${\Phi_2} \in [\beta:\alpha]$ ($[-\Omega_2:\Omega_1]$).

\end{itemize}
Moreover it is worth noticing that there is a region of coexistence between the two stable branches for $ -\Omega_{min} \le \Omega \le \Omega_{min}$,
where oscillators with the same frequency $\Omega$ can be locked to 2 different phases (see  red solid curves in Fig. \ref{fig:potential}).

The self-consistent equation for the order parameter in the thermodynamic limit
is analogous to the one for the Kuramoto model \cite{Kuramoto2012} and it can be  written as
\begin{equation}
R_1 {\rm e}^{j \Psi_1} =  {\rm e}^{j \Psi_1} \enskip \int \int \enskip d \phi \enskip d \Omega 
\enskip \rho(\phi|\Omega) \enskip {\rm e}^{j \phi} g(\Omega)
\label{self}
\end{equation}
where $g(\Omega)$ is the distribution of the natural frequencies and $\rho(\phi|\Omega)$ the conditional
probability distribution function characterizing the oscillator population. 
The oscillators can be divided in drifting or locked depending on their natural frequency.
By following \cite{komarov2015} we denote with $Y_{max}$ ($Y_{min}$) the maximum (minimum)
of the function  $Y(\phi, R_1)$ \eqref{locked}. According to this notation, the locked (drifting) oscillators
will have $Y_{min} \le \Omega \le Y_{max}$ ($\Omega > Y_{max}$ and $\Omega < Y_{min}$).
The probability distribution $\rho_d$ for the drifting oscillators is proportional
to the inverse of their phase velocity \cite{Kuramoto2012}:
\begin{equation}
\rho_d (\phi|\Omega) = \frac{C}{|\Omega - Y(\phi, R_1)|} \enskip;\enskip C = \frac{1}{\int_{-\pi}^{\pi} d \phi |\Omega - Y|}
\enskip .
\label{rhod}
\end{equation}

The distribution $\rho_l$ of the locked oscillators in the region of coexistence can be written as
\begin{equation}
\rho_l (\phi|\Omega) = \sigma \delta(\phi - \Phi_1(\Omega)) + (1- \sigma) \delta(\phi - \Phi_2(\Omega))
\enskip ;
\label{rhol_bi}
\end{equation}
where $ 0 \le \sigma \le 1$ is the redistribution factor of the oscillators among the two branches.
Outside the coexistence region, the distribution $\rho_l$ is simply given by
\begin{equation}
\rho_l (\phi|\Omega) =  \delta(\phi - \Phi_x(\Omega))   \quad,\enskip x=1,2
\enskip ,
\label{rhol_uni}
\end{equation}
depending if the oscillator characterized by the phase $\phi$ is locked to the first or second branch.

For $\lambda_2 >0$, by assuming the frequency distribution to be even ($g(\Omega) = g (-\Omega)$) 
and by taking into account the symmetry of the integrand, the expression for the Kuramoto order parameter takes the form
\begin{eqnarray}
R_1 &=& 2 \left[ \int_0^\alpha  d Y(\phi) S_1(Y(\phi)) g(Y(\phi)) \cos(\phi) \right. 
\nonumber \\
&+& \left. \int_\beta^\pi  d Y(\phi) S_2(Y(\phi)) g(Y(\phi)) \cos(\phi) \right] 
\nonumber \\
&+& \int_{U}  \int_{-\pi}^\pi d \Omega d \phi \frac{C g(\Omega) \cos(\phi)}{|\Omega - Y(\phi, R_1)|} \enskip , 
\label{R_pos}
\end{eqnarray}
where $S_1 = \sigma$ and $S_2 = 1 - \sigma$ in the coexistence regions, $S_1 = S_2 = 1$ outside the coexistence region and
\begin{equation}
d Y(\phi) =   [ \lambda_1 R_1 \cos(\phi) + \lambda_2 R_1^2 \cos(2 \phi )] d\phi \enskip    .
\label{dom}
\end{equation}
The integration region $U = (-\infty, Y_{min}) \cup (Y_{max}, +\infty)$ that appears in the third integral in \eqref{R_pos}, is the frequency domain where the oscillators are drifting.

For $\lambda_2  <0$ the expression for $R_1$ becomes
\begin{eqnarray}
R_1 &=&  \left[ \int_{-\beta}^{-\alpha}  d Y(\phi) S_1(Y(\phi)) g(Y(\phi)) \cos(\phi) \right. 
\nonumber \\
&+& \left. \int_\beta^\alpha  d Y(\phi) S_2(Y(\phi)) g(Y(\phi)) \cos(\phi) \right] 
\nonumber \\
&+& \int_{U}  \int_{-\pi}^\pi d \Omega d \phi \frac{C g(\Omega) \cos(\phi)}{|\Omega - Y(\phi, R_1)|} \enskip .
\label{R_neg}
\end{eqnarray}
In this case, due to the symmetry of the potential wells (see Fig. \ref{fig:potential} (c-d)), the value of $R_1$ does not depend on the value of the redistribution factor $\sigma$, as we have verified. Therefore we can set $S_1 = S_2 = 1/2$ inside 
the coexistence region and $S_1 = S_2 = 1$ outside, without any loss of generality. 
It should be noted that, for the parameter values considered in this paper, the contribution
of the drifting oscillators to \eqref{R_pos} and \eqref{R_neg} is usually negligible.

The self-consistent approach allows to find coherent solutions of the system in the thermodynamic limit
corresponding to finite values of the Kuramoto order parameter,
but not to determine their stability properties.

\subsection{Unimodal Distribution of the Natural Frequencies}

Let us first consider the case where natural frequencies are distributed according to an unimodal
Gaussian distribution $g_U(\omega)$ with $\Delta=0.25$.  In this case we have characterized the synchronization
transition in terms of the order parameter $R_1$, evaluated by performing simulations with protocols (I A) and (II A).
In the following, if not stated otherwise, we avoid explicitly indicating the time average of $R_1(t)$ by introducing a new symbol
and we intend for $R_1$ its time averaged value. Moreover, we fix the parameter $\lambda_2$ to a positive or negative value
and we vary quasi-adiabatically $\lambda_1$.

\begin{figure*}
\centering
    \includegraphics[width=0.34\linewidth, height=0.24\linewidth]{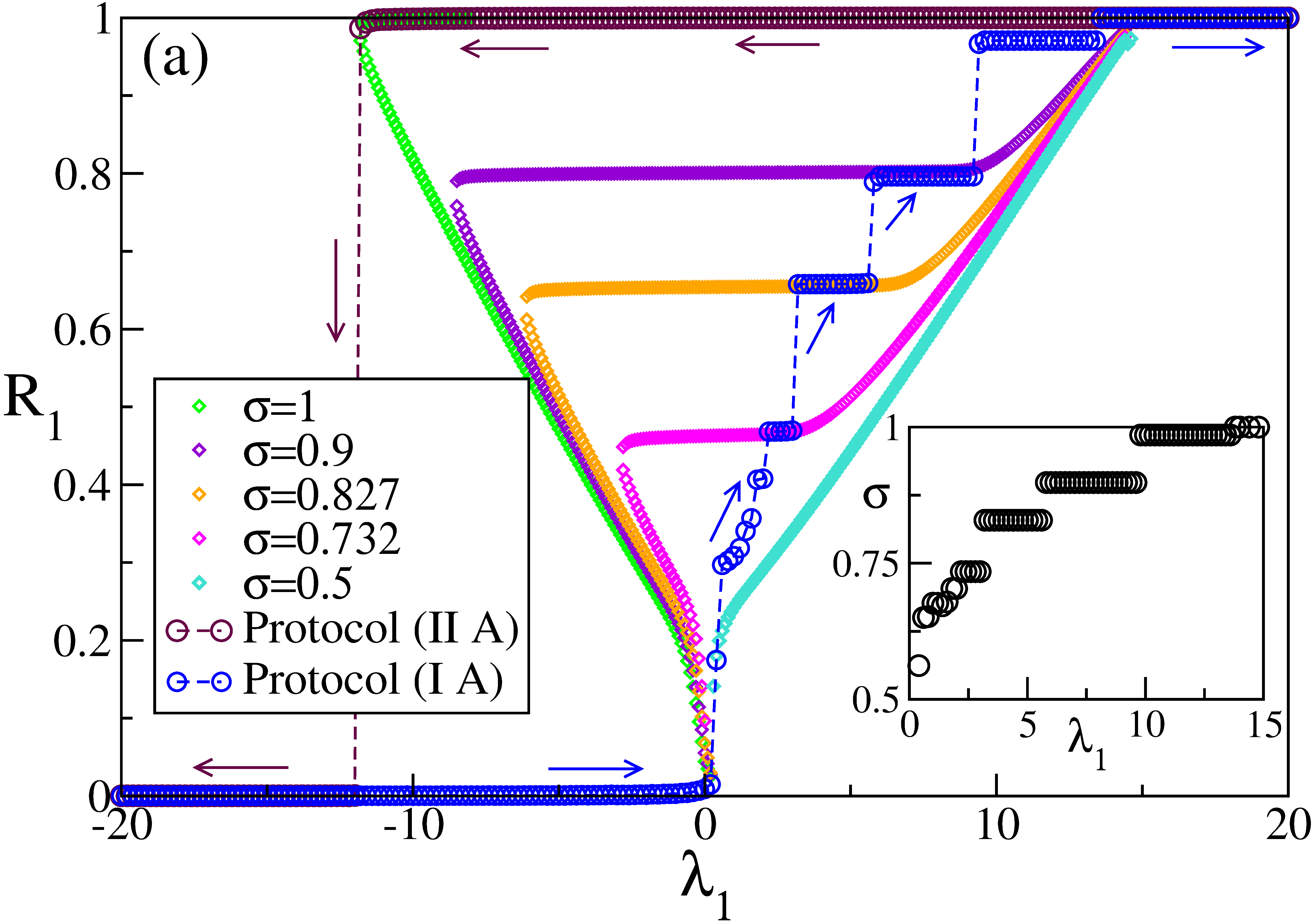}
    \includegraphics[width=0.32\linewidth]{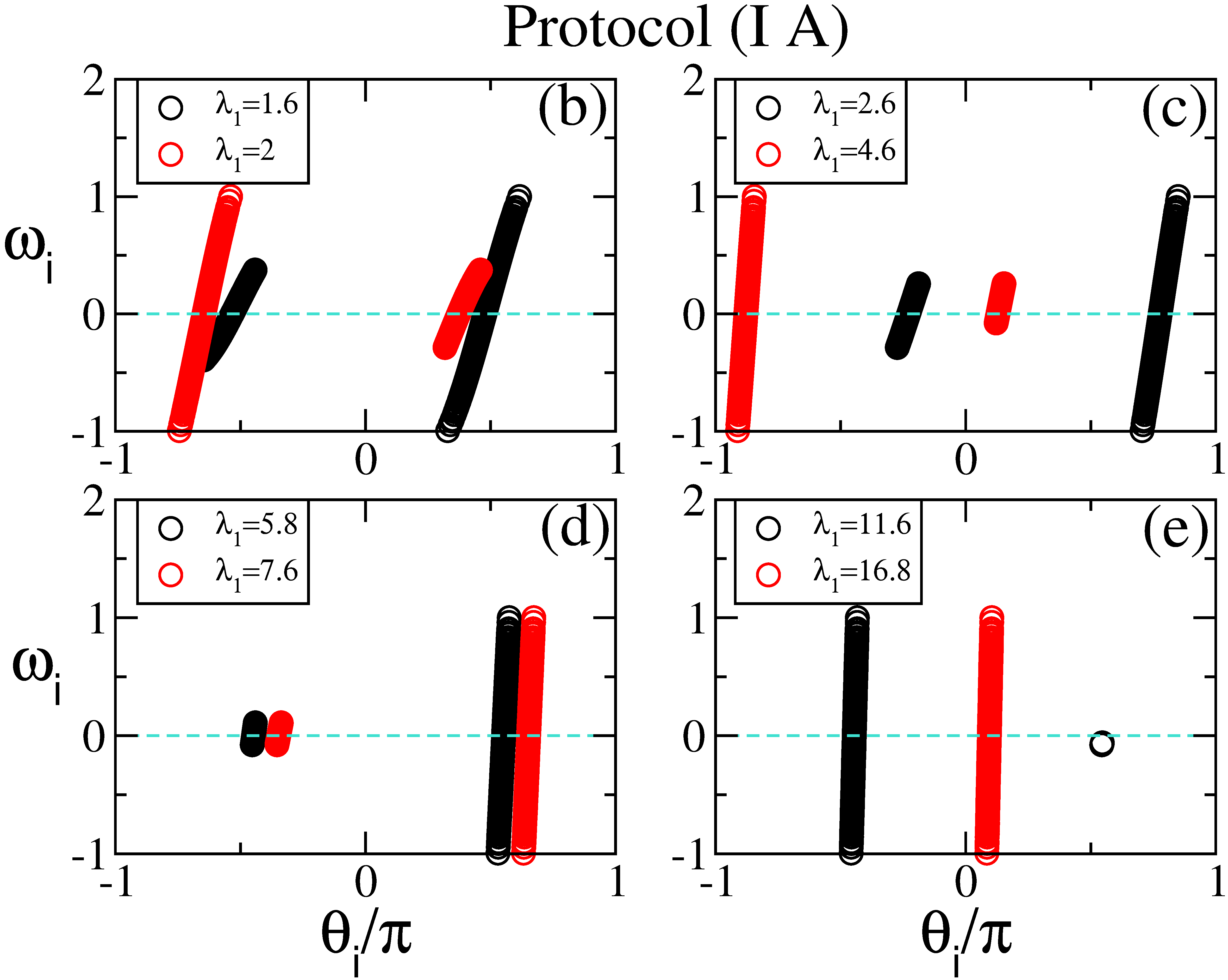}
    \includegraphics[width=0.28\linewidth, height=0.24\linewidth]{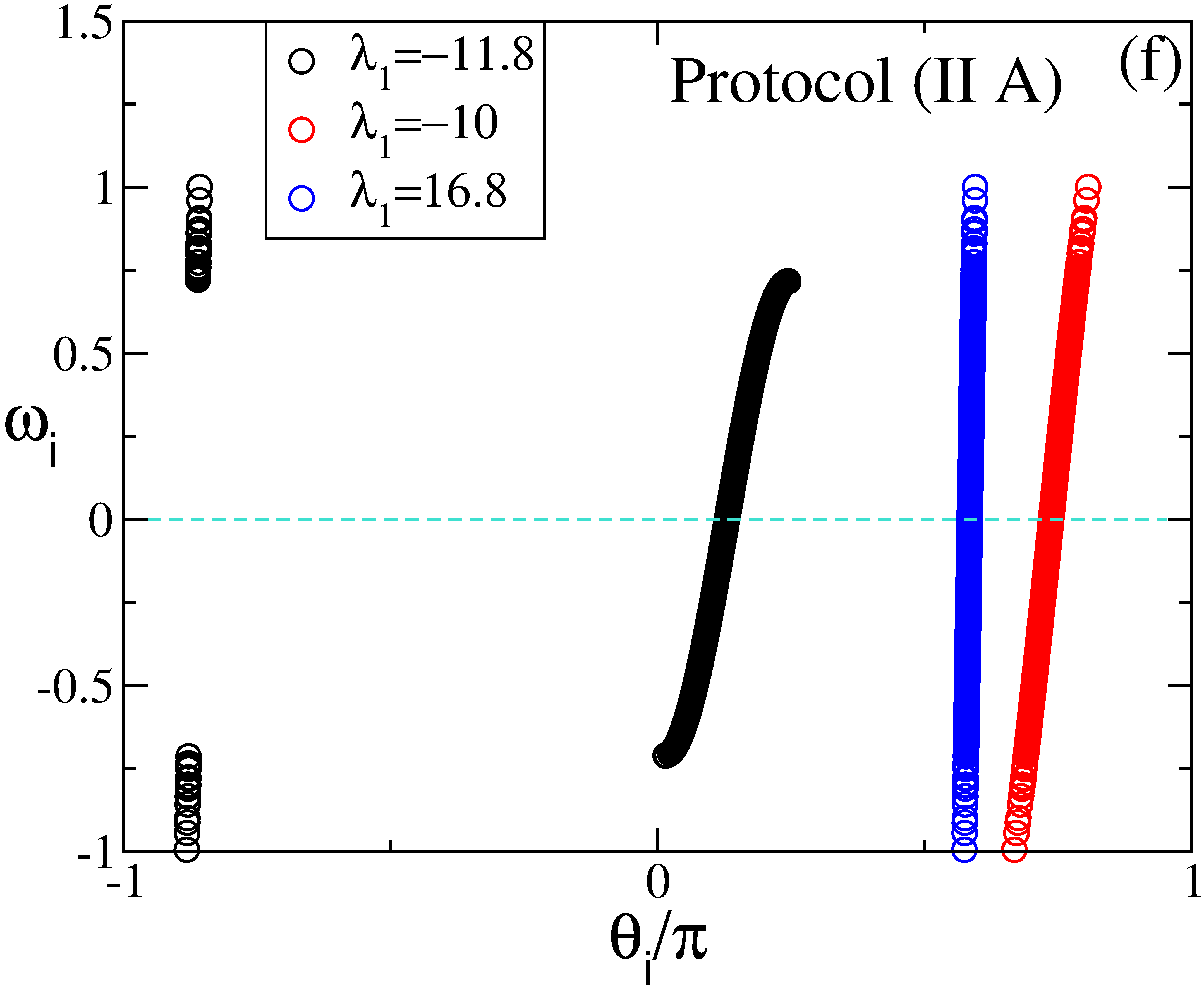}
 \caption{Unimodal Gaussian natural frequencies distribution and $\lambda_2 > 0$.  (a) $R_1$ versus $\lambda_1$ : the blue and maroon circles refer to simulations performed by following protocols (I A) and (II A), respectively. The value reported for $R_1$ represents the average in time over a simulation time $T_s$. The diamonds of different colours denote the self consistent evaluation of the order parameter $R_1$, for different $\sigma$ values, obtained by employing \eqref{R_pos}.  
 In the inset it is shown $\sigma$ versus $\lambda_1$ as estimated by following protocol (I A). 
 In panels (b-e) are reported snapshots of the natural frequencies $\omega_i$ versus their phases $\theta_i$ for different
 values of $\lambda_1$ for simulations obtained by following protocol (I A). Analogous snapshots are shown
 in (f) for simulations done by following protocol (II A). When necessary, the snapshots are arbitrarly shifted along the x-axis to improve readability.   
 The displayed results refer to  $\lambda_2=15$, $\Delta=0.25$ for $g_U(\omega)$, $\Delta \lambda_1 = 0.2$, $T_t = 10$, $T_s=200$, integration step $dt=0.001$ 
 for the adiabatic simulations, and  a network size $N=10000$.}
    \label{fig:uni_pos}
\end{figure*}

\subsubsection{Positive $\lambda_2$}
The simulation results for $\lambda_2 = 15$ and $N=10000$ are shown in Fig. \ref{fig:uni_pos} (a).
For protocol (I A) the partially synchronized state emerges for $\lambda_1^{(AS)} \simeq 0.4 $,
corresponding to the destabilization of the asynchronous regime of the usual Kuramoto model \cite{Kuramoto2012},
i.e. $\lambda_1^{(AS)}  = \frac{2}{\pi g(0)}$.
In the present case the higher order interactions are proportional to $R_1^2$, as shown in \eqref{eq:mfbon}, therefore they cannot affect the stability of the incoherent regime, which is
usually analysed in proximity of the transition to coherence via an expansion of the model at the first order in $R_1$  \cite{strogatz1991}. 
For this reason we obtain a critical $\lambda_1^{(AS)}$ value corresponding to the one for the usual Kuramoto model.  
Moreover the system becomes fully synchronized ($R_1=1$) for $\lambda_1 \simeq 13.7$
and the approach to the fully synchronized regime occurs via a series of plateaus where
$R_1$ remains constant when varying $\lambda_1$. The partially synchronized  regime is always characterized 
by two phase clusters differently populated, as shown in Figs. \ref{fig:uni_pos} (b-e), which
correspond to the two minima of the potential at $\phi=0$ and $\phi=\pm\pi$ displayed in Fig. \ref{fig:potential} (a). By increasing $\lambda_1$ the main cluster, corresponding to the minimum at $\phi=0$ of $V(\phi)$, becomes more and more populated at the expenses of the secondary cluster. Finally only this cluster remains.
This can be clearly appreciated by measuring the redistribution factor $\sigma$ directly from the simulations,
where $\sigma$ corresponds to the fraction of oscillators present in the minimum located
at $\phi=0$ (see the inset of Fig.  \ref{fig:uni_pos} (a)). In particular we observe that $\sigma \ge 0.5$,
meaning that the cluster at $\phi=0$ is always more populated than that at $\phi=\pm \pi$.
For protocol (II A), the system is initialized with all the phases equal to $\theta_i=0$.
Therefore the oscillators are all located in the minimum $\phi=0$ of the potential and they
remain fully synchronized until $\lambda_1^{(PS)}(\sigma=1) \equiv \lambda_1^{(FS)} \simeq -11.8$, when they abruptly de-synchronize. In particular, $\lambda_1^{(PS)}(\sigma)$ indicates the critical value at which the coherent regime characterized
by $\sigma N$ ($(1-\sigma) N$) oscillators in the minimum $\phi=0$ ($\phi=\pm \pi$) looses stability.
A peculiar difference with the usual Kuramoto model is that, as soon as $\lambda_1 > \lambda_1^{(AS)}$,
two phase clusters emerge, corresponding to the two minima in the potential shown in Fig. \ref{fig:potential} (a), at an angular distance
of $\pi$.  This is confirmed by the snapshots of the phase oscillators
presented in  Figs. \ref{fig:uni_pos} (b-e), which always reveal two clusters in the partially synchronized regime
located at a distance $\pi$ in phase.  

To get some more understanding we estimated the self-consistent solutions \eqref{R_pos}
for different $\sigma$-values, thus obtaining the evolution of $R_1$ as a function of $\lambda_1$, shown 
in Fig. \ref{fig:uni_pos} (a) for a few $\sigma$-values. These curves for $\sigma > 0.5$ have all a
similar shape. The partially synchronized state emerge abruptly via a bifurcation, that we can
identify as a saddle-node, at a value $\lambda_1^{(PS)}(\sigma) < 0$, with a finite order parameter value
$$
R_1(\sigma) = 2 \sigma -1 \quad ,
$$
corresponding to a 2 cluster state with $\sigma N$ oscillators in $\phi=0$
and $(1-\sigma) N$ oscillators in $\phi = \pm \pi$.
The critical value $\lambda_1^{(PS)}(\sigma)$ should be larger than $\lambda_1 = - \lambda_2 R_1(\sigma)$,
where the minimum at $\phi=0$ in $V(\phi)$ emerges.
By increasing $\lambda_1$, the value of the order parameter
remains equal to $R_1(\sigma)$ within a certain interval, giving rise to a plateau and finally it approaches
$R_1 =1$ for $\lambda_1 = \lambda_2$, corresponding to the value where the minimum in $V(\phi)$ at $\phi=\pm \pi$ disappears and all the oscillators are fully synchronized at $\phi=0$. 
We have estimated a few self-consistent curves
corresponding to $\sigma$-values actually measured during the finite size quasi-adiabatic simulation.
The plateaus observed during the  simulation performed via protocol (I A) coincide with those
obtained by the self-consistent approach for the corresponding $\sigma$-value, as clearly evident in
Fig. \ref{fig:uni_pos} (a). Furthermore, the self-consistent results for $\sigma=1$ perfectly coincide with the
simulations obtained with protocol (II A) and the de-synchronization transition occurs at 
$\lambda_1^{(FS)} = -11.8$ as in the finite size simulations.

For the case $\sigma=0.5$, corresponding to equally populated clusters, the self-consistent approach
reveals that this state emerges at $\lambda_1^{(PS)}(\sigma=0.5) \simeq 0.25$ with an order
parameter value that is quite small ($R_1 \simeq 0.04$). Moreover, according to the self-consistent approach for $\sigma=0.5$,
no plateau is observable (see the cyan line in  Fig. \ref{fig:uni_pos} (a)).
Similar results have been found for $\sigma < 0.5$ with curves $R_1=R_1(\lambda_1)$ lying slightly below
the one obtained for $\sigma=0.5$.

To better clarify the situation, we expect  the incoherent regime to  become unstable at $\lambda_1^{AS} = \frac{2}{\pi g(0)} \simeq 0.4$
in the thermodynamic limit. However there is a wide region of multistability
in the interval $\lambda_1 \in [\lambda_1^{(FS)} ; \lambda_2] = [-11.8;15]$, where 
two cluster state solutions coexist, although populated with a different percentage $\sigma$ of oscillators.  Therefore we expect the synchronization transition to be generically hysteretic,
and to observe a scenario somehow similar to that already observed for the Kuramoto model with inertia \cite{olmi2014}.
 The hysteretic nature of the transition is evident by considering the simulation performed by employing protocol (II A).

\begin{figure*}
\centering
    \includegraphics[width=0.32\linewidth]{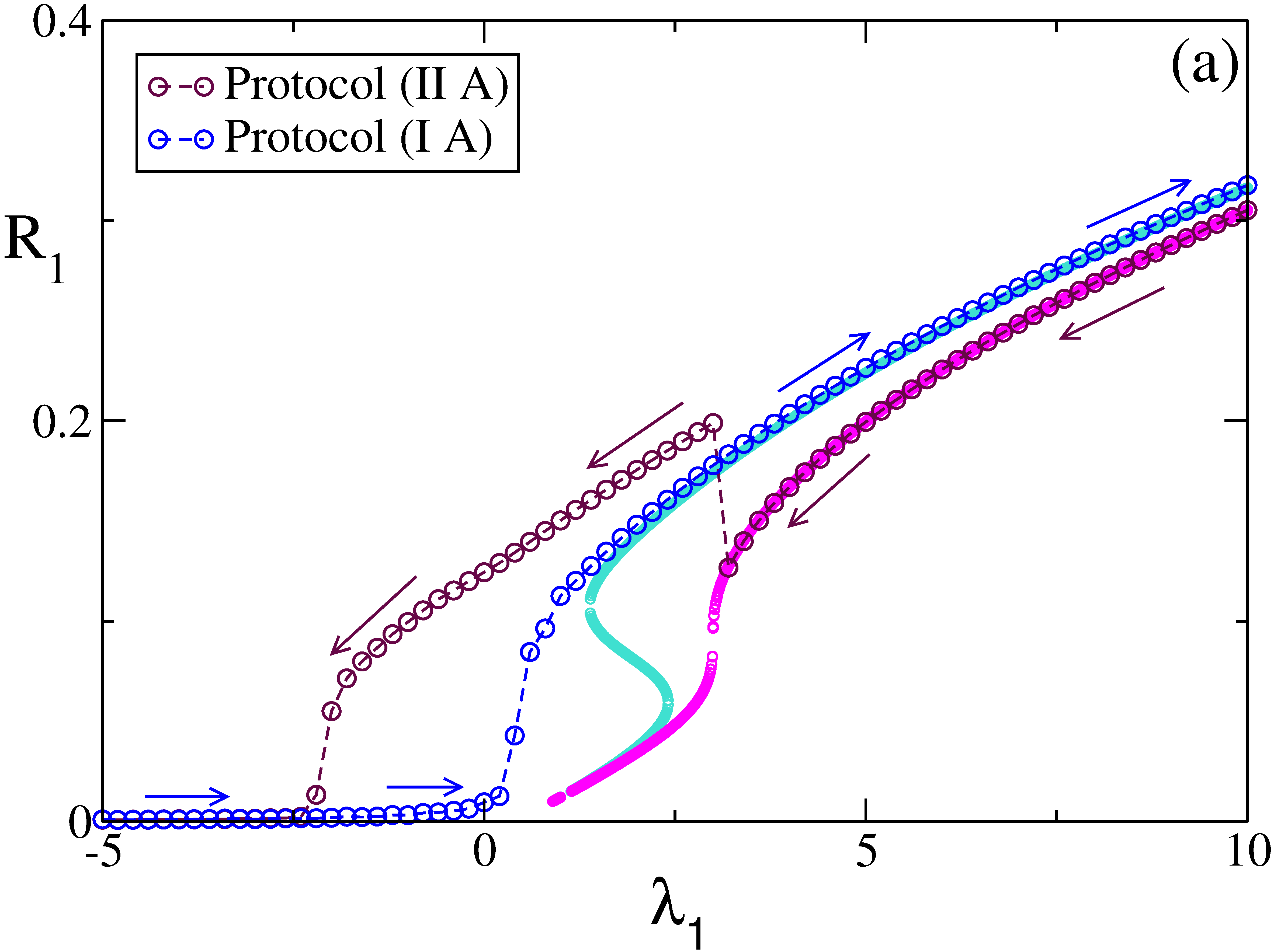}
    \includegraphics[width=0.32\linewidth]{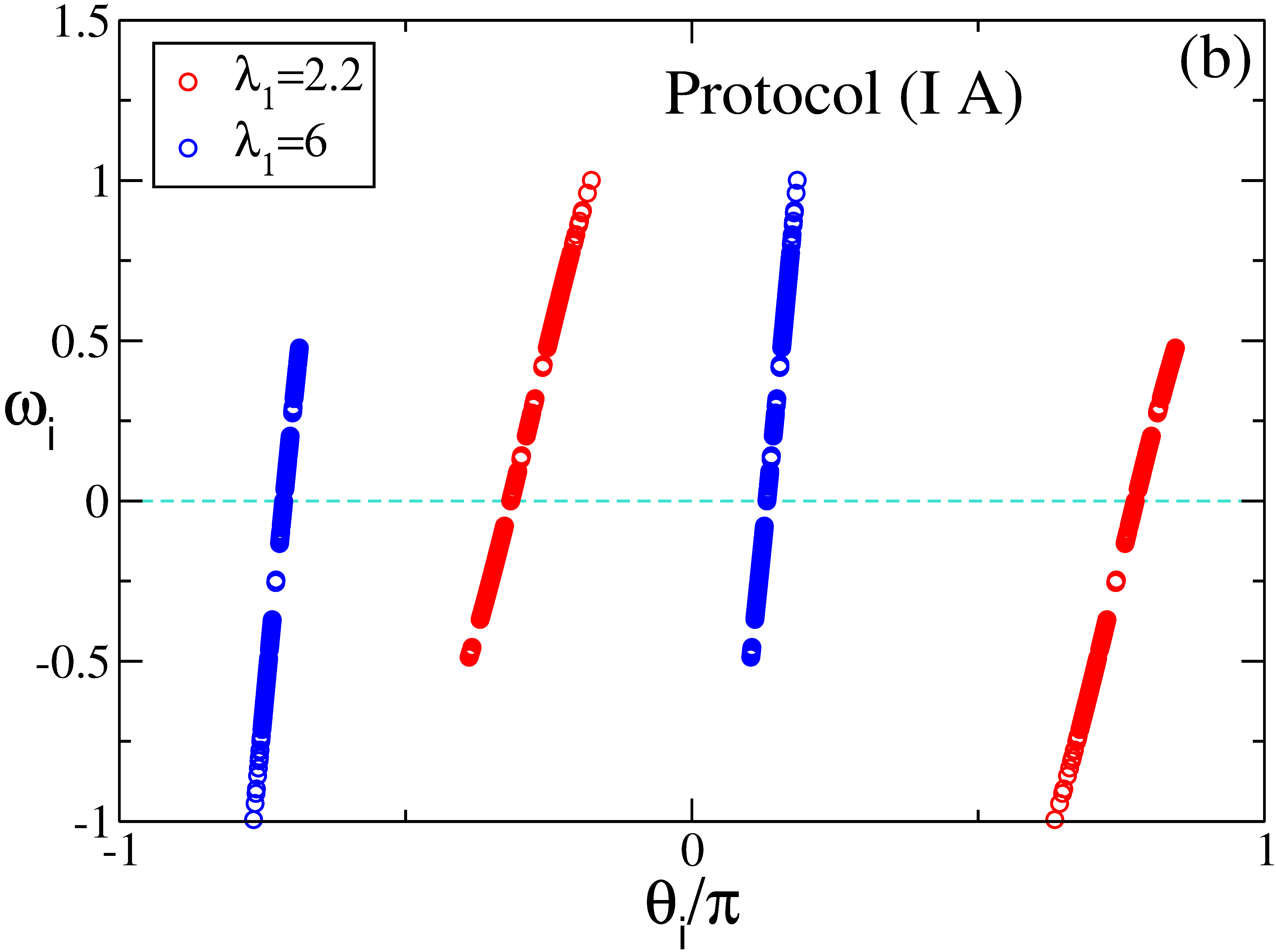}
    \includegraphics[width=0.32\linewidth]{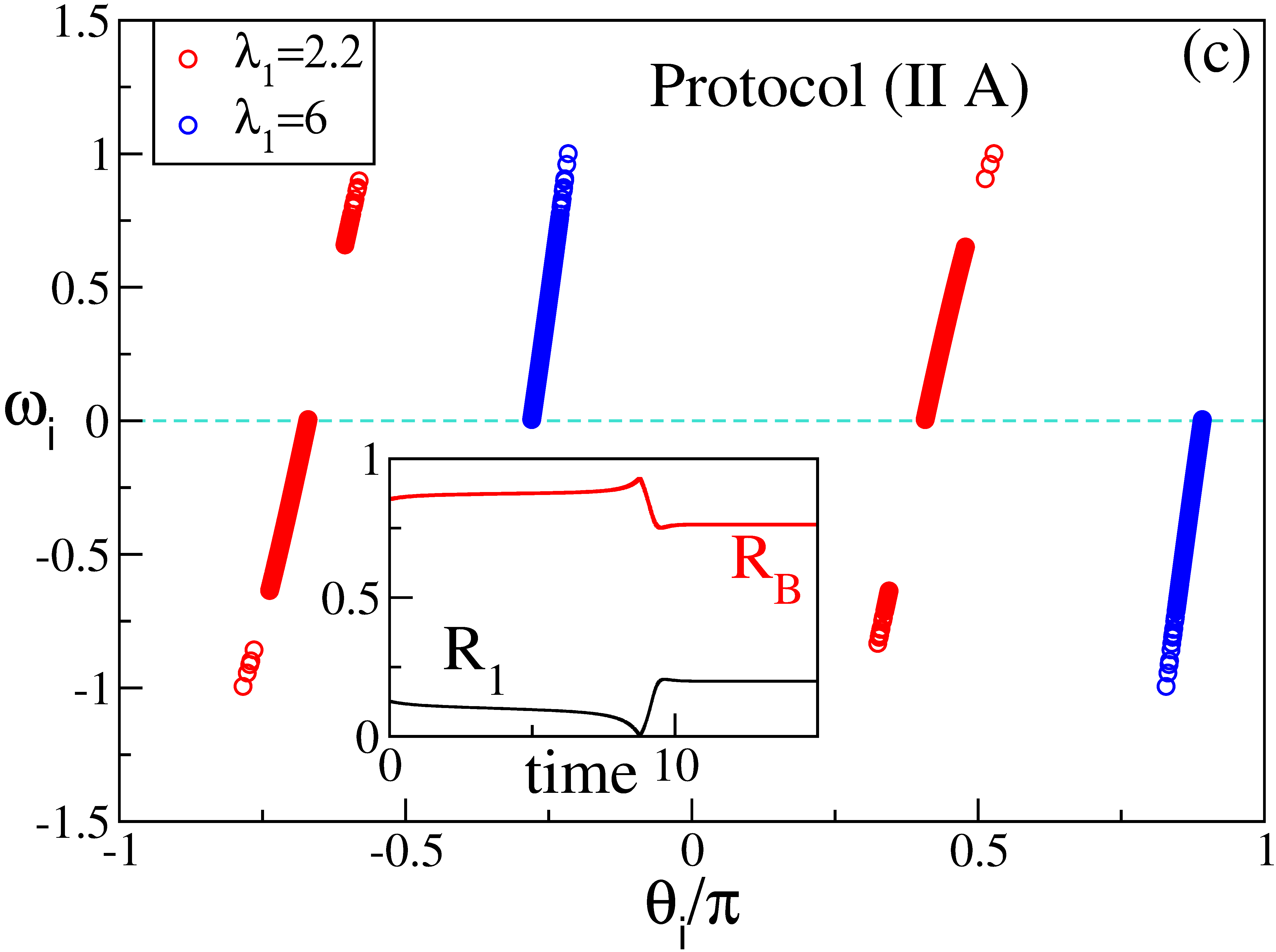}
	\caption{Natural frequencies distributed according to an unimodal Gaussian distribution and $\lambda_2 < 0$. (a) $R_1$ versus $\lambda_1$ : the blue and maroon circles
 refer to simulations performed by following protocols (I A) and (II A), respectively. The value reported for $R_1$ represents the average in time over a simulation
 time $T_s$. The cyan (magenta) line refers to the self consistent evaluation of the order parameter $R_1$ by employing \eqref{R_neg} estimated in the whole range
 of frequencies (restricted to positive frequencies). In panels (b) and (c) are reported snapshots of the natural frequencies $\omega_i$ versus their phases $\theta_i$ for different
 values of $\lambda_1$ for simulations obtained by following protocols (I A) and (II A), respectively. In the inset of panel (c) we report the instantaneous values of $R_1(t)$ (black curve) and $R_B (t)$ (red curve) versus time for $\lambda_1=3$ obtained by following protocol (II A).  The displayed results refer to 
 $\lambda_2=-100$, $\Delta=0.25$ for $g_U(\omega)$, $\Delta \lambda_1 = 0.2$ $T_t = 10$, $T_s=200$, integration step $dt=0.001$ for the adiabatic simulations, and  a network size $N=10000$.
	}
    \label{fig:uni_neg}
\end{figure*}

\subsubsection{Negative $\lambda_2$}

In the case $\lambda_2=-100$ and $N=10000$ oscillators, by following protocol (I A),
the asynchronous regime looses stability via a continuous transition in the proximity of
$\lambda_1^{(AS)} = \frac{2}{\pi g(0)}$, as expected (see blue circles in Fig. \ref{fig:uni_neg} (a)).
In the partially synchronized regime two equally populated clusters are always observable
and the increase of $\lambda_1$ renders the two clusters more synchronized, i.e. the oscillators within
each cluster tend to have more and more similar phases, as shown in Fig. \ref{fig:uni_neg} (b).
This behaviour was expected from the shape of the potential $V(\phi)$, that in this case  presents
two minima of equal heights as observable in Fig. \ref{fig:potential} (b).

By following protocol (II A) we observe that, when decreasing $\lambda_1$, the Kuramoto order parameter $R_1$ decreases
smoothly with the control parameter and then, at $\lambda_1^{(\pi)} \simeq 3 $, it
presents an abrupt jump to a higher value (see maroon circles in Fig. \ref{fig:uni_neg} (a)). Here we indicate with $\lambda_1^{(\pi)}$ the critical
value at which a discontinous transition is observable when performing protocol (II A), where
the system jumps from a lower value of $R_1$ to a higher one. Once performed the jump, by decreasing $\lambda_1$,
the value of $R_1$ continuously diminishes towards zero
and the asynchronous regime is smoothly achieved for $\lambda_1^{(PS)} \simeq -2.1$.
This discontinuity in $R_1$ has been previously reported in \cite{Bi2016}, but its origin is still unclear.
An important aspect to notice is that, before the jump, the oscillators are organised in two equally populated clusters,
characterized by either positive or negative natural frequencies while, after the jump, the two clusters still present mainly
positive (negative) natural frequencies, but together with a small group of oscillators with opposite $\omega_i$ sign
(as visible in  Fig. \ref{fig:uni_neg} (c)). According to protocol (II A), the oscillators are all
initialized with the same phase $\theta_i=0$, that now corresponds to a local maximum of the potential,
see Fig. \ref{fig:potential} (c-d). Therefore, depending on the sign of their natural frequencies, the oscillators
split in 2 clusters characterized, each one, by the same sign of $\omega_i$. In this way, the oscillators
migrate towards one of the two minima of $V(\phi)$ located in correspondence of the positive or the negative phases.

Also in the present case we can estimate the possible values attained by $R_1$ as a function of $\lambda_1$ 
in the mean field limit corresponding to protocol (I A), by solving the self-consistent equation Eq. \eqref{R_neg}. As already mentioned, in this case, due to the symmetries of the potential $V(\phi)$, we do not observe any dependence on $\sigma$, therefore we set $\sigma=0.5$ in \eqref{R_neg}. The results are shown as a cyan line in Fig. \ref{fig:uni_neg} (a):
the self-consistent solution for $R_1$ reveals two successive saddle-node bifurcations at $\lambda_1 \simeq 2.41$ and $\lambda_2 \simeq 1.38$. However a good agreement
with the numerical data is observable for $\lambda_1 \ge 2.0$ only. This is due to the fact that the phase clusters emerge only above  $\lambda_1 \ge 2.0$ while,
below such a value, the system do not reveal any clear cluster.

To estimate the results corresponding to protocol (II A) via the self-consistent approach, we solved Eq. \eqref{R_neg}
by integrating only on positive (negative) natural frequencies depending if the oscillators
are in the cluster with positive (negative) phases. Once more the agreement with the simulations is quite good
from large $\lambda_1$ values down to the discontinuous transition value $\lambda_1^{(\pi)}$. The self-consistent solution (magenta line in Fig. \ref{fig:uni_neg} (a)) reveals  a sharp transition at  $\lambda_1^{(\pi)}\simeq 3.0$, that we have termed $\pi$-transition for reasons that will become evident in the following.
Since we have no information about the stability of the solutions themselves, we can argue
that the solution is stable (unstable) for $\lambda_1 > \lambda_1^{(\pi)}$ ($\lambda_1 < \lambda_1^{(\pi)}$)
and that the 2 solutions annihilate in a saddle-node bifurcation at  $\lambda_1^{(\pi)}$. 
The destabilization of the phase cluster states characterized by either positive or negative ${\omega_i}$
can be understood by following the evolution in time of the order parameters $R_1$ and $R_B$ at  $\lambda_1 = 3.0$, as
shown in the inset of Fig. \ref{fig:uni_neg} (c). The value of $R_1$ ($R_B$) is initially constant then, abruptely,
$R_1$ vanishes ($R_B$ reaches one) and, afterwards, the value of $R_1$ ($R_B$) becomes larger (smaller). This behaviour
can be interpreted as follows: the two cluster states are initially separated by an angle $\gamma < \pi$, 
then they recede and $\gamma$ reaches exactly $\pi$, corresponding to an anti-phase situation. The vanishing
of $R_1$ implies a vanishing of the potential $V(\phi)$, therefore the 2 clusters destabilize
and reform with a smaller angle $\gamma$ between them. Moreover they reform with a different composition inside each cluster,
which now includes oscillators with positive and negative natural frequencies. 
Therefore, the $\pi$-transition was at the origin of the discontinuity in $R_1$ reported in \cite{Bi2016}, and in the following,
we will examine this transition in more detail.

\begin{figure}
\centering
    \includegraphics[width=0.95\linewidth]{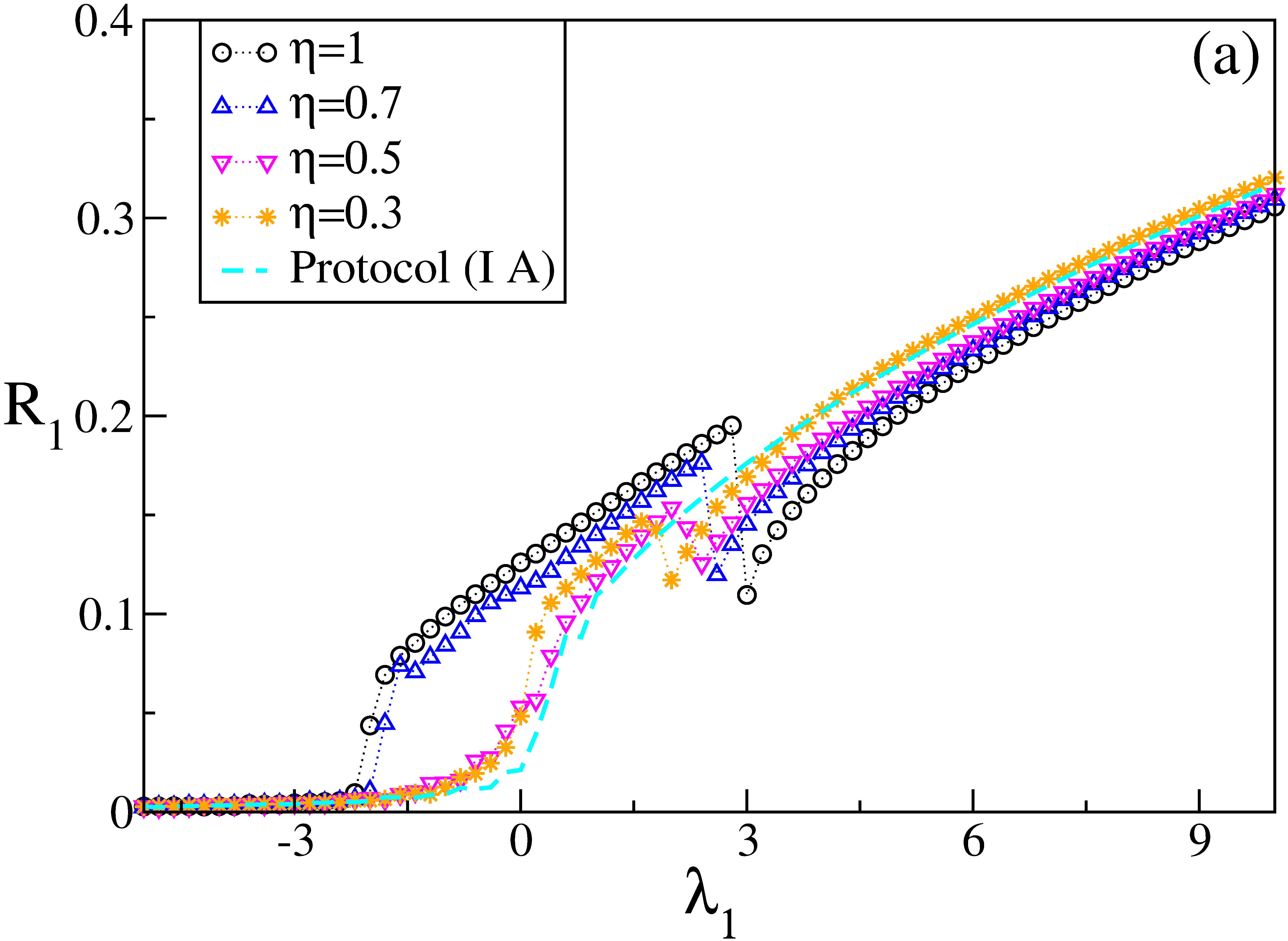}
    \includegraphics[width=0.95\linewidth]{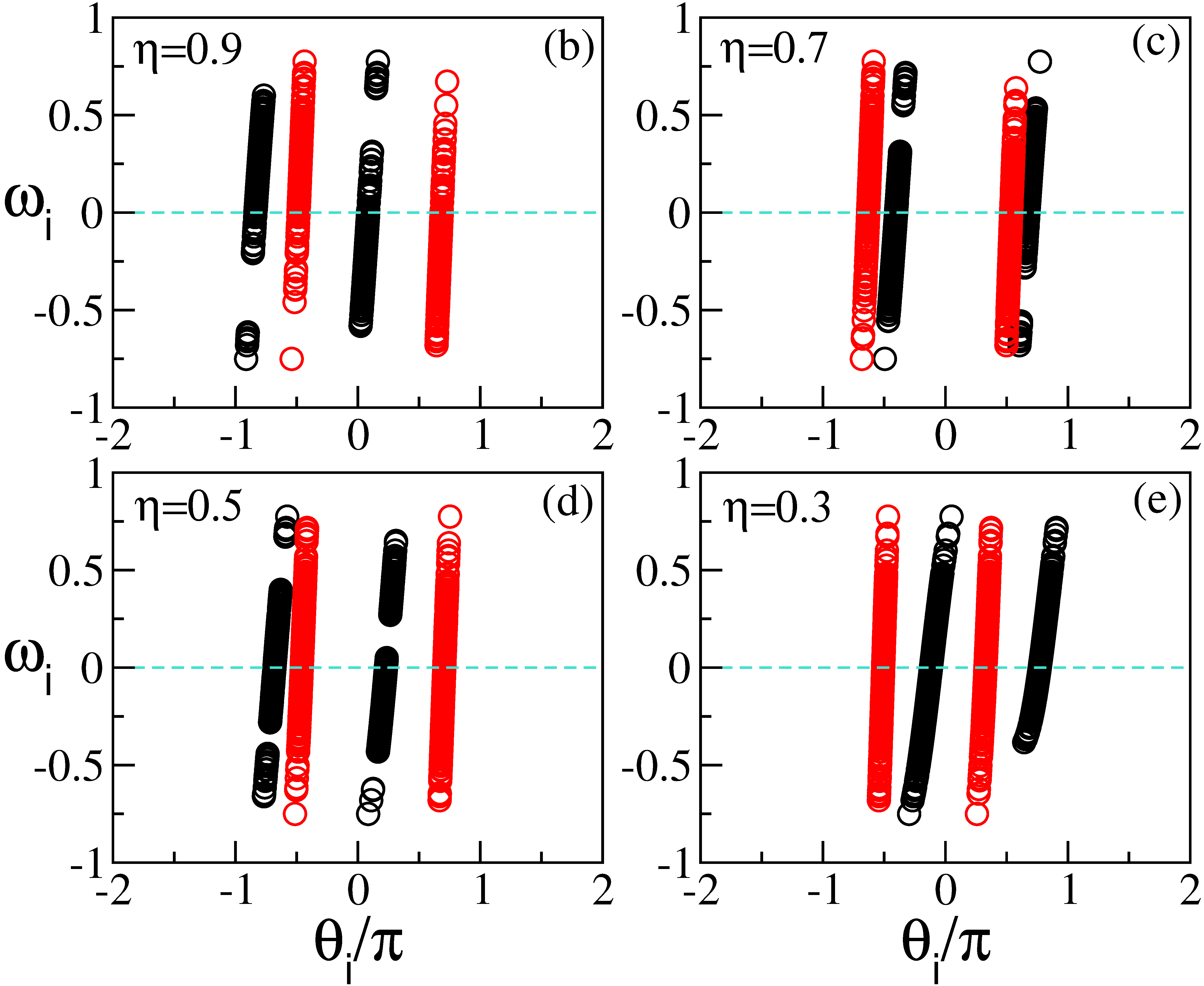}
 \caption{The $\pi$ transition. (a) $R_1$ versus $\lambda_1$ 
 for different initial $\eta$ values (i.e. different percentage of initially synchronized oscillators): the black dots, blue and magenta triangles and orange stars refer to simulations performed by following protocol (II A). The value reported for $R_1$ represents the average in time over a simulation time $T_s$. The cyan line refers to the simulations performed by following protocol (I A) and already shown in Fig. \ref{fig:uni_neg}. In panels (b)-(e) are reported snapshots of the natural frequencies $\omega_i$ versus their phases $\theta_i$ for different initial
 values of $\eta$, for simulations obtained by following protocols (II A). In all panels black dots correspond to $\lambda_1=2.2$, red dots to $\lambda_1=6$.
 The displayed results refer to $\lambda_2=-100$, $\Delta=0.25$ for $g_U(\omega)$, $\Delta \lambda_1 = 0.2$ $T_t = 10$, $T_s=200$, integration step $dt=0.001$ for the adiabatic simulations, and  a network size $N=10000$.}
    \label{fig:uni_neg_vari_eta}
\end{figure}

\subsubsection{The $\pi$-transition: dependence on $\eta$}

The same analysis to the one reported in the previous sub-section for protocol (IIA), has been performed here by varying the different percentage $\eta$ of initially phase-locked oscillators. In particular we have calculated the time average value of the order parameter $R_1$ by following protocol (II A), here adapted to take into account different initial states where only a percentage $\eta$ of the oscillators' phases are set to the same value. The results of these simulations, shown in Fig. \ref{fig:uni_neg_vari_eta} (a), reveal that, irrespectively of the chosen $\eta$ value,  the order parameter $R_1$ decreases smoothly with the control parameter and then, almost at the same critical value, it presents an abrupt jump to a higher value. By decreasing $\eta$ the jump
occurs at smaller and smaller $\lambda^{(\pi)}$-values while, at the same time, the entity of the jump in $R_1$ decreases.
Finally the asynchronous regime is smoothly achieved at a $\lambda_1^{(PS)}$ value that increases for decreasing $\eta$.
 From the numerical data, we can infer that, in the vanishing $\eta$ limit, the curve will become continuous and 
 coincident with the one obtained by following the protocol (I A), shown as a dashed cyan line in Fig. \ref{fig:uni_neg_vari_eta} (a).  
 An interesting aspects is that, for $\eta < 1$, the two cluster states
 are always composed by oscillators with positive and negative $\{\omega_i\}$ before and after the transition,
 as shown in Fig. \ref{fig:uni_neg_vari_eta} (b-e). The main difference that we observe in the 2 clusters before and after
 the $\pi-$transition is that, once they reform, the distribution of the $\{\omega_i\}$ in each cluster does not cover any
 more a compact interval, but there are {\it holes} in the support of the distributions.
 
It should be remarked that all the discontinuous transitions observed for finite $\eta$ are associated to
the fact that the two clusters reach an anti-phase configuration, as we have verified by inspecting the
time evolution of $R_1$ and $R_B$ at the transition value.

\subsection{Bimodal Distribution of the Natural Frequencies}
\label{bimodal}

We will now examine the bimodal case, where the natural frequencies are
distributed by following $g_B(\omega)$ with $\omega_0=1$ and $\Delta=0.25$.  
Also in this case we will compare numerical simulation results for 
protocols (I A) and (II A) with mean-field results obtained through the self-consistent approach.

\subsubsection{Positive $\lambda_2$}

\begin{figure*}
\centering
    \includegraphics[width=0.34\linewidth, height=0.24\linewidth]{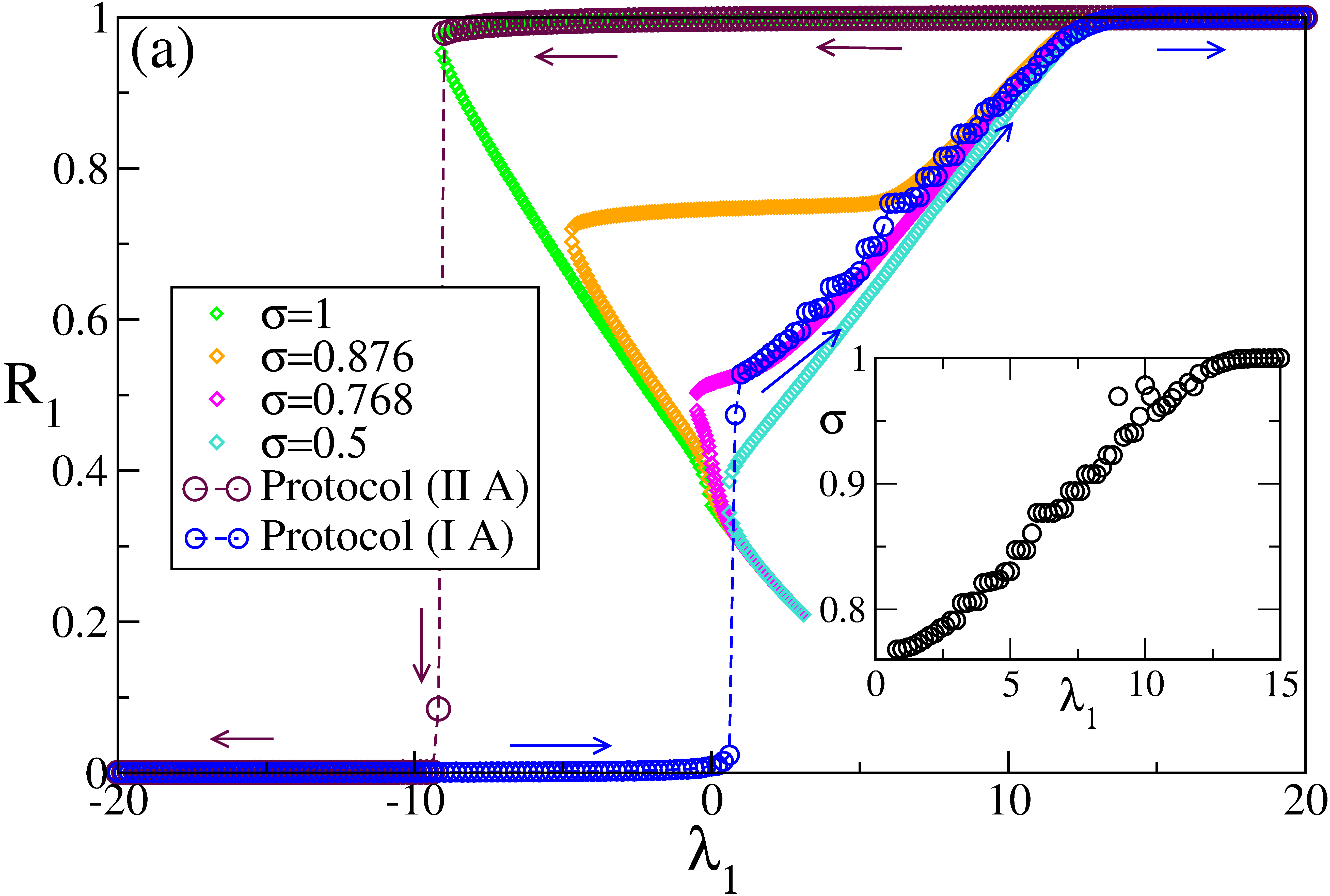}
    \includegraphics[width=0.32\linewidth]{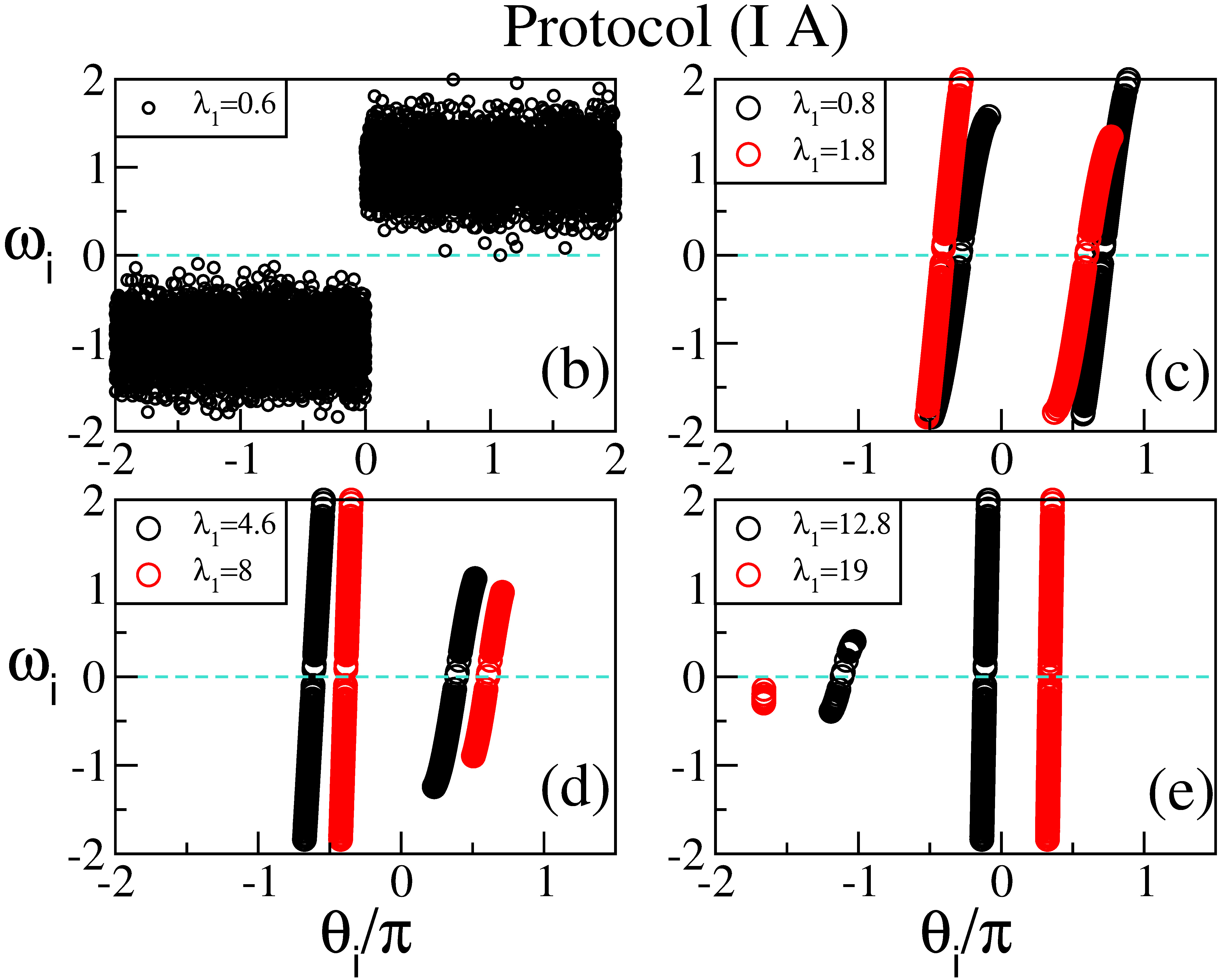}
    \includegraphics[width=0.28\linewidth, height=0.24\linewidth]{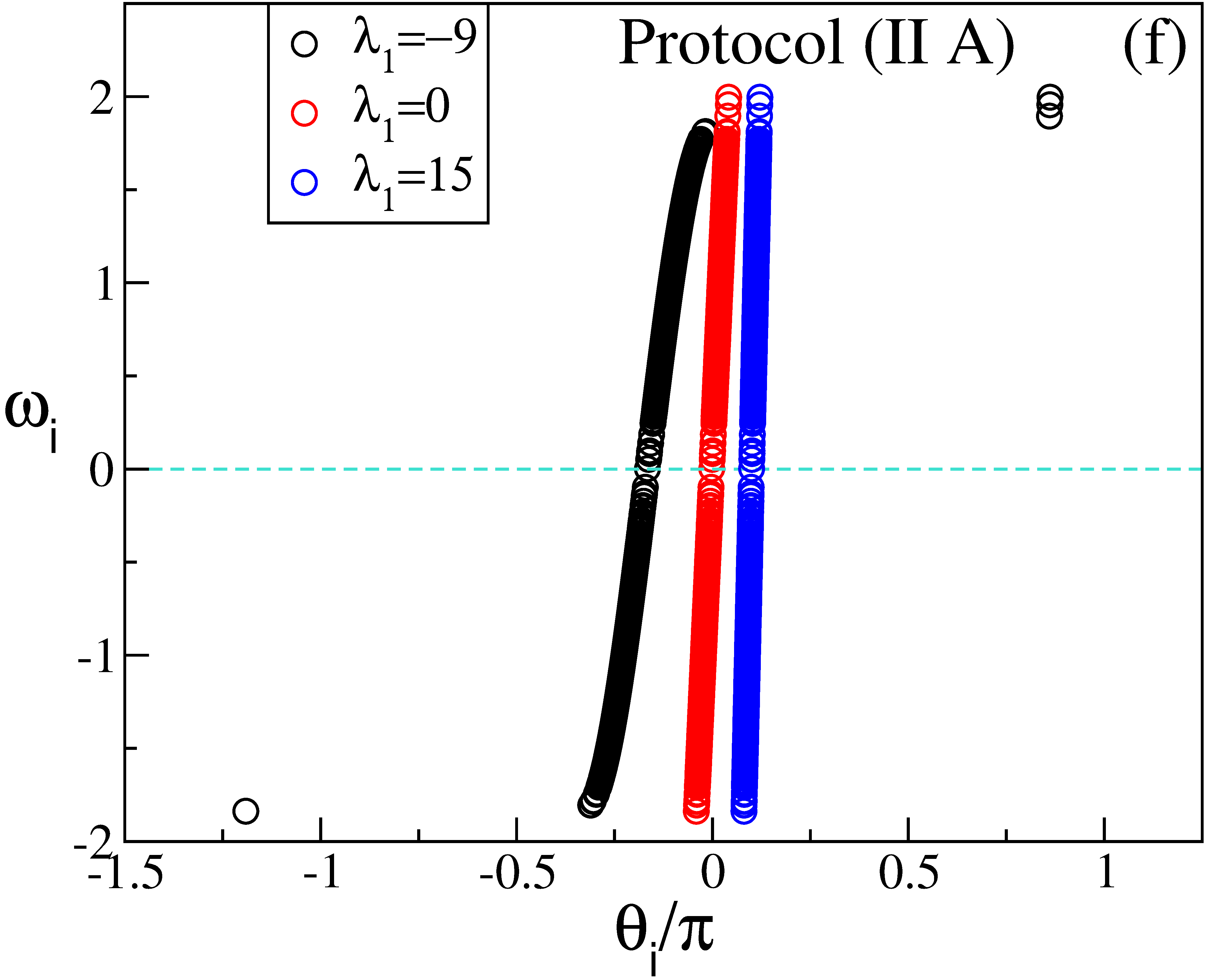}
 \caption{Natural frequencies distributed according to a bimodal Gaussian distribution and $\lambda_2 > 0$.
   (a) $R_1$ versus $\lambda_1$ : the blue and maroon circles refer to simulations performed by following protocols (I A) and (II A), respectively. The value reported for $R_1$ represents the average in time over a simulation time $T_s$. The diamonds of different colours denote the self consistent evaluation of the order parameter $R_1$, for different $\sigma$ values, obtained by employing \eqref{R_pos}. In the inset it is shown $\sigma$ versus $\lambda_1$ as estimated by following protocol (I A). 
   In panels (b-e) are reported snapshots of the natural frequencies $\omega_i$ versus their phases $\theta_i$ for different
 values of $\lambda_1$ for simulations obtained by following protocol (I A). Analogous snapshots are shown
 in (f) for simulations done by following protocol (II A). The displayed results refer to 
 $\lambda_2=15$, $\omega_0=1$ and $\Delta=0.25$ for $g_B(\omega)$, $\Delta \lambda_1 = 0.2$, $T_t = 10$, $T_s=200$, integration step $dt=0.001$ for the adiabatic simulations, and  a network size $N=10000$.}
    \label{fig:bi_pos}
\end{figure*}

Due to the bimodal frequency distribution, the asynchronous regime is characterized by two groups of oscillators rotating in opposite directions. On one hand, in the unimodal framework, the oscillators synchronize around the common frequency $\Omega=0$, which corresponds to the most probable natural frequency value, and form a phase cluster that smoothly grows  with $\lambda_1$. On the other hand, here we pass from 2 groups of oscillators rotating with opposite frequencies around $\pm \omega_0$ to locked oscillators with $\Omega =0$ divided in two phase clusters. 
Therefore, we expect the partial synchronization to emerge via a discontinuous transition for a coupling $\lambda_1^{(AS)}$ larger than in the unimodal case \cite{Pazo2009}. Indeed, as shown in Fig. \ref{fig:bi_pos} (a) for $\lambda_2=15$, 
the simulations for $N=10000$ and protocol (I A) show a discontinuous transition at $\lambda_1^{(AS)} \simeq 0.7$ (blue circles in Fig. \ref{fig:bi_pos} (a)). The 2 cluster states formed after the transition find themselves at a distance $\pi$, as in the unimodal case. However they are now quite asymmetric, as evident from the
value of $\sigma$, measured during the simulations and shown in the inset of Fig. \ref{fig:bi_pos} (a). By increasing
$\lambda_1$ the main cluster at $\phi=0$ becomes more and more populated and, for $\lambda_1 \ge \lambda_2 = 15$, all the oscillators reside in such a cluster (see Fig. \ref{fig:bi_pos} (b-e)). Therefore, once the partially synchronized regime is emerged, the scenario turns out to be quite similar to that observed for the unimodal frequency distribution with $\lambda_2 > 0$. Indeed, also in this case a small plateau is observable for $\sigma=0.876$. 

The self-consistent approach estimated by employing Eq. \eqref{R_pos} for $\sigma > 0.5$ reveals, in complete
analogy with the unimodal case, that a partially synchronized regime emerges
via a saddle-node bifurcation at $\lambda_1^{(PS)}(\sigma)$, characterized by a finite value of the order 
parameter $R_1(\sigma) = 2 \sigma - 1$. In this regard we have reported a few of these curves in Fig. \ref{fig:bi_pos} (a).
The solution shown as magenta diamonds, corresponds to $\sigma=0.768$, which is the minimal $\sigma$ value
measured via the simulations in the PS regime. The magenta solution reproduces quite well the numerical simulations 
for $\lambda_1 > 0.7$ up to the complete synchronization, apart for the plateau, which is instead captured by the
self-consistent solution for $\sigma=0.876$ (orange symbols). Finally, the curve for $\sigma=1$
(green symbols) perfectly reproduces the simulations obtained by following protocol (II A) (maroon circles)
at the associated de-synchronization transition. At $\lambda_1 \simeq -9$ the evaporation of few oscillators towards the second minimum results in the reduction of $R_1$, 
with a consequent abrupt system de-synchronization, as shown in Fig. \ref{fig:bi_pos} (e). 

The self-consistent solution for $\sigma=0.5$ (cyan symbols) shows the formation of a PS state via a saddle-node bifurcation
for $\lambda_1 > \lambda_1^{(PS)}(\sigma=0.5) \simeq 0.6$ with a finite value of the order parameter $R_1 \simeq 0.36$.
The value $\lambda_1^{(PS)}(\sigma=0.5)$ can represent a reasonable approximation of the numerically observed transition.
As in the unimodal case, we have a wide interval of $\lambda_1$-values where we can have coexistence
of incoherent and partially synchronized states. Moreover, also in this case, hysteretic synchronization transitions are 
observable.

From the analysis reported in \cite{Martens2009}, one expects the transition from the asynchronous to the partially synchronous regime to be mediated by the emergence of SWs. However, for the present choice of parameters, we do not observe stable SWs.
Indeed for $\omega_0=\pm 1$ and $\Delta=0.25$, SWs are observable only in the range $ \lambda_2 \in [-11.7;2]$, as we will discuss in the following.

\begin{figure*}
\centering
    \includegraphics[width=0.32\linewidth]{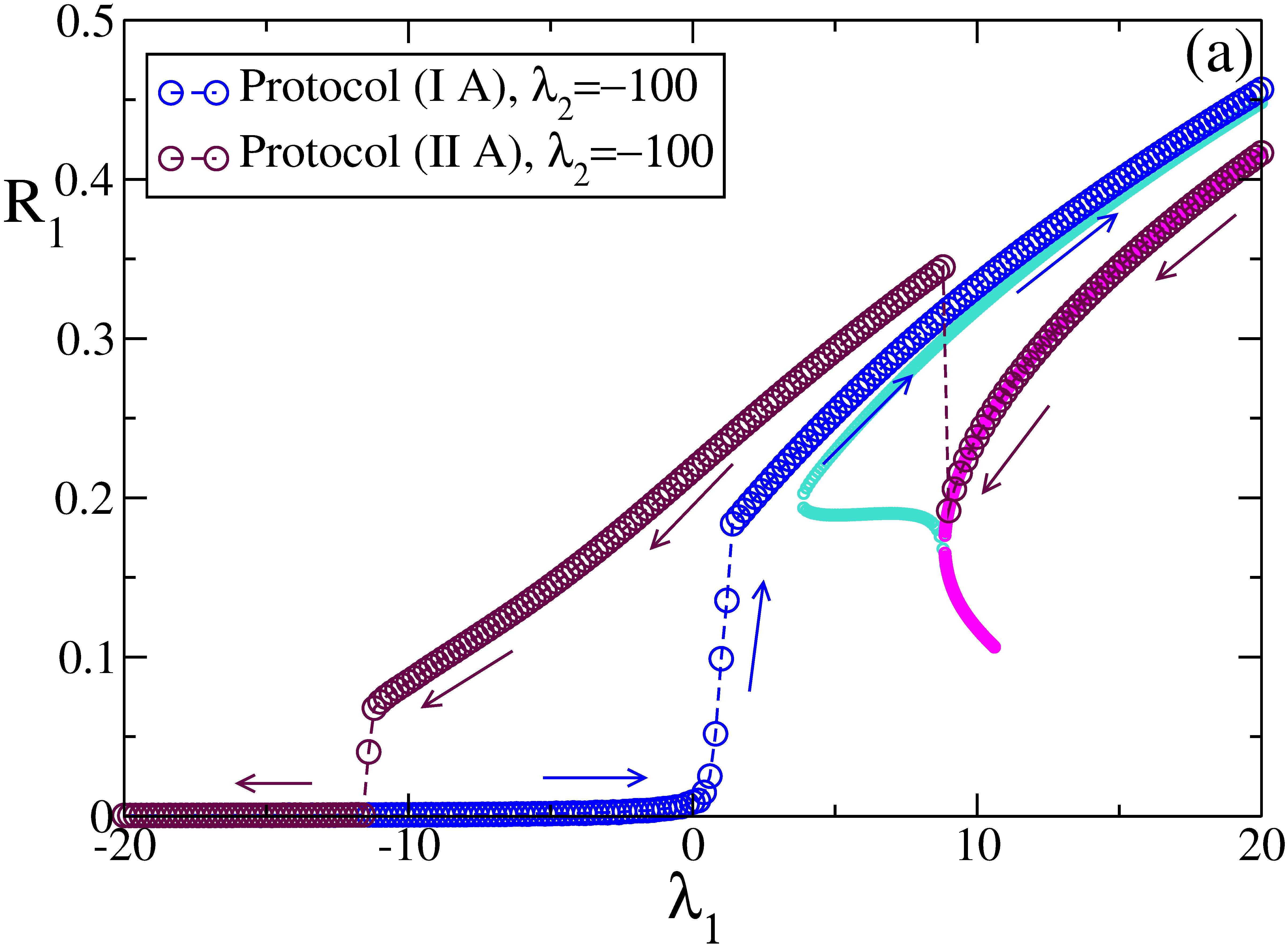}
    \includegraphics[width=0.32\linewidth]{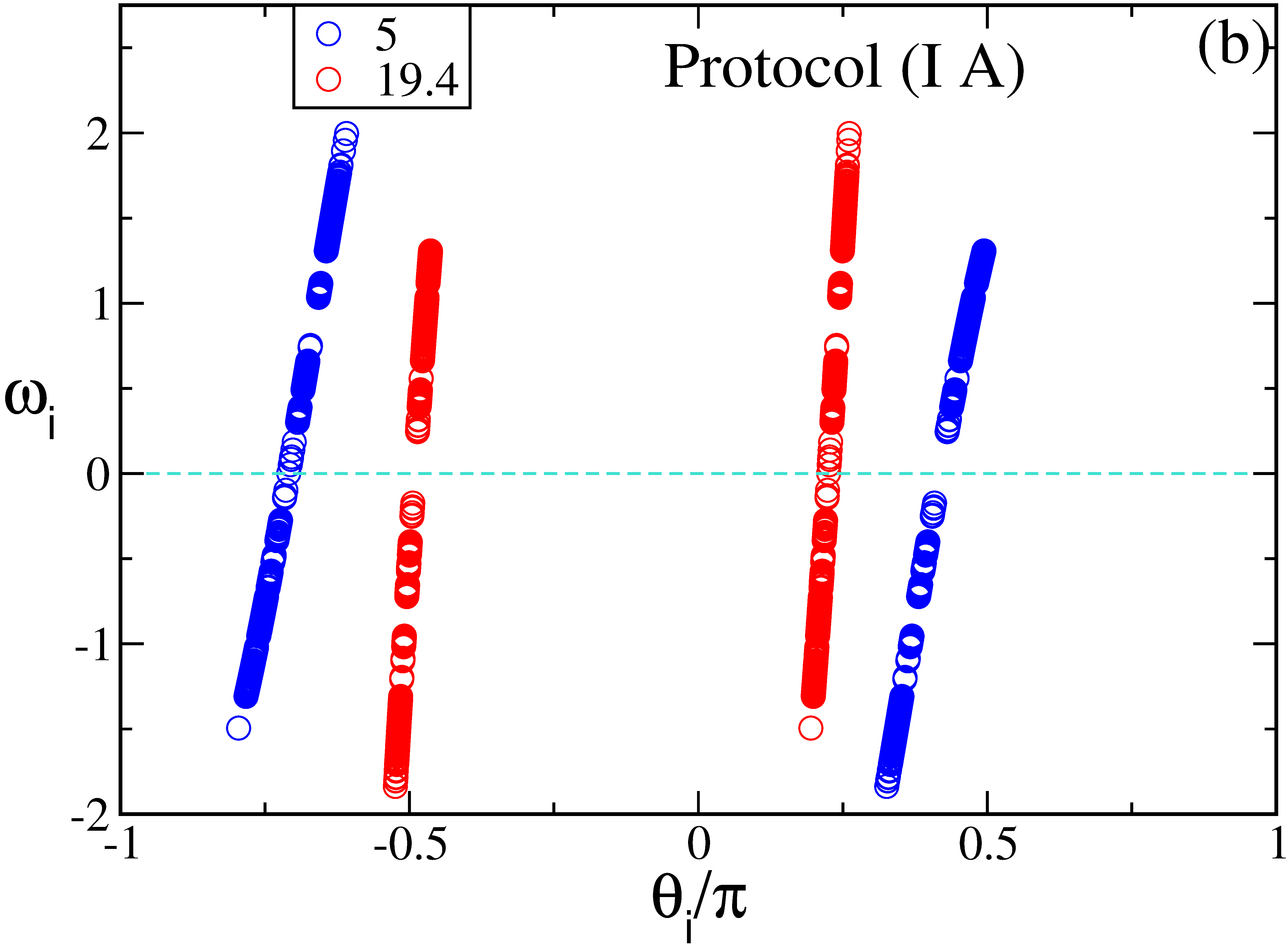}
    \includegraphics[width=0.32\linewidth]{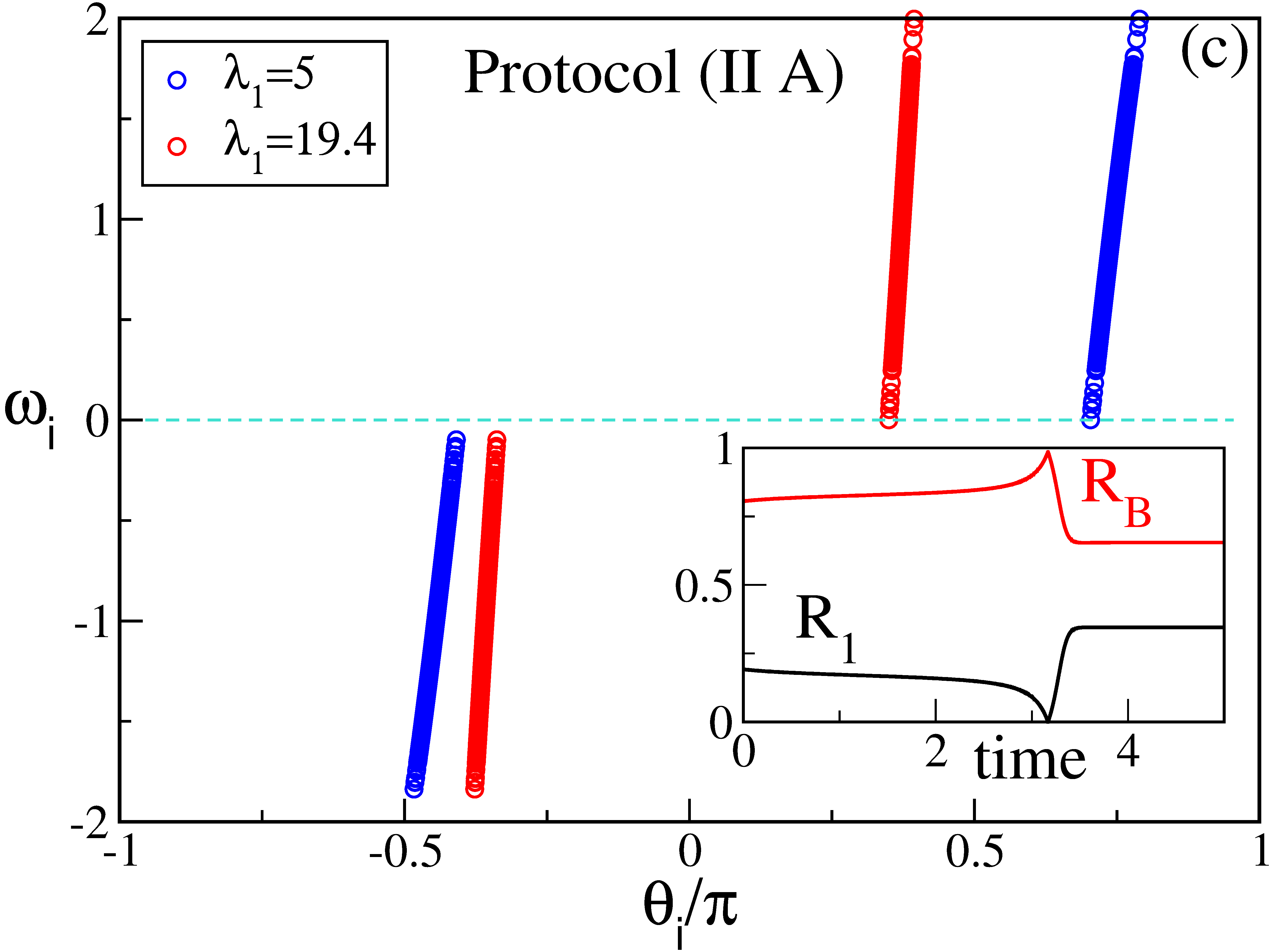}
 \caption{Natural frequencies distributed according to a bimodal Gaussian distribution and $\lambda_2 < 0$. (a) $R_1$ versus $\lambda_1$ : the blue and maroon circles
 refer to simulations performed by following protocols (I A) and (II A), respectively. The value reported for $R_1$ represents the average in time over a simulation
 time $T_s$. The cyan (magenta) line refers to the self consistent evaluation of the order parameter $R_1$ by employing \eqref{R_neg} estimated in the whole range
 of frequencies (restriced to positive frequencies). In panels (b) and (c) are reported snapshots of the natural frequencies $\omega_i$ versus their phases $\theta_i$ for different
 values of $\lambda_1$ for simulations obtained by following protocols (I A) and (II A), respectively. 
In the inset of panel (c) we report the instantaneous values of $R_1(t)$ (black curve) and $R_B (t)$ (red curve) versus time for $\lambda_1=8.8$ obtained
by following protocol (II A). The displayed results refer to $\lambda_2=-100$, $\omega_0=1$ and $\Delta=0.25$ for $g_B(\omega)$, with $\Delta \lambda_1 = 0.2$,  $T_t = 10$, $T_s=200$, integration step $dt=0.001$ for the adiabatic simulations, and  a network size $N=10000$.}
    \label{fig:bi_neg}
\end{figure*}

\subsubsection{Negative $\lambda_2$}

For $\lambda_2=-100$, we observe that, at variance with the unimodal case,
the transition from the asynchronous to the partially synchronized state is discontinuous. In more details, the network simulations for protocol (I A) show that the transition is steep but occurring in a finite interval
$\lambda_1^{(AS)}\in [0.6:1.4]$, probably due to finite size and finite time effects. The self-consistent estimation of the order parameter $R_1$ is obtained by solving  Eq. \eqref{R_neg}. Analogously to the unimodal case we observe that, due to the symmetric shape of the potential, the integral
\eqref{R_neg} does not depend on $\sigma$, therefore we fix $\sigma=0.5$. 
From the numerical simulations the PS emerges at $\lambda_1^{(PS)} \simeq 3.8$. Despite this time the two cluster state is
already observable immediately after the transition, the mean-field results and numerical simulations are in reasonable agreement only for $\lambda_1 > 5$, 
as shown in Fig. \ref{fig:bi_neg} (b).

The analysis of the network simulations performed following
protocol (II A) reveals that (as in the unimodal case) the oscillators
are in a two cluster state characterized by only positive (negative) natural frequencies (see Fig. \ref{fig:bi_neg} (c)).  By estimating  the self-consistent integral
in \eqref{R_neg}, limited to either positive or negative frequencies,
we obtain two branches of solutions that merge in a saddle-node bifurcation at
$\lambda_1^{(\pi)} \simeq 8.80$ (magenta line in Fig. \ref{fig:bi_neg} (a)).
From the direct simulations, by decreasing $\lambda_1$, we observe a very good agreement with the
mean-field results down to the saddle-node bifurcation. As shown in the inset of Fig. \ref{fig:bi_neg} (c),  where
the time evolutions of $R_1$ and $R_B$ are reported at $\lambda_1=8.8$, 
$R_1$ ($R_B$) decreases (increases) in time until it reaches the zero (one) value
corresponding to two clusters in phase opposition (i.e. with an angular distance $\gamma=\pi$).
Immediately after, the clusters rearrange at a different angle distance $\gamma < \pi$.
Each cluster is, in this case, still populated by oscillators with all positive or all negative natural frequencies,
but the system is now characterized by a higher level of synchronization $R_1$ than
before the transition. 
Therefore we can affirm that this is another example of $\pi$-transition, analogous to the one reported
for the unimodal distribution.
By further decreasing $\lambda_1$, the system becomes less and less synchronized until it becomes completely asynchronous at $\lambda_1^{(PS)}\simeq -12$, due to an abrupt transition. Unfortunately, with the employed self-consistent approach we cannot predict the new clusterized state appearing below $\lambda_1^{(\pi)}$ for protocol (II A).  We leave this
analysis for future work. It is important to notice that, for larger $\lambda_2 \ge -35$,
the state formed after the saddle-node bifurcation is made by clusters containing mostly oscillators with 
positive (negative) natural frequencies plus a few oscillators with negative (positive) $\omega_i$, analogously to what
observed for positive $\lambda_2$.

\begin{figure*}
\centering
    \includegraphics[width=0.99\linewidth]{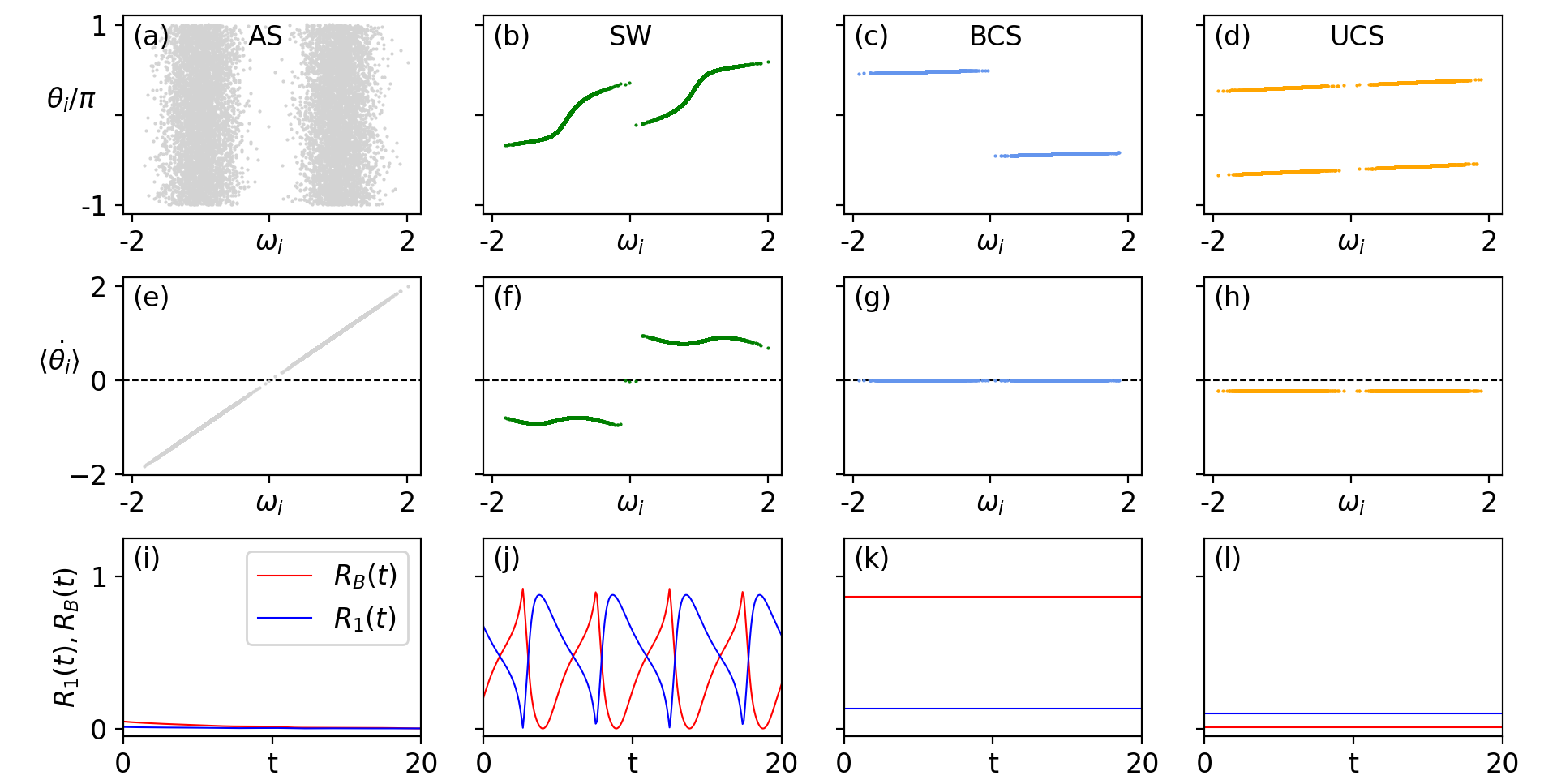}
    \caption{Dynamical regimes: asynchronous state (AS) obtained for ($\lambda_1$,$\lambda_2$)=(-10,-10); standing wave (SW) for ($\lambda_1$,$\lambda_2$)=(3,-5); bimodal cluster (BC) for ($\lambda_1$,$\lambda_2$)=(10,-1000) and uniform cluster (UC) coexisting with the BC for the same parameters. Row-wise, instantaneous phases $\theta_i$ (a-d) and time-averaged phase velocities $\langle \dot \theta_i \rangle$ (e-h) versus the natural frequencies $\omega_i$ of the oscillators, together with the corresponding time evolution of the order parameters $R_1$ and $R_B$ (i-l), for the considered dynamical state. UCs (BCs) are obtained by employing the simulation protocol (II) (protocol (I)). For all cases the bimodal distribution $g_B(\omega)$ with  $\omega_0 = 1.0$ and $\Delta = 0.25$ is considered and the system size is $N=10000$.}
    \label{fig:CS_state}
\end{figure*}

\section{Numerical Investigations}

This Section will be devoted to the numerical analysis of the dynamical regimes observable for the model
\eqref{eq:2} by considering mostly natural frequencies bimodally distributed as $g_B(\omega)$.

\subsection{Dynamical regimes}

In this sub-section we will characterize all the states identified in our model
by direct numerical simulations, apart the fully synchronized regime 
(FS) that corresponds to a group of oscillators locked to the same phase
and traveling with the same angular velocity, for which $R_1=1$.
The other observed states are presented in Fig. \ref{fig:CS_state} and can be classified as follows:
\begin{itemize}

\item{{\it Asynchronous State} (AS)}: as shown in panel (a), this state presents uniformly 
distributed phases, while the oscillators are drifting, driven by their natural frequencies
(e). The order parameters $R_1$ and $R_B$ are essentially zero apart finite size fluctuations (i).
 
\item{{\it Standing Waves} (SWs)}: these solutions correspond to two clusters of
oscillators (b) moving with opposite angular velocity (f), thus inducing oscillations in the order parameters, as shown in panel (j).

\item{{\it Bimodal Clusters} (BCs)}: these clusters are made up of two oscillators groups
located in correspondence to the peaks of the bimodal distribution. Even though each cluster is locked to a different phase value (c), 
they are rotating with the same angular velocity, that, for symmetrically populated clusters, is zero, as shown in panel (g).
The value of $R_1$ is tipically finite and not zero, since
the two clusters are generically not located in phase opposition, while
$R_B$ has a finite value that depends on the phase shift $\gamma$ between the clusters
(k). As we have seen in the previous Section, these solutions emerge by performing simulations with
protocol (II) for negative $\lambda_2$. In addition to this, for not too negative  $\lambda_2$
(e. g. $\lambda_2=-30$) and $\lambda_1 < \lambda_1^{(\pi)}$, are observable clusters composed not only of oscillators associated to
one peak of the distribution, but also a few associated to the other peak.

\item{{\it Uniform Clusters} (UCs)}: they are constituted by two groups of 
oscillators phase locked at two different phases, where the oscillators
of each cluster can have both positive and negative natural frequencies (d). 
Furthermore the two groups are usually not equally populated
and this can give rise to a finite common angular velocity (h).
Finally $R_B$ is essentially zero, since in this case it is not possible
to identify two clusters on the basis of their natural frequencies, while $R_1$
is usually small but not zero since the angle $\gamma$ between the clusters is
tipically not $\pi$ (l). These solutions usually emerge by performing
simulations with protocol (I), as seen in the previous Section.

\end{itemize}

The typical solutions that emerge due to the bimodal Gaussian distribution are
the SWs, observable also in the usual Kuramoto model in the absence of many body interactions 
\citep{Martens2009}, while the UCs and BCs have been already observed for unimodal distributions
in presence either of a second harmonic or a three body interaction with a sufficiently negative $\lambda_2$.

\subsection{Phase Diagrams for Synchronized and Random Initial States}

We will now numerically investigate the phase diagram of the model \eqref{eq:2} as
a function of the coupling parameters $\lambda_1$ and $\lambda_2$,
for bimodal Gaussian distributions centered at values $\pm \omega_0$ with a fixed distribution width $\Delta = 0.25$. 
In particular, we have compared the results for the unimodal case (corresponding
to $\omega_0=0$) with two bimodal cases (centered in $\omega_0 =\pm 1$ and $\omega_0=\pm 3$ respectively), 
where the overlap of the two Gaussians is definitely negligible for the chosen $\Delta$-value.
Both these distributions satisfy the criterion reported in \cite{Martens2009} to discriminate unimodal from bimodal
distributions: i.e. $\omega_0 > \Delta/\sqrt{3}$.

In order to characterize the macroscopic dynamics of the network we
consider the modulus $R_1$ of the Kuramoto order parameter averaged in time and we measure the level of synchronization in the system
by starting with two different sets of initial conditions corresponding to
protocol (I) and (II), respectively. 

\begin{figure*}
    \centering
	\includegraphics[width=0.99\linewidth]{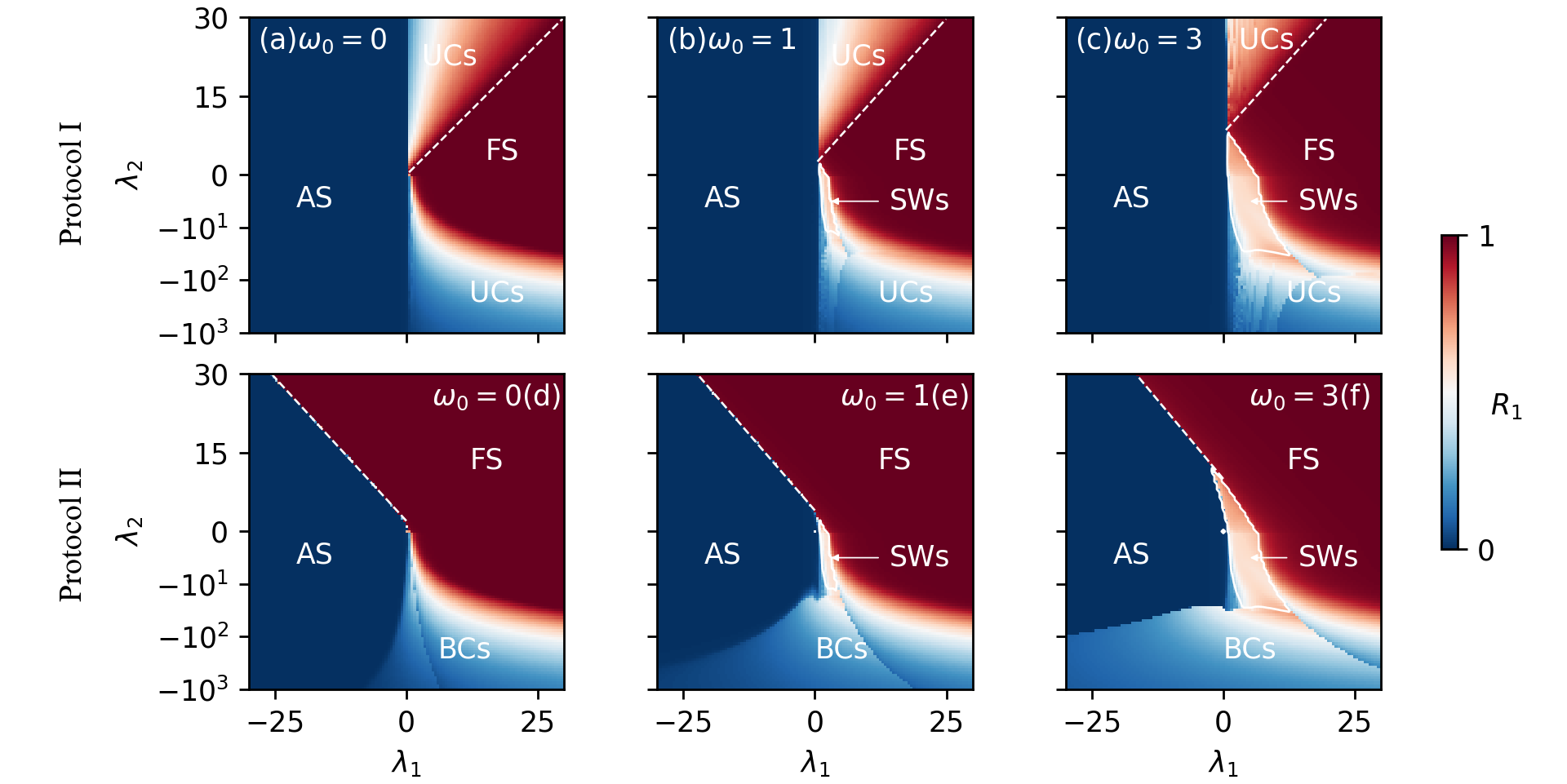}      
\caption{Modulus of the Kuramoto order parameter $R_1$ versus the coupling parameters
$\lambda_1$ and $\lambda_2$ for three different cases : unimodal Gaussian distribution 
of the natural frequencies with $\omega_0=0$  (a, b) and bimodal Gaussian distribution with $\omega_0=1$ (c, d) and $\omega_0 = 3$ (e, f).
In all cases the standard deviation of the Gaussians is fixed to $\Delta=0.25$. The upper (lower) row refers to results obtained
by following the protocol (I) (protocol (II)). The reported abbreviations refer to
different dynamical regimes : asynchronous state (AS); fully synchronous case (FS);
standing waves (SW); bimodal cluster (BC) and uniform cluster (UC).
The white dashed lines mark the onst of the FS regime.
The size of the network has been fixed to $N=5000$ and the parameter $R_1$  has
been averaged  over a time $T_s =200$ once discarding
a transient time $T_t=100$.
     }
    \label{fig:bimodalDiagrams}
\end{figure*}

As a general result, from Fig. \ref{fig:bimodalDiagrams} it is quite evident that asynchronous states (ASs)
are observable for sufficiently negative $\lambda_1$, for both unimodal and bimodal distributions,
while large positive $\lambda_1$ values favour the emergence of fully synchronized regimes (FS).
Furthermore, performing simulations with protocol (I) or (II) leads to large regions of the phase diagram in the plane $(\lambda_1,\lambda_2)$ where coexisting solutions are observable.

For protocol (I) (upper row in Fig. \ref{fig:bimodalDiagrams}) we observe that in general, for the unimodal frequency distribution,
we have a transition from an AS regime  towards a PS one via the emergence of UCs that occurs for $\lambda_1^{(AS)} = \frac{2}{\pi g(0)}$. 
From the mean-field analysis, for sufficiently positive $\lambda_1$,
a FS regime corresponding to all oscillators in the same minimum of the potential $V(\phi)$ at $\phi=0$
is expected to occur for $\lambda_2^{(FS)} = \lambda_1$:
this prediction is well satisfied for the unimodal distribution
(dashed white line in panel (a)). For bimodal distributions
we observe that $\lambda_2^{(FS)} \simeq 1.1 \lambda_1 + C$,
with $C=2$ ($C=8$) for $\omega_0=1$ ($\omega_0=3$).
Therefore, the transition towards the FS regime occurs when
the potential $V(\phi)$ still displays 2 minima.
For negative $\lambda_2$ the transition fo FS become steeper.
Furthermore, for bimodal frequency distributions, the emergence of synchronized states is now mediated by SWs, observable for not too negative, slightly positive $\lambda_2$ and in regions whose size increases for increasing $\omega_0$ values, as expected from \cite{Crawford1994}.
Moreover, the UCs observable for sufficiently positive (negative) $\lambda_2$
are indeed different, since one has not equally populated (equally populated) UCs due to the shape of the potential $V(\phi)$, as discussed in detail in Section \ref{mean}.

For protocol (II) (lower row in Fig. \ref{fig:bimodalDiagrams}) we observe 2 main differences with respect to protocol (I) simulations.
The first one is detectable for positive $\lambda_2$: the FS regime extends to quite negative $\lambda_1$ values where it abruptly de-synchronizes 
for values $(\lambda_2^{(FS)},\lambda_1^{(FS)})$, in agreement with the mean-field results, as shown in the previous section. 
The numerical simulations show a linear dependence of the critical $\lambda_2^{(FS)}$ value for the destabilization of the FS state versus the corresponding $\lambda_1^{(FS)}$, in particular
$\lambda_2^{(FS)} = s \lambda_1^{(FS)} + C$ with $s \simeq 1.1 -1.2$
increasing with the value of $\omega_0$ (white dashed lines in Fig. \ref{fig:bimodalDiagrams}). 

The second difference concerns the emergence of BCs at sufficiently negative $\lambda_2$. As already discussed in
detail in Section \ref{mean}, these states emerge perfectly symmetric at large $\lambda_1$ and destabilize at some smaller $\lambda_1^{(\pi)}$ value of the parameter. 
The $\pi$-transition lines are clearly visible in the
panels (d,e,f), at sufficiently negative $\lambda_2$ and positive $\lambda_1$, as a discontinous variation in $R_1$
leading from a lower to a higher value of the order parameter by decreasing $\lambda_1$. This is due to the fact
that the BCs find themselves in phase opposition at the $\pi$-transition (i.e. with an angular distance $\gamma=\pi$). Immediately after, 
the clusters rearrange at a different angle distance in a configuration characterized by a higher level of synchrony
that can eventually include a few oscillators belonging to the other peak of the bimodal distribution.

\subsection{Clustered regimes}

\begin{figure}
\centering
    \includegraphics[width=0.99\linewidth]{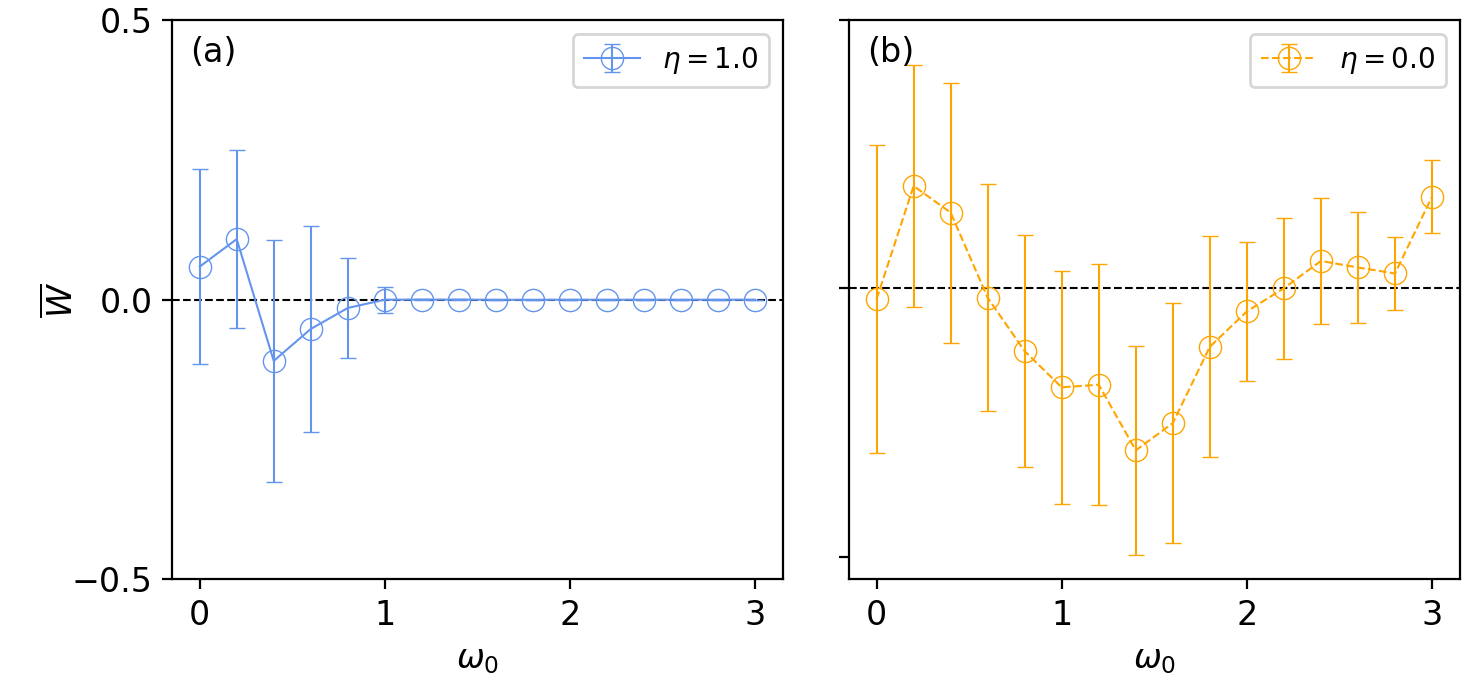}
    \caption{Average angular velocity $W = 1/N\sum^N_{j=1} \langle \dot{\theta_j} \rangle$ versus the distance of the peaks from the center of the distribution. The velocity has been averaged over many different system realizations ($m = 100$): the symbol $\overline{ \cdot}$ refers to this average. The blue points correspond to simulations started from synchronous initial conditions (protocol (II), $\eta = 1$) while the orange points correspond to simulations started from asynchronous ones (protocol (I), $\eta = 0$). For each point, $(\lambda_1,\lambda_2) = (10,-1000)$, $\Delta = 0.25$, $N = 5000$.
   }
    \label{fig:varW0}
\end{figure}

To better characterize the two coexisting cluster states BCs and UCs, we have investigated the
dependence of the average angular velocity $W = \frac{1}{N} \sum_{j=1}^N
\langle  {\dot \theta}_j  \rangle$ on the distance of the peaks  from the center of the frequency distribution.
The BCs emerge when starting with fully synchronized initial conditions (protocol (II)),
as shown in Fig. \ref{fig:varW0} (a). In this case we observe that $W$ becomes essentially zero whenever $\omega_0 \ge 1$,
i.e. whenever the 2 peaks are sufficiently separated. On the other hand, for uniformly
distributed initial phases (corresponding to $\eta=0$ and protocol (I)), the system ends
up in  UCs and $W$ exhibits large oscillations depending on the initial conditions. Moreover $W$
remains finite even for large $\omega_0 \simeq 6 \times \Delta$, see Fig. \ref{fig:varW0} (b).
This behaviour finds its explanation in the fact that, for $\eta=1$, for a sufficiently large separation between the 
peaks, one observes the emergence of two equally populated clusters  with opposite natural
frequencies while, for random initial conditions, the oscillators belong to one of the two clusters irrespectively of the sign of their natural frequency. 
This imbalance is responsible for a finite common angular velocity ${W}$ for all the oscillators.
However, in both cases, all the oscillators are frequency locked, i.e. they can move coherently all together
around the circle or they can be at rest, but these two situations can be considered as identical when examined in a comoving 
reference frame moving with velocity $W$.

\begin{figure}
\centering
    \includegraphics[width=0.99\linewidth]{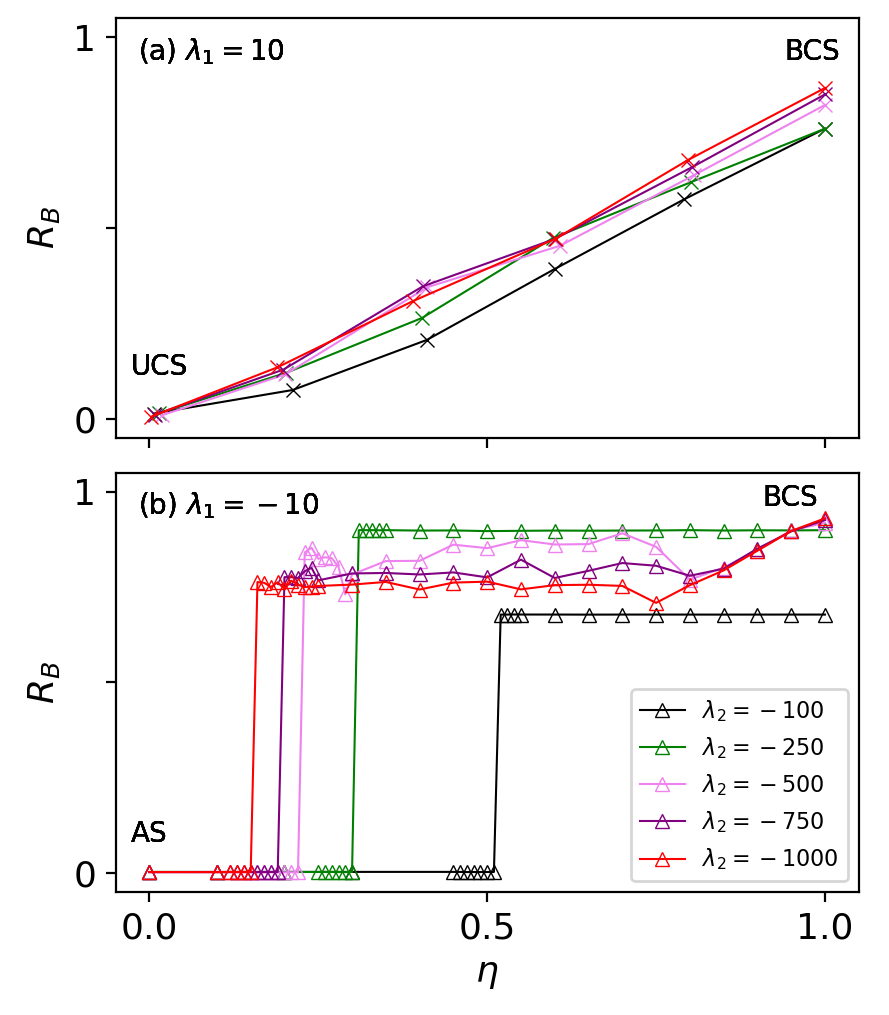}
    \caption{Clustering order parameter $R_B$ for increasing values of the initial degree of phase-locking $\eta$ and for different values of the triadic coupling $\lambda_2 \in [-1000:-100]$. In (a), $\lambda_1=10$. The extreme points correspond to the panels (c) and (d) of Figure \ref{fig:CS_state}. In (b), where $\lambda_1=-10$, we pass from an AS to a BC regime at different critical points 
    $\eta_c$ as a function of the $\lambda_2$ value. Parameters are $\omega_0 = 1.0$, $\Delta = 0.25$ and $N = 5000$.}
    \label{fig:varIC}
\end{figure}

To further explore how the cluster states depend on the initial conditions, we perform several simulations with different initial degree of the phase-locking $\eta$, for definitely negative $\lambda_2$. In this region, for $\lambda_1=10$, we have previously identified UCs for protocol (I) ($\eta=0$) and BCs for protocol (II) ($\eta=1$).
As shown in Fig. \ref{fig:varIC} (a), by sweeping  $\eta$ from $0$ to $1$, the clustering order parameter $R_B$ (Eq. \eqref{eq:Rs}) 
grows from zero to one. Moreover this grow does not depend particularly on $\lambda_2 \in [-1000:-100]$, thus indicating that we pass smoothly from a situation where the phase clusters are not associated
to the frequency distribution of the $\omega_i$ (as observable for the UCs, see panel (d) in Fig. \ref{fig:CS_state})
to a situation where each cluster state involves only oscillators with positive or negative $\omega_i$ (panel (c) in Fig. \ref{fig:CS_state}). By increasing the level of the initial synchronization
of the phase oscillators $\eta$, we favour the emergence of phase clusters consisting of oscillators that
are in prevalence associated to positive or negative $\omega_i$.

Let us now consider negative $\lambda_1$ values. For negative $\lambda_1 = -10$ we have previously identified an AS (BC) regime for $\eta=0$ ($\eta=1$). 
By varying $\eta$ we expect to find a transition between these two regimes, that indeed
occurs for the considered negative $\lambda_2$ values, as show in Fig. \ref{fig:varIC} (b).
The transition is abrupt and it occurs at some $\eta_c >0$: this value decreases
noticeably when increasing $\lambda_2$. Therefore this parameter has a de-stabilising effect
on the asyncronous regime, favouring the emergence of the BCs, since the depth of the two symmetric minima
in the potential increases with $|\lambda_2|$ (see Fig. \ref{fig:potential} (b)).

\begin{figure*}
\centering
	\includegraphics[width=0.99\linewidth]{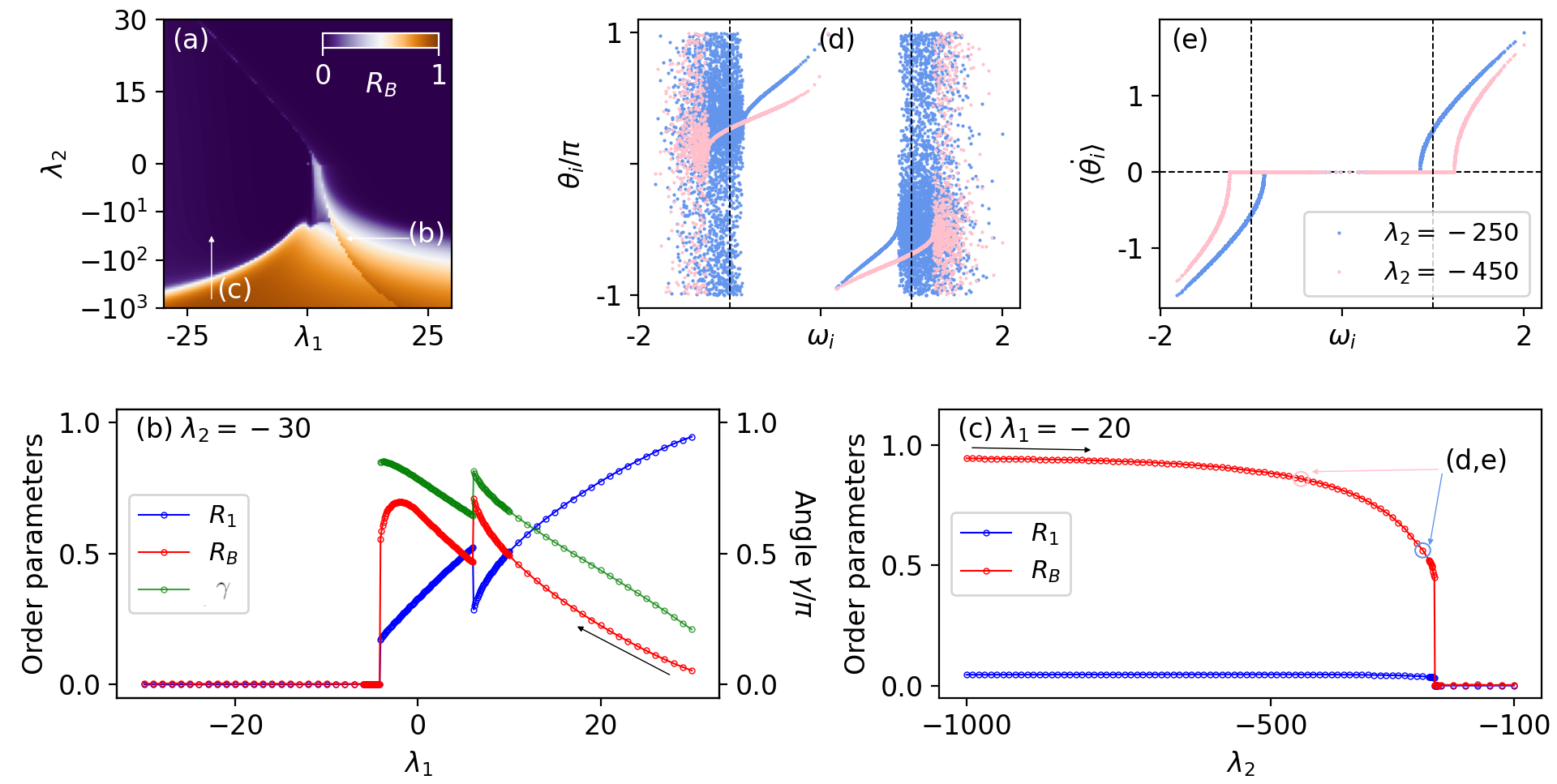}		
    \caption{Numerical characterization of the $\pi$-transition and of the BC state. In (a), the order parameter $R_B$ is a function of the coupling strengths $\lambda_1$ and $\lambda_2$, only for the simulation protocol (II A). The white arrows give the sense and direction to the de-synchronization transitions shown in panels (b) and (c), obtained by varying quasi-adiabatically the coupling parameters. 
    The value reported for $R_1, R_B, \gamma$ in panels (b) and (c) represent the average in time over a simulation time $T_s$. The angle $\gamma$ is rescaled by a factor $\pi$ to plot all variables on the same scale. For (b) $\lambda_2 = -30$, while for (c) $\lambda_1 = -20$. The pink and blue circles highlighted in panel (c) refer to ($\lambda_1$, $\lambda_2$)=(-20, -450) and ($\lambda_1$, $\lambda_2$)=(-20, -250) respectively. Using the same colorcode (pink and blue), these states are analyzed in panels (d,e) in terms of their instantaneous phases and averaged velocities. The snapshot presented in (d) is related to the instantaneous phases against their natural frequencies. In (e) the time-averaged velocity of each oscillator is reported against its natural frequency. For all panels, the bimodal parameters were fixed at $\omega_0 = 1$ and $\Delta = 0.25$.
  }
    \label{fig:R2}
\end{figure*}

As a final aspect, we wish to better characterize the emergence of BCs in terms of the clustering order parameter $R_B$
and the angle $\gamma$ between the phase clusters, introduced in the Sub-section \ref{coh}. As already shown,
the BCs emerge for protocol (II A) simulations, therefore we limit to these ones.
In particular, we have redone the phase diagram reported in Fig. \ref{fig:bimodalDiagrams} (e) by employing the clustering order
parameter $R_B$, as shown in Fig. \ref{fig:R2} (a). This panel reveals that, across the negative semi-plane of $\lambda_2$, we have indeed a very strong level of cluster-synchronicity associated to BC states and that, despite the presence of the SN bifurcation line (still clearly visible), the BCs remain well defined before and after traversing such a bifurcation line. 
 The white arrows drawn in the diagram tell the position and sense of the cuts shown in panels (b) and (c), where we display the 
 de-synchronization transitions obtained via quasi-adibatic simulations by varying $\lambda_1$ and $\lambda_2$, respectively. 
As a matter of fact, we started with clusterized initial conditions with $\eta = 1$ and varied the coupling parameters adiabatically, in contrast to the simulations shown in panel (a) where the initial conditions are newly generated for each pixel. 

In panel (b) we report $R_1$, $R_B$ and $\gamma$ versus $\lambda_1$ for the case $\lambda_2=-30$. This simulation 
corresponds to the protocol (II A) previously examined. We observe that, while $R_1$ decreases for decreasing $\lambda_1$, the parameter
$R_B$ increases, thus indicating that the unique cluster present at $\lambda_1=20$, corresponding to $\gamma=0$, begins
to split in two for decreasing $\lambda_1$, as indicated by the growing value of $\gamma$. This grows is abruptly interrupted
at $\lambda_1^{(\pi)}$ when the two clusters reach the phase opposition, corresponding to $\gamma=\pi$. Immediately after, they rearrange in a configuration with a smaller angle $\gamma$, while
the new BC state is characterized by a higher $R_1$ and a smaller  $R_B$. By further decreasing $\lambda_1$ the angle tends to increase
as $R_B$, while $R_1$ decreases. The BC state destroys abruptly at some negative $\lambda_1$ value and the system becomes asynchronous. 
This latter transition occurs also when one maintains $\lambda_1$ constant and
increases $\lambda_2$, as suggested by the vertical white arrow in panel (a). This transition is reported in more detail
in panel (c).  As $\lambda_2$ increases, the clusters begin to de-synchronize slowly, thus $R_2$ decreases monotonically, until a critical value is reached and both clusters disappear abuptly. An interesting remark is that the oscillators that are dropping out from the clusters are not random, but they are the ones with the natural frequencies furthest from the center of the distribution. 
This can be verified looking both at the snapshots of the phases and the time-averaged velocities for two different $\lambda_2$ values, corresponding to the points circled in pink and blue in panel (c), along the knee of the transition. Specifically, to the  $\lambda_2$ values highlighted in pink and blue in panel (c), correspond the dynamical behaviors shown in panel (d, e), where the color code is maintained:
the snapshots of the phases (the average velocities) are reported in panel (d) ((e)). From these figures it is clear that only the oscillators with natural frequencies in proximity of $\omega_i=0$ are synchronized and their number decrease by approaching the transition towards the incoherent state.

\section{Summary and Outlook}

In this work, we have considered a higher-order version of the Kuramoto model combining the usual pair-wise first harmonic
interaction term with a symmetric three body interaction for unimodal and bimodal Gaussian distributions of the 
natural frequencies, with a particular emphasis on the latter distribution.
We have characterized, via a self-consistent approach, the different synchronization (de-synchronization) transitions occurring by increasing (decreasing)
the parameter $\lambda_1$ controlling the pair-wise interactions, starting from de-synchronized (fully synchronized) 
initial conditions. To better understand the origin of the transitions at the mean-field level we have analyzed the dynamics of a reference oscillator as that of an overdamped particle moving in a potential landscape $V(\phi)$ \eqref{eq:pot}.
Depending on the sign of the parameter $\lambda_2$ controlling the higher order interactions, the potential has 
striking different characteristics: for positive $\lambda_2$ the potential shows the existence of two 
minima with different depth (controlled by $\lambda_1$), corresponding to two clusters in anti-phase;
for negative $\lambda_2$ we have two symmetric minima at a distance $\gamma$ in phase, which is controlled by
$\lambda_1$. 

For unimodal Gaussian distribution and positive $\lambda_2$ the scenario of the possible 
transitions is reported in Fig. \ref{fig:uni_pos}.  In particular, we observe that the incoherent
state looses stability via a super-critical Hopf bifurcation by giving rise to two asymmetric clusters at the same critical coupling found for the usual Kuramoto model \cite{Kuramoto2012}. The 2 clusters are usually
not equally populated, tipically a fraction $\sigma$ of the oscillators will be in one minimum and
a fraction $1 -  \sigma$ in the other one.
In this context the FS regime, corresponding to all the oscillators located in the main minimum of the potential ($\sigma=1$),
remains stable even for very negative $\lambda_1$ values and it looses stability abruptly exhibiting a discontinous
transition to the incoherent state at $\lambda_1^{(FS)}$. 
In the mean-field context, such transition is expected to occur at $\lambda_1^{(FS)}  > -\lambda_2$,
whenever the minimum of $V(\phi)$, corresponding to $\phi=0$, disappears. In a wide range of parameters
we observe the coexistence between the incoherent regime and the coherent states characterized by 
different values of $\sigma$. The synchronization transition is clearly hysteretic in this case.

As shown in Fig. \ref{fig:uni_neg},
for negative $\lambda_2$ and unimodal Gaussian distribution, we observe that the incoherent regime looses
stability via a smooth transition  giving rise to two equally populated
phase clusters composed by oscillators with positive and negative natural frequencies,
termed Uniform Clusters (UCs). On the other hand, starting from a synchonized initial condition at large $\lambda_1$, a two
cluster state arises with equally populated clusters, each including oscillators with either positive or negative 
natural frequencies: the so-called Bimodal Clusters (BCs). The BCs are stable  for $\lambda_1 > \lambda_1^{(\pi)}$,
and disappear at $\lambda_1^{(\pi)}$ via the so-called $\pi$-transition : a SN bifurcation, where the two clusters reach a perfect phase opposition.
This induces a vanishing of the order parameter as well as of the mean-field potential $V(\phi)$,
thus giving rise to the abrupt transition. Immediately after, a two cluster state is reformed but it cannot be considered anymore a BC, since now the oscillators of the two previous clusters are mixed up.
In particular, in each cluster we have mostly oscillators with $\omega_i$ of a specific sign,
but there are also a few oscillators with $\omega_i$ of opposite signs (see Fig. \ref{fig:uni_neg} (c)).
This new state becomes unstable at a definitely negative $\lambda_1$ value. In this case,
depending on the level of synchronization $\eta$ of the initial conditions, we observe 
cluster states, which present $\pi$-transitions at smaller and smaller $\lambda_1^{(\pi)} >0$ values for decreasing $\eta$, followed by a two cluster state disappearing at $\lambda_1^{(PS)}$ that approaches $\lambda_1^{(AS)}$ for $\eta \to 0$ (as shown in Fig. \ref{fig:uni_neg_vari_eta}). Indeed for $\eta \to 0$ the $\pi$-transition disappears and the continuous transition obtained starting from the incoherent state is recovered, suggesting that no hysteretic behaviour is present in this context.

Whenever we introduce bimodal Gaussian distributions we observe, with respect to the previous scenario, that, on one side,
for sufficiently positive or negative $\lambda_2$, the incoherent state looses stability at $\lambda_1^{(AS)}$ via discontinuous transitions. On the other side,  at intermediate values of $\lambda_2$, 
the transition from the AS to the coherent regime is mediated by the emergence of SWs, observable in a region that enlarges with increasing distance between the two peaks of the bimodal distributions, as displayed in Fig. \ref{fig:bimodalDiagrams}.
On the contrary, the scenario describing the loss of stability of the FS states initialized with $\eta=1$ is essentially identical
to the one observed for the unimodal distribution, with the only difference that, for sufficiently negative $\lambda_2$,
the clusterized state that is reformed, after the $\pi$-transition involving the BC state, has indeed a higher level of coherence, but it can be still identified as a BC (see Fig. \ref{fig:bi_neg} (c)).

In this paper we have put special emphasis on the negative (repulsive) coupling since this situation is still far from being fully explored. This was previously analysed in \cite{Kovalenko2021} for unimodal Gaussian distributions, where the authors observed that, when both couplings become repulsive, synchrony still persists in the system in the form of BC states. Here we have shown that BCs emerge in a much wider region of the negative  half-plane for bimodal Gaussian distributions and that the $\pi$-transition can be better characterized by introducing a new order parameter  $R_B$ (Eq. \eqref{eq:Rs}) that takes into account the level of synchronization within each single cluster. In particular, $R_B$ growths smoothly with the angle $\gamma$ between the two clusters, while
the Kuramoto-Daido order parameter $R_2$, usually employed to characterize a two cluster state, 
does not show a monotonic dependence on $\gamma$, as shown in Fig. \ref{fig:angles}. 
By employing this new indicator we observe that the $\pi$-transition is characterized by a de-synchronization of the system (in terms of $R_1$), which is due to
the fact that the 2 clusters have reached an angular distance $\gamma=\pi$, thus being in anti-phase. 
After the transition, the clusters find themselves at an angular distance $\gamma$ smaller than before the transition. 
The level of synchronization within each cluster can remain unchanged,
what changes abruptly is the angular distance $\gamma$ between the two clusters, and this explains the
jump in $R_1$ at the $\pi$-transition shown in Fig. \ref{fig:R2}.

Another interesting result of our work, is that the negative higher-order interactions render
the SWs observable in a wider range of parameters with respect to the usual Kuramoto model.
We must note that, in \cite{Manoranjani2022}, it has been carried out an analysis similar to the one here reported. In particular in \cite{Manoranjani2022} the authors 
consider four body asymmetric interactions with Lorentzian bimodal distributions of the natural frequencies,
thus being able to obtain analytically the mean-field evolution via the OA approach.
Their main result is that the higher order interactions essentially enlarge the bi-stable regions found in \cite{Martens2009}. 
Nonetheless, our higher order term is quite different from their one; moreover we have analysed repulsive interactions, while in \cite{Manoranjani2022} they limited themselves
to attractive ones. In the future it would be worth  examining how the presence of additional peaks in the
natural frequency distribution would affect the emergence of the SWs, as well as the
scenario here reported. 

Furthermore, another general aspect that we have highlighted by performimg quasi-adiabatic simulations with a simplified version of the original model (Eq. \eqref{eq:simp}), is that, for attractive higher order interactions, the synchronization transitions are hysteretic with a large region of coexistence, while this is not the case for repulsive higher order ones,
as clearly shown in Fig. \ref{fig:simp2}.
 
Recently, the analysis of an {\it enlarged Kuramoto model} encompassing
a first harmonic pair-wise interaction plus symmetric and anti-symmetric
three body interactions has revelead the emergence of collective chaos \cite{leon2022}. In particular, in \cite{leon2022}, the authors have obtained mean-field results
for unimodal Gaussian frequency distributions by employing a truncated Hermite-Fourier decomposition of the oscillator density. 
It would be worth performing an analysis to interpret the observed 
transition to collective chaos in terms of our mean-field potential landscape $V(\phi)$ and to compare the results of the self-consistent approach
with those reported in \cite{leon2022}.

Finally, it would be interesting to combine the usual pair-wise first harmonic interaction term with three body interactions also for the Kuramoto model with inertia,
thus extending the results obtained in \cite{olmi2014, olmi2016}, where it has been investigated in detail the hysteretic transition to synchronization for a network of Kuramoto oscillators with inertia, in presence of the usual pair-wise first harmonic interaction term, for unimodal and bimodal Gaussian distributions of the natural frequencies. 
 

\acknowledgments
We acknowledge useful interactions with  Nina La Miciotta,  Arkady Pikovsky and Diego Paz\'o.
AT received financial support by the Labex MME-DII (Grant No ANR-11-LBX-0023-01), by the ANR Project ERMUNDY (Grant No ANR-18-CE37-0014), and  by CY Generations (Grant No ANR-21-EXES-0008),  all part of the French programme ``Investissements d'Avenir''.    
APM gratefully acknowledges ﬁnancial support by the Spanish Ministerio de Economía y Competitividad and European Regional Development Fund under contract PID2022-138322OB-I00 AEI/FEDER, UE, and by Xunta de Galicia (together with AC) under Research Grant No. 2021-PG036. All the programs acknowledged by APM are co-funded by FEDER (UE). 
\appendix*

\section{Simplified Version of the Model}

\begin{figure*}
\centering
    \includegraphics[width=0.99\linewidth]{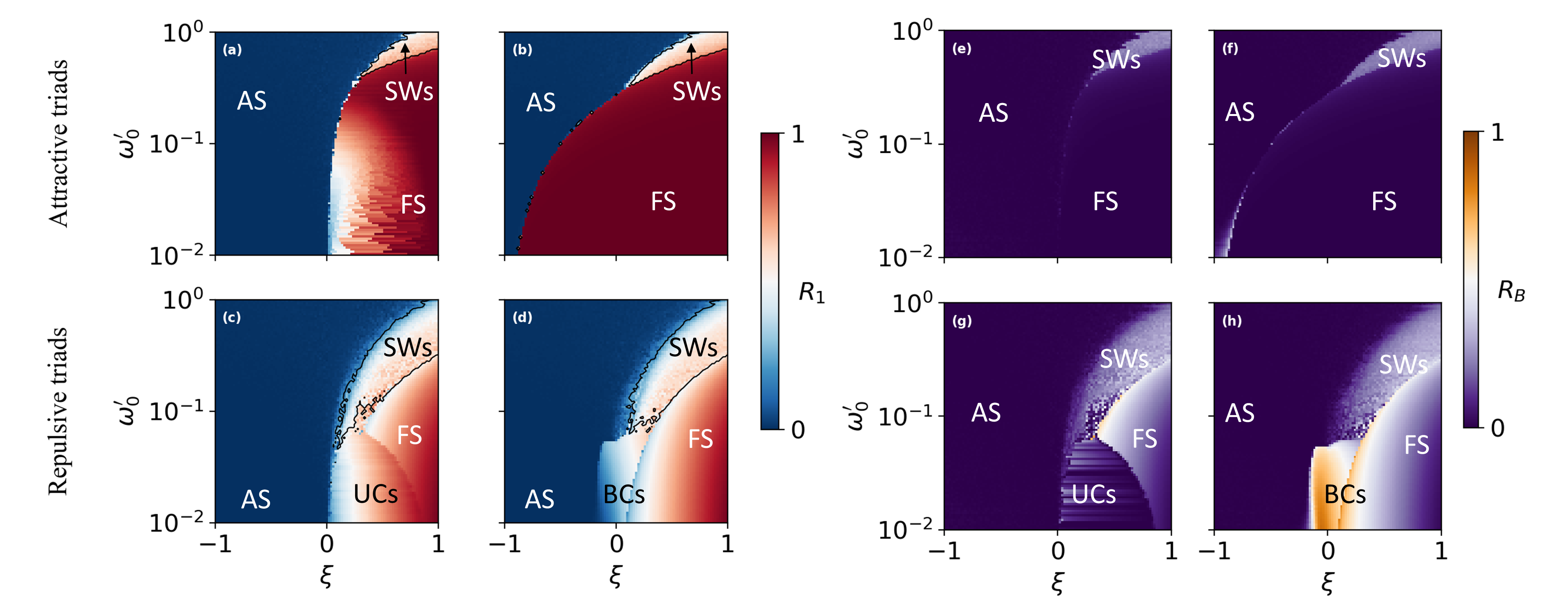}
    \caption{Kuramoto order parameter $R_1$ (a-d) and clustering order parameter $R_B$ (e-h) as a function of the coupling parameter $\xi$ 
    and of the peak position of the bimodal Gaussian distribution $\omega_0'$ in rescaled units.
    Upper row refers to attractive triadic interactions, while the lower row to repulsive ones. These simulations were performed quasi-adiabatically as explained in the text,
    by either increasing $\xi$, as in panels (a,c,e,g), or decreasing $\xi$, as in panels (b,d,f,h).
 For each value of $\omega_0^\prime$, we fixed $\Delta^\prime = 0.25 \omega_0^\prime $. The total number of oscillators is $N=5000$. For every simulation $T_s=200-1000$ and $T_t=100-500$ with modification of $\xi$ with a  step $\Delta \xi = 0.02$ in the quasi-adiabatic simulations.    
    }
    \label{fig:simp1}
\end{figure*} 

In this Appendix, we present a simplified version of the Kuramoto model with higher order interactions and show that it recovers all the transitions and dynamical behaviors that we have reported so far.
In particular, we notice that one of the parameters entering Eq. \eqref{eq:2} 
is actually redundant and the formulation of the model can be simplified by
rescaling the time as ${t}^\prime = t\cdot |\lambda_2|$ (the frequencies
as $\omega_i^\prime = \omega_i /|\lambda_2|$) and by introducing an unique
coupling parameter $\xi = \lambda_1/|\lambda_2|$.
The new version of the model reads as  :
\begin{eqnarray}
&& \frac{d\theta_i(t)}{d{t}^\prime} = \omega_i'+\frac{\xi}{N}\sum^N_{j=1}\sin{(\theta_j-\theta_i)}  \nonumber \\
&+& {\rm sign}(\lambda_2) \frac{1}{2N^2}\sum^N_{j=1}\sum^N_{k=1}\sin{(\theta_j+\theta_k-2\theta_i)} \enskip ,  
    \label{eq:simp}
\end{eqnarray} 
which corresponds to two distinct ordinary differential equations depending on the sign of $\lambda_2$, 
one with attractive higher order interactions (plus sign) and another with repulsive one (minus sign).

\begin{figure}
\centering
    \includegraphics[width=0.99\linewidth]{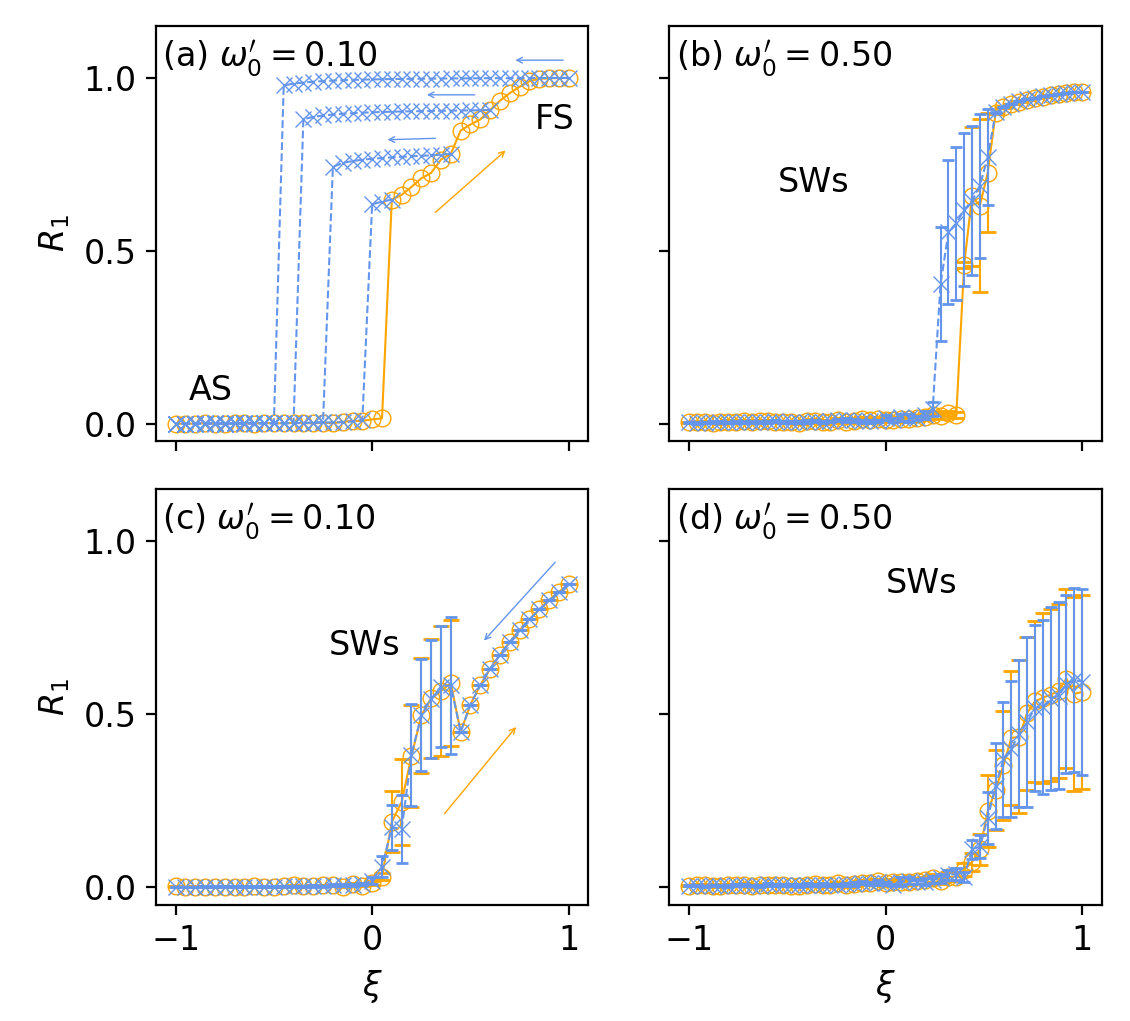}
    \caption{Kuramoto order parameter $R_1$ as a function of the coupling parameter $\xi$, for attractive (a,b) and repulsive (c,d) triadic interactions. 
    The value reported for $R_1$ represents the average in time over a simulation time $T_s$; the standard deviation of $R_1$ during the simulation time is also shown as error bars.
    The re-scaled bimodal distribution frequencies are centered at $\pm\omega_0'=0.1$ (a,c) and $\pm\omega_0'=0.5$ (b,d) with dispersion $\Delta^\prime=0.25 \enskip \omega_0^\prime$ equal for every panel. The simulations are run quasi-adiabatically starting at $\eta = 0$ with a total number of $N=5000$ oscillators with $T_s=200-1000$ and $T_t=100-500$ with $\Delta \xi = 0.02$.}
    \label{fig:simp2}
\end{figure}

A detailed description of the phase diagram of the system according to the actual parameters is shown in Figure \ref{fig:simp1}, where the top panels correspond to attractive three-body interactions and the lower ones to repulsive interactions. In particular in such a figure  are shown the  order parameters $R_1$ (panels a-d) and $R_B$ (e-h) for quasi-adiabatic simulations, where we start at $\eta = 0$ and $\xi=-1.0$ and we increase quasi-adiabatically $\xi$ up to $\xi=1.0$. Then $\xi$ is decreased quasi-adiabatically, starting from the state at
$\xi=1.0$ obtained from the previous simulations, down to $\xi=-1.0$. The parameter is always varied in steps of $\Delta \xi = 0.02$.
The forward synchronizing transitions are shown in panels (a,c,e,g) while the backward de-synchronizing ones are shown in panels (b,d,f,h). Note that $R_1$ and $R_B$ are measured from the same simulations and thus they are complementary. In each panel the dynamical states found with the two-parametric model are tagged, similarly to what done in Figure \ref{fig:bimodalDiagrams}. For the 
acronyms reported in the panels we refer to the states described in the main text.
For each value of the symmetric peaks centers, we re-scaled the standard deviation $\Delta$ accordingly, so that the relation $\omega_0'/\Delta'=4$ is fulfilled. This scaling choice was mainly motivated  in order to
compare the results of this Appendix with the ones presented in the previous sections.

In Figure \ref{fig:simp2} we show two cuts of the two dimensional phase diagrams for $\omega_0^\prime = 0.1$ and $0.5$. Again, top panels account for positive, attractive triadic interactions and lower panels for negative, repulsive ones. 
In (a), for $\omega_0^\prime = 0.1$, we observe a direct abrupt synchronization transition analogous to the one
shown in Fig. \ref{fig:bi_pos} (a). The new feature to highlight here is that, if the backward simulations (obtained by decreasing $\xi$)
are initialized with some clustered state with finite $R_1$ value, this state is maintained until $\xi$ becomes sufficiently negative 
to destroy such a state. The critical negative value of $\xi$ is larger and larger for higher level of synchronization
of the initial state, since the depth of the minimum of the potential where the synchronized state resides depends on $R_1$,
as previously shown. These results indicate that there is a large region of multistability where clusters of different
level of synchronization can coexist. For larger $\omega_0^\prime = 0.5$ the synchronization transition is preceeded
by the emergence of SW, as expected (see panel (b)). The SWs are characterized by a  large value of the
standard deviation of $R_1$, since $R_1$ is now oscillating in time. Also in the present case we have a clear hysteretic
region.

For repulsive triadic interaction, the SW regime occupies a quite large portion of the phase diagram
for sufficiently distant peaks, i.e. sufficiently large $\omega_0^\prime$. This is evident both in
panels (c) and (d). In particular for $\omega_0^\prime = 0.5$, the partially synchronized regimes are 
always characterized by standing waves and we are not able to observe a FS regime. For negative triadic interactions no hysteresis is observable
for sufficiently large $\omega_0^\prime$.

\begin{figure}
\centering
    \includegraphics[width=0.99\linewidth]{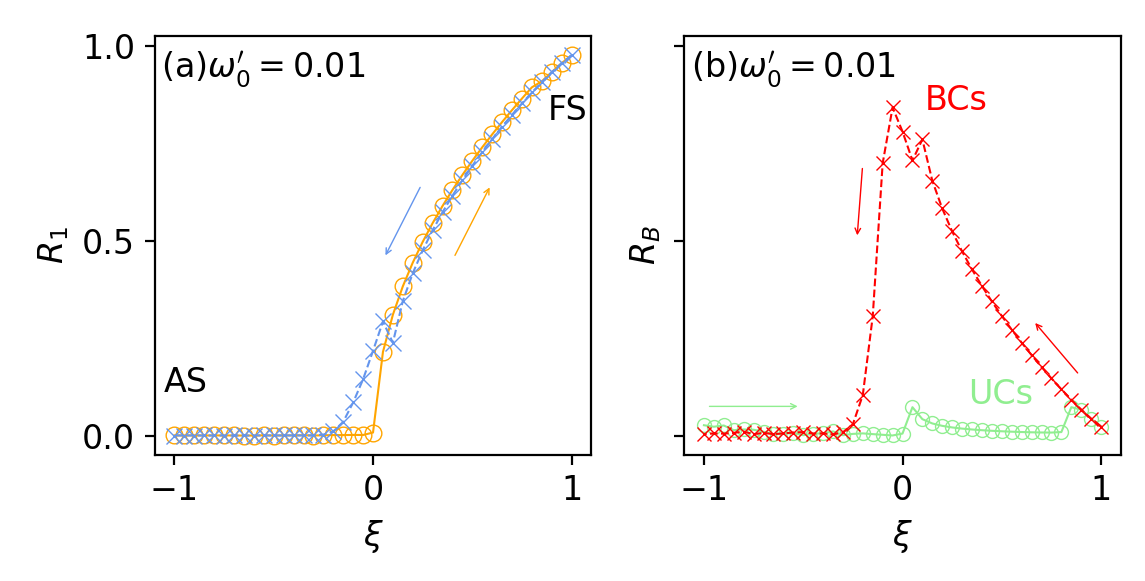}
    \caption{The $\pi$-transition. Kuramoto order parameter $R_1$ (a) and clustering order parameter $R_B$ (b) versus $\xi$
    for repulsive triadic interactions 
    and $\omega_0^\prime = 0.01$. The simulation is run adiabatically starting at $\eta = 0$ with a total number of $N=5000$ oscillators.}
    \label{fig:simp3}
\end{figure}

At lower $\omega_0^\prime$ we can find a coexisting region for the clustered regimes, UCs and BCs, as in the standard model.
This is shown in Fig.  \ref{fig:simp3} for $\omega_0^\prime = 0.01$, where the order parameters
$R_1$ and $R_B$ are reported as a function of $\xi$ for quasi-adiabatic simulations. 
As displayed in  Fig. \ref{fig:simp3} (a),  we observe a smooth transition in $R_1$ by increasing $\xi$, that leads to an UC regime, as it is evident from the value of the clustering order parameter  $R_B$ in panel (b).
In the backward simulations, the BC is clearly present and it de-synchronizes via the $\pi$-transition, before reforming. As shown in panel (b), the order parameter $R_B$ tends to one approaching the $\pi$-transition where the BC de-synchronizes. After the $\pi$-transition, the cluster states are reformed and both $R_1, R_B$ smoothly de-synchronize for decreasing $\xi$-values.

\bibliographystyle{apsrev4-1}
\bibliography{biblio}

\begin{thebibliography}{34}%
\makeatletter
\providecommand \@ifxundefined [1]{%
 \@ifx{#1\undefined}
}%
\providecommand \@ifnum [1]{%
 \ifnum #1\expandafter \@firstoftwo
 \else \expandafter \@secondoftwo
 \fi
}%
\providecommand \@ifx [1]{%
 \ifx #1\expandafter \@firstoftwo
 \else \expandafter \@secondoftwo
 \fi
}%
\providecommand \natexlab [1]{#1}%
\providecommand \enquote  [1]{``#1''}%
\providecommand \bibnamefont  [1]{#1}%
\providecommand \bibfnamefont [1]{#1}%
\providecommand \citenamefont [1]{#1}%
\providecommand \href@noop [0]{\@secondoftwo}%
\providecommand \href [0]{\begingroup \@sanitize@url \@href}%
\providecommand \@href[1]{\@@startlink{#1}\@@href}%
\providecommand \@@href[1]{\endgroup#1\@@endlink}%
\providecommand \@sanitize@url [0]{\catcode `\\12\catcode `\$12\catcode
  `\&12\catcode `\#12\catcode `\^12\catcode `\_12\catcode `\%12\relax}%
\providecommand \@@startlink[1]{}%
\providecommand \@@endlink[0]{}%
\providecommand \url  [0]{\begingroup\@sanitize@url \@url }%
\providecommand \@url [1]{\endgroup\@href {#1}{\urlprefix }}%
\providecommand \urlprefix  [0]{URL }%
\providecommand \Eprint [0]{\href }%
\providecommand \doibase [0]{http://dx.doi.org/}%
\providecommand \selectlanguage [0]{\@gobble}%
\providecommand \bibinfo  [0]{\@secondoftwo}%
\providecommand \bibfield  [0]{\@secondoftwo}%
\providecommand \translation [1]{[#1]}%
\providecommand \BibitemOpen [0]{}%
\providecommand \bibitemStop [0]{}%
\providecommand \bibitemNoStop [0]{.\EOS\space}%
\providecommand \EOS [0]{\spacefactor3000\relax}%
\providecommand \BibitemShut  [1]{\csname bibitem#1\endcsname}%
\let\auto@bib@innerbib\@empty
\bibitem [{\citenamefont {Kuramoto}(2012)}]{Kuramoto2012}%
  \BibitemOpen
  \bibfield  {author} {\bibinfo {author} {\bibfnamefont {Y.}~\bibnamefont
  {Kuramoto}},\ }\href
  {https://www.ebook.de/de/product/33663784/y_kuramoto_chemical_oscillations_waves_and_turbulence.html}
  {\emph {\bibinfo {title} {Chemical Oscillations, Waves, and Turbulence}}}\
  (\bibinfo  {publisher} {Springer Berlin Heidelberg},\ \bibinfo {year}
  {2012})\BibitemShut {NoStop}%
\bibitem [{\citenamefont {Winfree}(1967)}]{Winfree1967}%
  \BibitemOpen
  \bibfield  {author} {\bibinfo {author} {\bibfnamefont {A.~T.}\ \bibnamefont
  {Winfree}},\ }\href {\doibase 10.1016/0022-5193(67)90051-3} {\bibfield
  {journal} {\bibinfo  {journal} {Journal of Theoretical Biology}\ }\textbf
  {\bibinfo {volume} {16}},\ \bibinfo {pages} {15} (\bibinfo {year}
  {1967})}\BibitemShut {NoStop}%
\bibitem [{\citenamefont {Buck}(1988)}]{Buck1988}%
  \BibitemOpen
  \bibfield  {author} {\bibinfo {author} {\bibfnamefont {J.}~\bibnamefont
  {Buck}},\ }\href {\doibase 10.1086/415929} {\bibfield  {journal} {\bibinfo
  {journal} {The Quarterly Review of Biology}\ }\textbf {\bibinfo {volume}
  {63}},\ \bibinfo {pages} {265} (\bibinfo {year} {1988})}\BibitemShut
  {NoStop}%
\bibitem [{\citenamefont {White}\ \emph {et~al.}(1998)\citenamefont {White},
  \citenamefont {Chow}, \citenamefont {Rit}, \citenamefont {Soto-Treviño},\
  and\ \citenamefont {Kopell}}]{White1998}%
  \BibitemOpen
  \bibfield  {author} {\bibinfo {author} {\bibfnamefont {J.~A.}\ \bibnamefont
  {White}}, \bibinfo {author} {\bibfnamefont {C.~C.}\ \bibnamefont {Chow}},
  \bibinfo {author} {\bibfnamefont {J.}~\bibnamefont {Rit}}, \bibinfo {author}
  {\bibfnamefont {C.}~\bibnamefont {Soto-Treviño}}, \ and\ \bibinfo {author}
  {\bibfnamefont {N.}~\bibnamefont {Kopell}},\ }\href {\doibase
  10.1023/A:1008841325921} {\bibfield  {journal} {\bibinfo  {journal} {Journal
  of Computational Neuroscience}\ }\textbf {\bibinfo {volume} {5}},\ \bibinfo
  {pages} {5} (\bibinfo {year} {1998})}\BibitemShut {NoStop}%
\bibitem [{\citenamefont {Waters}\ and\ \citenamefont
  {Bassler}(2005)}]{Waters2005}%
  \BibitemOpen
  \bibfield  {author} {\bibinfo {author} {\bibfnamefont {C.~M.}\ \bibnamefont
  {Waters}}\ and\ \bibinfo {author} {\bibfnamefont {B.~L.}\ \bibnamefont
  {Bassler}},\ }\href {\doibase 10.1146/annurev.cellbio.21.012704.131001}
  {\bibfield  {journal} {\bibinfo  {journal} {Annual Review of Cell and
  Developmental Biology}\ }\textbf {\bibinfo {volume} {21}},\ \bibinfo {pages}
  {319} (\bibinfo {year} {2005})}\BibitemShut {NoStop}%
\bibitem [{\citenamefont {Néda}\ \emph {et~al.}(2000)\citenamefont {Néda},
  \citenamefont {Ravasz}, \citenamefont {Brechet}, \citenamefont {Vicsek},\
  and\ \citenamefont {Barabási}}]{Neda2000}%
  \BibitemOpen
  \bibfield  {author} {\bibinfo {author} {\bibfnamefont {Z.}~\bibnamefont
  {Néda}}, \bibinfo {author} {\bibfnamefont {E.}~\bibnamefont {Ravasz}},
  \bibinfo {author} {\bibfnamefont {Y.}~\bibnamefont {Brechet}}, \bibinfo
  {author} {\bibfnamefont {T.}~\bibnamefont {Vicsek}}, \ and\ \bibinfo {author}
  {\bibfnamefont {A.-L.}\ \bibnamefont {Barabási}},\ }\href {\doibase
  10.1038/35002660} {\bibfield  {journal} {\bibinfo  {journal} {Nature}\
  }\textbf {\bibinfo {volume} {403}},\ \bibinfo {pages} {849} (\bibinfo {year}
  {2000})}\BibitemShut {NoStop}%
\bibitem [{\citenamefont {Acebrón}\ \emph {et~al.}(2005)\citenamefont
  {Acebrón}, \citenamefont {Bonilla}, \citenamefont {Pérez~Vicente},
  \citenamefont {Ritort},\ and\ \citenamefont {Spigler}}]{Acebron2005}%
  \BibitemOpen
  \bibfield  {author} {\bibinfo {author} {\bibfnamefont {J.~A.}\ \bibnamefont
  {Acebrón}}, \bibinfo {author} {\bibfnamefont {L.~L.}\ \bibnamefont
  {Bonilla}}, \bibinfo {author} {\bibfnamefont {C.~J.}\ \bibnamefont
  {Pérez~Vicente}}, \bibinfo {author} {\bibfnamefont {F.}~\bibnamefont
  {Ritort}}, \ and\ \bibinfo {author} {\bibfnamefont {R.}~\bibnamefont
  {Spigler}},\ }\href {\doibase 10.1103/RevModPhys.77.137} {\bibfield
  {journal} {\bibinfo  {journal} {Reviews of Modern Physics}\ }\textbf
  {\bibinfo {volume} {77}},\ \bibinfo {pages} {137} (\bibinfo {year}
  {2005})}\BibitemShut {NoStop}%
\bibitem [{\citenamefont {Arenas}\ \emph {et~al.}(2008)\citenamefont {Arenas},
  \citenamefont {Díaz-Guilera}, \citenamefont {Kurths}, \citenamefont
  {Moreno},\ and\ \citenamefont {Zhou}}]{Arenas2008}%
  \BibitemOpen
  \bibfield  {author} {\bibinfo {author} {\bibfnamefont {A.}~\bibnamefont
  {Arenas}}, \bibinfo {author} {\bibfnamefont {A.}~\bibnamefont
  {Díaz-Guilera}}, \bibinfo {author} {\bibfnamefont {J.}~\bibnamefont
  {Kurths}}, \bibinfo {author} {\bibfnamefont {Y.}~\bibnamefont {Moreno}}, \
  and\ \bibinfo {author} {\bibfnamefont {C.}~\bibnamefont {Zhou}},\ }\href
  {\doibase 10.1016/j.physrep.2008.09.002} {\bibfield  {journal} {\bibinfo
  {journal} {Physics Reports}\ }\textbf {\bibinfo {volume} {469}},\ \bibinfo
  {pages} {93} (\bibinfo {year} {2008})}\BibitemShut {NoStop}%
\bibitem [{\citenamefont {Ott}\ and\ \citenamefont {Antonsen}(2008)}]{ott2008}%
  \BibitemOpen
  \bibfield  {author} {\bibinfo {author} {\bibfnamefont {E.}~\bibnamefont
  {Ott}}\ and\ \bibinfo {author} {\bibfnamefont {T.~M.}\ \bibnamefont
  {Antonsen}},\ }\href@noop {} {\bibfield  {journal} {\bibinfo  {journal}
  {Chaos: An Interdisciplinary Journal of Nonlinear Science}\ }\textbf
  {\bibinfo {volume} {18}},\ \bibinfo {pages} {037113} (\bibinfo {year}
  {2008})}\BibitemShut {NoStop}%
\bibitem [{\citenamefont {Crawford}(1994)}]{Crawford1994}%
  \BibitemOpen
  \bibfield  {author} {\bibinfo {author} {\bibfnamefont {J.~D.}\ \bibnamefont
  {Crawford}},\ }\href {\doibase 10.1007/BF02188217} {\bibfield  {journal}
  {\bibinfo  {journal} {Journal of Statistical Physics}\ }\textbf {\bibinfo
  {volume} {74}},\ \bibinfo {pages} {1047} (\bibinfo {year}
  {1994})}\BibitemShut {NoStop}%
\bibitem [{\citenamefont {Martens}\ \emph {et~al.}(2009)\citenamefont
  {Martens}, \citenamefont {Barreto}, \citenamefont {Strogatz}, \citenamefont
  {Ott}, \citenamefont {So},\ and\ \citenamefont {Antonsen}}]{Martens2009}%
  \BibitemOpen
  \bibfield  {author} {\bibinfo {author} {\bibfnamefont {E.~A.}\ \bibnamefont
  {Martens}}, \bibinfo {author} {\bibfnamefont {E.}~\bibnamefont {Barreto}},
  \bibinfo {author} {\bibfnamefont {S.~H.}\ \bibnamefont {Strogatz}}, \bibinfo
  {author} {\bibfnamefont {E.}~\bibnamefont {Ott}}, \bibinfo {author}
  {\bibfnamefont {P.}~\bibnamefont {So}}, \ and\ \bibinfo {author}
  {\bibfnamefont {T.~M.}\ \bibnamefont {Antonsen}},\ }\href {\doibase
  10.1103/PhysRevE.79.026204} {\bibfield  {journal} {\bibinfo  {journal}
  {Physical Review E}\ }\textbf {\bibinfo {volume} {79}},\ \bibinfo {pages}
  {026204} (\bibinfo {year} {2009})}\BibitemShut {NoStop}%
\bibitem [{\citenamefont {Pazó}\ and\ \citenamefont
  {Montbrió}(2009)}]{Pazo2009}%
  \BibitemOpen
  \bibfield  {author} {\bibinfo {author} {\bibfnamefont {D.}~\bibnamefont
  {Pazó}}\ and\ \bibinfo {author} {\bibfnamefont {E.}~\bibnamefont
  {Montbrió}},\ }\href {\doibase 10.1103/PhysRevE.80.046215} {\bibfield
  {journal} {\bibinfo  {journal} {Physical Review E}\ }\textbf {\bibinfo
  {volume} {80}},\ \bibinfo {pages} {046215} (\bibinfo {year}
  {2009})}\BibitemShut {NoStop}%
\bibitem [{\citenamefont {Bi}\ \emph {et~al.}(2016)\citenamefont {Bi},
  \citenamefont {Hu}, \citenamefont {Boccaletti}, \citenamefont {Wang},
  \citenamefont {Zou}, \citenamefont {Liu},\ and\ \citenamefont
  {Guan}}]{Bi2016}%
  \BibitemOpen
  \bibfield  {author} {\bibinfo {author} {\bibfnamefont {H.}~\bibnamefont
  {Bi}}, \bibinfo {author} {\bibfnamefont {X.}~\bibnamefont {Hu}}, \bibinfo
  {author} {\bibfnamefont {S.}~\bibnamefont {Boccaletti}}, \bibinfo {author}
  {\bibfnamefont {X.}~\bibnamefont {Wang}}, \bibinfo {author} {\bibfnamefont
  {Y.}~\bibnamefont {Zou}}, \bibinfo {author} {\bibfnamefont {Z.}~\bibnamefont
  {Liu}}, \ and\ \bibinfo {author} {\bibfnamefont {S.}~\bibnamefont {Guan}},\
  }\href {\doibase 10.1103/PhysRevLett.117.204101} {\bibfield  {journal}
  {\bibinfo  {journal} {Physical Review Letters}\ }\textbf {\bibinfo {volume}
  {117}},\ \bibinfo {pages} {204101} (\bibinfo {year} {2016})}\BibitemShut
  {NoStop}%
\bibitem [{\citenamefont {Winfree}(1980)}]{winfree1980}%
  \BibitemOpen
  \bibfield  {author} {\bibinfo {author} {\bibfnamefont {A.~T.}\ \bibnamefont
  {Winfree}},\ }\href@noop {} {\emph {\bibinfo {title} {The geometry of
  biological time}}},\ Vol.~\bibinfo {volume} {2}\ (\bibinfo  {publisher}
  {Springer},\ \bibinfo {year} {1980})\BibitemShut {NoStop}%
\bibitem [{\citenamefont {Daido}(1996)}]{daido1996}%
  \BibitemOpen
  \bibfield  {author} {\bibinfo {author} {\bibfnamefont {H.}~\bibnamefont
  {Daido}},\ }\href {\doibase 10.1103/PhysRevLett.77.1406} {\bibfield
  {journal} {\bibinfo  {journal} {Physical Review Letters}\ }\textbf {\bibinfo
  {volume} {77}},\ \bibinfo {pages} {1406} (\bibinfo {year}
  {1996})}\BibitemShut {NoStop}%
\bibitem [{\citenamefont {Battiston}\ \emph {et~al.}(2020)\citenamefont
  {Battiston}, \citenamefont {Cencetti}, \citenamefont {Iacopini},
  \citenamefont {Latora}, \citenamefont {Lucas}, \citenamefont {Patania},
  \citenamefont {Young},\ and\ \citenamefont {Petri}}]{Battiston2020}%
  \BibitemOpen
  \bibfield  {author} {\bibinfo {author} {\bibfnamefont {F.}~\bibnamefont
  {Battiston}}, \bibinfo {author} {\bibfnamefont {G.}~\bibnamefont {Cencetti}},
  \bibinfo {author} {\bibfnamefont {I.}~\bibnamefont {Iacopini}}, \bibinfo
  {author} {\bibfnamefont {V.}~\bibnamefont {Latora}}, \bibinfo {author}
  {\bibfnamefont {M.}~\bibnamefont {Lucas}}, \bibinfo {author} {\bibfnamefont
  {A.}~\bibnamefont {Patania}}, \bibinfo {author} {\bibfnamefont {J.-G.}\
  \bibnamefont {Young}}, \ and\ \bibinfo {author} {\bibfnamefont
  {G.}~\bibnamefont {Petri}},\ }\href {\doibase 10.1016/j.physrep.2020.05.004}
  {\bibfield  {journal} {\bibinfo  {journal} {Physics Reports}\ }\bibinfo
  {series} {Networks beyond pairwise interactions: {Structure} and dynamics},\
  \textbf {\bibinfo {volume} {874}},\ \bibinfo {pages} {1} (\bibinfo {year}
  {2020})}\BibitemShut {NoStop}%
\bibitem [{\citenamefont {Boccaletti}\ \emph {et~al.}(2023)\citenamefont
  {Boccaletti}, \citenamefont {De~Lellis}, \citenamefont {del Genio},
  \citenamefont {Alfaro-Bittner}, \citenamefont {Criado}, \citenamefont
  {Jalan},\ and\ \citenamefont {Romance}}]{boccaletti2023}%
  \BibitemOpen
  \bibfield  {author} {\bibinfo {author} {\bibfnamefont {S.}~\bibnamefont
  {Boccaletti}}, \bibinfo {author} {\bibfnamefont {P.}~\bibnamefont
  {De~Lellis}}, \bibinfo {author} {\bibfnamefont {C.}~\bibnamefont {del
  Genio}}, \bibinfo {author} {\bibfnamefont {K.}~\bibnamefont
  {Alfaro-Bittner}}, \bibinfo {author} {\bibfnamefont {R.}~\bibnamefont
  {Criado}}, \bibinfo {author} {\bibfnamefont {S.}~\bibnamefont {Jalan}}, \
  and\ \bibinfo {author} {\bibfnamefont {M.}~\bibnamefont {Romance}},\
  }\href@noop {} {\bibfield  {journal} {\bibinfo  {journal} {Physics Reports}\
  }\textbf {\bibinfo {volume} {1018}},\ \bibinfo {pages} {1} (\bibinfo {year}
  {2023})}\BibitemShut {NoStop}%
\bibitem [{\citenamefont {Daido}(1995)}]{daido1995}%
  \BibitemOpen
  \bibfield  {author} {\bibinfo {author} {\bibfnamefont {H.}~\bibnamefont
  {Daido}},\ }\href@noop {} {\bibfield  {journal} {\bibinfo  {journal} {Journal
  of Physics A: Mathematical and General}\ }\textbf {\bibinfo {volume} {28}},\
  \bibinfo {pages} {L151} (\bibinfo {year} {1995})}\BibitemShut {NoStop}%
\bibitem [{\citenamefont {Tanaka}\ and\ \citenamefont
  {Aoyagi}(2011)}]{Tanaka2011}%
  \BibitemOpen
  \bibfield  {author} {\bibinfo {author} {\bibfnamefont {T.}~\bibnamefont
  {Tanaka}}\ and\ \bibinfo {author} {\bibfnamefont {T.}~\bibnamefont
  {Aoyagi}},\ }\href {\doibase 10.1103/PhysRevLett.106.224101} {\bibfield
  {journal} {\bibinfo  {journal} {Physical Review Letters}\ }\textbf {\bibinfo
  {volume} {106}},\ \bibinfo {pages} {224101} (\bibinfo {year}
  {2011})}\BibitemShut {NoStop}%
\bibitem [{\citenamefont {Komarov}\ and\ \citenamefont
  {Pikovsky}(2014)}]{komarov2014}%
  \BibitemOpen
  \bibfield  {author} {\bibinfo {author} {\bibfnamefont {M.}~\bibnamefont
  {Komarov}}\ and\ \bibinfo {author} {\bibfnamefont {A.}~\bibnamefont
  {Pikovsky}},\ }\href@noop {} {\bibfield  {journal} {\bibinfo  {journal}
  {Physica D: Nonlinear Phenomena}\ }\textbf {\bibinfo {volume} {289}},\
  \bibinfo {pages} {18} (\bibinfo {year} {2014})}\BibitemShut {NoStop}%
\bibitem [{\citenamefont {Komarov}\ and\ \citenamefont
  {Pikovsky}(2015)}]{komarov2015}%
  \BibitemOpen
  \bibfield  {author} {\bibinfo {author} {\bibfnamefont {M.}~\bibnamefont
  {Komarov}}\ and\ \bibinfo {author} {\bibfnamefont {A.}~\bibnamefont
  {Pikovsky}},\ }\href@noop {} {\bibfield  {journal} {\bibinfo  {journal}
  {Physical Review E}\ }\textbf {\bibinfo {volume} {92}},\ \bibinfo {pages}
  {020901} (\bibinfo {year} {2015})}\BibitemShut {NoStop}%
\bibitem [{\citenamefont {Ashwin}\ and\ \citenamefont
  {Rodrigues}(2016)}]{ashwin2016}%
  \BibitemOpen
  \bibfield  {author} {\bibinfo {author} {\bibfnamefont {P.}~\bibnamefont
  {Ashwin}}\ and\ \bibinfo {author} {\bibfnamefont {A.}~\bibnamefont
  {Rodrigues}},\ }\href@noop {} {\bibfield  {journal} {\bibinfo  {journal}
  {Physica D: Nonlinear Phenomena}\ }\textbf {\bibinfo {volume} {325}},\
  \bibinfo {pages} {14} (\bibinfo {year} {2016})}\BibitemShut {NoStop}%
\bibitem [{\citenamefont {Le{\'o}n}\ and\ \citenamefont
  {Paz{\'o}}(2019)}]{leon2019}%
  \BibitemOpen
  \bibfield  {author} {\bibinfo {author} {\bibfnamefont {I.}~\bibnamefont
  {Le{\'o}n}}\ and\ \bibinfo {author} {\bibfnamefont {D.}~\bibnamefont
  {Paz{\'o}}},\ }\href@noop {} {\bibfield  {journal} {\bibinfo  {journal}
  {Physical Review E}\ }\textbf {\bibinfo {volume} {100}},\ \bibinfo {pages}
  {012211} (\bibinfo {year} {2019})}\BibitemShut {NoStop}%
\bibitem [{\citenamefont {Skardal}\ and\ \citenamefont
  {Arenas}(2020)}]{Skardal2020}%
  \BibitemOpen
  \bibfield  {author} {\bibinfo {author} {\bibfnamefont {P.~S.}\ \bibnamefont
  {Skardal}}\ and\ \bibinfo {author} {\bibfnamefont {A.}~\bibnamefont
  {Arenas}},\ }\href {\doibase 10.1038/s42005-020-00485-0} {\bibfield
  {journal} {\bibinfo  {journal} {Communications Physics}\ }\textbf {\bibinfo
  {volume} {3}},\ \bibinfo {pages} {1} (\bibinfo {year} {2020})}\BibitemShut
  {NoStop}%
\bibitem [{\citenamefont {Skardal}\ and\ \citenamefont
  {Arenas}(2019)}]{Skardal2019}%
  \BibitemOpen
  \bibfield  {author} {\bibinfo {author} {\bibfnamefont {P.~S.}\ \bibnamefont
  {Skardal}}\ and\ \bibinfo {author} {\bibfnamefont {A.}~\bibnamefont
  {Arenas}},\ }\href {\doibase 10.1103/PhysRevLett.122.248301} {\bibfield
  {journal} {\bibinfo  {journal} {Physical Review Letters}\ }\textbf {\bibinfo
  {volume} {122}},\ \bibinfo {pages} {248301} (\bibinfo {year}
  {2019})}\BibitemShut {NoStop}%
\bibitem [{\citenamefont {Xu}\ \emph {et~al.}(2020)\citenamefont {Xu},
  \citenamefont {Wang},\ and\ \citenamefont {Skardal}}]{Xu2020}%
  \BibitemOpen
  \bibfield  {author} {\bibinfo {author} {\bibfnamefont {C.}~\bibnamefont
  {Xu}}, \bibinfo {author} {\bibfnamefont {X.}~\bibnamefont {Wang}}, \ and\
  \bibinfo {author} {\bibfnamefont {P.~S.}\ \bibnamefont {Skardal}},\ }\href
  {\doibase 10.1103/PhysRevResearch.2.023281} {\bibfield  {journal} {\bibinfo
  {journal} {Physical Review Research}\ }\textbf {\bibinfo {volume} {2}},\
  \bibinfo {pages} {023281} (\bibinfo {year} {2020})}\BibitemShut {NoStop}%
\bibitem [{\citenamefont {Dai}\ \emph {et~al.}(2021)\citenamefont {Dai},
  \citenamefont {Kovalenko}, \citenamefont {Molodyk}, \citenamefont {Wang},
  \citenamefont {Li}, \citenamefont {Musatov}, \citenamefont {Raigorodskii},
  \citenamefont {Alfaro-Bittner}, \citenamefont {Cooper}, \citenamefont
  {Bianconi},\ and\ \citenamefont {Boccaletti}}]{Dai2021}%
  \BibitemOpen
  \bibfield  {author} {\bibinfo {author} {\bibfnamefont {X.}~\bibnamefont
  {Dai}}, \bibinfo {author} {\bibfnamefont {K.}~\bibnamefont {Kovalenko}},
  \bibinfo {author} {\bibfnamefont {M.}~\bibnamefont {Molodyk}}, \bibinfo
  {author} {\bibfnamefont {Z.}~\bibnamefont {Wang}}, \bibinfo {author}
  {\bibfnamefont {X.}~\bibnamefont {Li}}, \bibinfo {author} {\bibfnamefont
  {D.}~\bibnamefont {Musatov}}, \bibinfo {author} {\bibfnamefont {A.~M.}\
  \bibnamefont {Raigorodskii}}, \bibinfo {author} {\bibfnamefont
  {K.}~\bibnamefont {Alfaro-Bittner}}, \bibinfo {author} {\bibfnamefont
  {G.~D.}\ \bibnamefont {Cooper}}, \bibinfo {author} {\bibfnamefont
  {G.}~\bibnamefont {Bianconi}}, \ and\ \bibinfo {author} {\bibfnamefont
  {S.}~\bibnamefont {Boccaletti}},\ }\href {\doibase
  10.1016/j.chaos.2021.110888} {\bibfield  {journal} {\bibinfo  {journal}
  {Chaos, Solitons \& Fractals}\ }\textbf {\bibinfo {volume} {146}},\ \bibinfo
  {pages} {110888} (\bibinfo {year} {2021})}\BibitemShut {NoStop}%
\bibitem [{\citenamefont {Kovalenko}\ \emph {et~al.}(2021)\citenamefont
  {Kovalenko}, \citenamefont {Dai}, \citenamefont {Alfaro-Bittner},
  \citenamefont {Raigorodskii}, \citenamefont {Perc},\ and\ \citenamefont
  {Boccaletti}}]{Kovalenko2021}%
  \BibitemOpen
  \bibfield  {author} {\bibinfo {author} {\bibfnamefont {K.}~\bibnamefont
  {Kovalenko}}, \bibinfo {author} {\bibfnamefont {X.}~\bibnamefont {Dai}},
  \bibinfo {author} {\bibfnamefont {K.}~\bibnamefont {Alfaro-Bittner}},
  \bibinfo {author} {\bibfnamefont {A.}~\bibnamefont {Raigorodskii}}, \bibinfo
  {author} {\bibfnamefont {M.}~\bibnamefont {Perc}}, \ and\ \bibinfo {author}
  {\bibfnamefont {S.}~\bibnamefont {Boccaletti}},\ }\href {\doibase
  10.1103/PhysRevLett.127.258301} {\bibfield  {journal} {\bibinfo  {journal}
  {Physical Review Letters}\ }\textbf {\bibinfo {volume} {127}},\ \bibinfo
  {pages} {258301} (\bibinfo {year} {2021})}\BibitemShut {NoStop}%
\bibitem [{\citenamefont {Strogatz}\ and\ \citenamefont
  {Mirollo}(1991)}]{strogatz1991}%
  \BibitemOpen
  \bibfield  {author} {\bibinfo {author} {\bibfnamefont {S.~H.}\ \bibnamefont
  {Strogatz}}\ and\ \bibinfo {author} {\bibfnamefont {R.~E.}\ \bibnamefont
  {Mirollo}},\ }\href@noop {} {\bibfield  {journal} {\bibinfo  {journal}
  {Journal of Statistical Physics}\ }\textbf {\bibinfo {volume} {63}},\
  \bibinfo {pages} {613} (\bibinfo {year} {1991})}\BibitemShut {NoStop}%
\bibitem [{\citenamefont {Olmi}\ \emph {et~al.}(2014)\citenamefont {Olmi},
  \citenamefont {Navas}, \citenamefont {Boccaletti},\ and\ \citenamefont
  {Torcini}}]{olmi2014}%
  \BibitemOpen
  \bibfield  {author} {\bibinfo {author} {\bibfnamefont {S.}~\bibnamefont
  {Olmi}}, \bibinfo {author} {\bibfnamefont {A.}~\bibnamefont {Navas}},
  \bibinfo {author} {\bibfnamefont {S.}~\bibnamefont {Boccaletti}}, \ and\
  \bibinfo {author} {\bibfnamefont {A.}~\bibnamefont {Torcini}},\ }\href@noop
  {} {\bibfield  {journal} {\bibinfo  {journal} {Physical Review E}\ }\textbf
  {\bibinfo {volume} {90}},\ \bibinfo {pages} {042905} (\bibinfo {year}
  {2014})}\BibitemShut {NoStop}%
\bibitem [{\citenamefont {Manoranjani}\ \emph {et~al.}(2022)\citenamefont
  {Manoranjani}, \citenamefont {Gopal}, \citenamefont {Senthilkumar},
  \citenamefont {Chandrasekar},\ and\ \citenamefont
  {Lakshmanan}}]{Manoranjani2022}%
  \BibitemOpen
  \bibfield  {author} {\bibinfo {author} {\bibfnamefont {M.}~\bibnamefont
  {Manoranjani}}, \bibinfo {author} {\bibfnamefont {R.}~\bibnamefont {Gopal}},
  \bibinfo {author} {\bibfnamefont {D.~V.}\ \bibnamefont {Senthilkumar}},
  \bibinfo {author} {\bibfnamefont {V.~K.}\ \bibnamefont {Chandrasekar}}, \
  and\ \bibinfo {author} {\bibfnamefont {M.}~\bibnamefont {Lakshmanan}},\
  }\href {\doibase 10.1103/PhysRevE.105.034307} {\bibfield  {journal} {\bibinfo
   {journal} {Physical Review E}\ }\textbf {\bibinfo {volume} {105}},\ \bibinfo
  {pages} {034307} (\bibinfo {year} {2022})}\BibitemShut {NoStop}%
\bibitem [{\citenamefont {Le{\'o}n}\ and\ \citenamefont
  {Paz{\'o}}(2022)}]{leon2022}%
  \BibitemOpen
  \bibfield  {author} {\bibinfo {author} {\bibfnamefont {I.}~\bibnamefont
  {Le{\'o}n}}\ and\ \bibinfo {author} {\bibfnamefont {D.}~\bibnamefont
  {Paz{\'o}}},\ }\href@noop {} {\bibfield  {journal} {\bibinfo  {journal}
  {Physical Review E}\ }\textbf {\bibinfo {volume} {105}},\ \bibinfo {pages}
  {L042201} (\bibinfo {year} {2022})}\BibitemShut {NoStop}%
\bibitem [{\citenamefont {Olmi}\ and\ \citenamefont
  {Torcini}(2016)}]{olmi2016}%
  \BibitemOpen
  \bibfield  {author} {\bibinfo {author} {\bibfnamefont {S.}~\bibnamefont
  {Olmi}}\ and\ \bibinfo {author} {\bibfnamefont {A.}~\bibnamefont {Torcini}},\
  }in\ \href {https://hal.archives-ouvertes.fr/hal-01264651} {\emph {\bibinfo
  {booktitle} {Control of {Self}-{Organizing} {Nonlinear} {Systems}}}},\
  \bibinfo {series and number} {Understanding {Complex} {Systems}},\ \bibinfo
  {editor} {edited by\ \bibinfo {editor} {\bibfnamefont {E.}~\bibnamefont
  {Schöll}}, \bibinfo {editor} {\bibfnamefont {S.~H.~L.}\ \bibnamefont
  {Klapp}}, \ and\ \bibinfo {editor} {\bibfnamefont {P.}~\bibnamefont
  {Hövel}}}\ (\bibinfo  {publisher} {Springer},\ \bibinfo {year} {2016})\ pp.\
  \bibinfo {pages} {25--45}\BibitemShut {NoStop}%
\bibitem [{\citenamefont {Iván~Le\'on}\ \emph {et~al.}(2023)\citenamefont
  {Iván~Le\'on}, \citenamefont {Muolo}, \citenamefont {Hata},\ and\
  \citenamefont {Nakao}}]{leon2023}%
  \BibitemOpen
  \bibfield  {author} {\bibinfo {author} {\bibfnamefont {I.}~\bibnamefont
  {Iván~Le\'on}}, \bibinfo {author} {\bibfnamefont {R.}~\bibnamefont {Muolo}},
  \bibinfo {author} {\bibfnamefont {S.}~\bibnamefont {Hata}}, \ and\ \bibinfo
  {author} {\bibfnamefont {H.}~\bibnamefont {Nakao}},\ }\href@noop {}
  {\bibfield  {journal} {\bibinfo  {journal} {preprint Arxiv}\ ,\ \bibinfo
  {pages} {2309.09265}} (\bibinfo {year} {2023})}\BibitemShut {NoStop}%
\end{thebibliography}%
\end{document}